\documentclass{jpp}
\usepackage{caption}
\usepackage{subcaption}
\usepackage{handytex/onimage}

\usepackage[makeroom]{cancel}  
\usepackage{color}
\usepackage{graphics}
\usepackage{graphicx}
\usepackage{float}  
\usepackage{epstopdf, epsfig}  
\usepackage{upgreek}  
\usepackage{amsmath}  
\usepackage{amssymb}  
\usepackage{mathrsfs}  
\usepackage[setpagesize=false,colorlinks=true,linkcolor=blue,citecolor=blue]{hyperref}  
\usepackage{cleveref}
\crefformat{section}{\S#2#1#3} 
\crefformat{subsection}{\S#2#1#3}
\crefformat{subsubsection}{\S#2#1#3}

\crefformat{figure}{figure~#2#1#3}
\Crefformat{figure}{Figure~#2#1#3}

\crefrangeformat{figure}{figures #3#1#4--#5#2#6}
\crefmultiformat{figure}{figures~#2#1#3}{ and~#2#1#3}{, #2#1#3}{, and~#2#1#3}

\Crefrangeformat{figure}{Figures #3#1#4--#5#2#6}
\Crefmultiformat{figure}{Figures~#2#1#3}{ and~#2#1#3}{, #2#1#3}{, and~#2#1#3}

\crefformat{equation}{(#2#1#3)}
\Crefformat{equation}{Equation~(#2#1#3)}

\crefrangeformat{equation}{(#3#1#4)--(#5#2#6)}
\crefmultiformat{equation}{(#2#1#3)}{ and~(#2#1#3)}{, (#2#1#3)}{, and~(#2#1#3)}

\Crefrangeformat{equation}{Equations~(#3#1#4)--(#5#2#6)}
\Crefmultiformat{equation}{Equations~(#2#1#3)}{ and~(#2#1#3)}{, (#2#1#3)}{, and~(#2#1#3)}

\usepackage{tikz}
\usepackage{circuitikz}
\usetikzlibrary{calc}
\usetikzlibrary{patterns}
\usetikzlibrary{arrows.meta}
\usetikzlibrary{shapes}
\usetikzlibrary{plotmarks}
\usetikzlibrary{decorations.markings}

\usepackage{pgfplots}
\usepgfplotslibrary{polar}
\usepgflibrary{shapes.geometric}
\usepackage[]{pstricks}

\usepackage{import}

\newcommand{\importpgf}[2]{\begingroup\import{#1}{#2}\endgroup}

\def\closesymbol{\!}

\ifdefined\handydir

\def\rmd{{\rm d}}

\def\rme{{\rm e}}
\def\rmE{{\rm E}}

\def\rmi{{\rm i}}
\def\rmI{{\rm I}}

\def\rmJ{{\rm J}}

\def\rmZ{{\rm Z}}

\else
	
\fi

\ifdefined\closesymbol
	\relax
\else
	\def\closesymbol{\relax}
\fi

\newcommand{\fourier}[1]{\tilde{#1}}  
\renewcommand{\fourier}[1]{#1}  

\ifdefined\laplace
	\renewcommand{\laplace}[1]{\hat{#1}}  
\else
	\newcommand{\laplace}[1]{\hat{#1}}  
\fi

\newcommand{\order}[1]{O\left(#1\right)}
\newcommand{\orderinline}[1]{O(#1)}
\newcommand{\der}[3][\relax]{\frac{\rmd^{#1} #2}{\rmd {#3}^{#1}}}
\newcommand{\re}{\text{Re}}
\newcommand{\im}{\text{Im}}
\newcommand{\abs}[1]{\left\rvert#1\right\rvert}
\newcommand{\absinline}[1]{\rvert#1\rvert}

\newcommand{\reals}{\mathbb{R}}
\newcommand{\integers}{\mathbb{Z}}

\ifdefined\partd
	\renewcommand{\partd}[3][]{\frac{{\partial^{#1} #2}}{{\partial #3}^{#1}}}
\else
	\newcommand{\partd}[3][]{\frac{{\partial^{#1} #2}}{{\partial #3}^{#1}}}
\fi

\ifdefined\vec  
	\renewcommand{\vec}[1]{\boldsymbol{#1}}
\else
	\newcommand{\vec}[1]{\boldsymbol{#1}}
\fi

\newcommand{\uvec}[1]{{\hat{\boldsymbol{#1}}}}
\providecommand{\bcdot}{\vec{\cdot}}

\newcommand{\grad}{{\vec{\nabla}}}
\newcommand{\gradperp}{\grad_\perp}
\ifdefined\div
	\renewcommand{\div}{{\vec{nabla}}}
\else
	\newcommand{\div}{{\vec{nabla}}}
\fi

\ifdefined\vr
	\renewcommand{\vr}{{\vec{r}}}
\else
	\newcommand{\vr}{{\vec{r}}}
\fi

\newcommand{\vk}{{\vec{k}}}

\ifdefined\vv
\renewcommand{\vv}{{\vec{v}}}
\else
\newcommand{\vv}{{\vec{v}}}
\fi

\newcommand{\vkperp}{{\vk_\perp}}
\newcommand{\kperp}{k_\perp}
\newcommand{\kpar}{k_\parallel}
\newcommand{\vperp}{v_\perp}
\newcommand{\vvperp}{\vv_\perp}
\newcommand{\vpar}{v_\parallel}

\newcommand{\intr}[1][3]{{\int \rmd^{#1} \vec{r} \ }}
\newcommand{\intrV}[1][3]{{\int \frac{\rmd^{#1} \vec{r}}{V}}}

\newcommand{\intv}[1][3]{{\int \rmd^{#1} \vec{v} \ }}
\newcommand{\rmdint}[2][]{{\rmd^{#1} #2 \: }}
\newcommand{\intinf}[1]{\int_{-\infty}^{+\infty} \rmd {#1} \ }
\newcommand{\inthalfinf}[1]{\int_{0}^{+\infty} \rmd {#1} \ }

\newcommand{\s}{s}

\newcommand{\vths}[1][\s]{v_{\text{th}{#1}}}
\newcommand{\vthi}{\vths[i]}
\newcommand{\vthe}{\vths[e]}

\newcommand{\rhos}[1][\s]{\rho_{#1}}
\newcommand{\rhoi}{\rhos[i]}
\newcommand{\rhoe}{\rhos[e]}
\newcommand{\ds}[1][\s]{d_{#1}}
\newcommand{\de}{\ds[e]}

\newcommand{\qs}[1][\s]{q_{#1}}
\newcommand{\qi}{\qs[i]}

\newcommand{\ms}[1][\s]{m_{#1}}
\newcommand{\me}{\ms[e]}
\newcommand{\mi}{\ms[i]}

\newcommand{\ns}[1][\s]{n_{#1}}

\newcommand{\betas}[1][\s]{\beta_{#1}}
\newcommand{\betae}{\betas[e]}

\renewcommand{\ne}{\ns[e]}

\renewcommand{\ni}{\ns[i]}

\newcommand{\dns}[1][\s]{\delta n_{#1}}
\newcommand{\dne}{\dns[e]}

\newcommand{\Ts}[1][\s]{T_{#1}}
\newcommand{\Te}{\Ts[e]}
\newcommand{\Ti}{\Ts[i]}

\newcommand{\dTpars}[1][\s]{\delta T_{\parallel {#1}}}
\newcommand{\dTpare}{\dTpars[e]}

\newcommand{\dTperps}[1][\s]{\delta T_{\perp {#1}}}
\newcommand{\dTperpe}{\dTperps[e]}

\newcommand{\dnsk}[1][\s]{\delta n_{#1\vk}}
\newcommand{\dnek}{\dnsk[e]}

\newcommand{\dTparsk}[1][\s]{\delta T_{\parallel {#1}\vk}}
\newcommand{\dTparek}{\dTparsk[e]}

\newcommand{\dTperpsk}[1][\s]{\delta T_{\perp {#1}\vk}}
\newcommand{\dTperpek}{\dTperpsk[e]}

\newcommand{\ps}[1][\s]{p_{#1}}
\newcommand{\pe}{\ps[e]}

\newcommand{\dppars}[1][\s]{\delta p_{\parallel {#1}}}
\newcommand{\dppare}{\dppars[e]}

\newcommand{\fs}{f_{\s}}

\newcommand{\Fs}{F_{\s}}

\newcommand{\dfs}{\delta \closesymbol f_{\s}}

\newcommand{\Fe}{F_{e}}

\newcommand{\Fi}{F_{i}}

\newcommand{\dfskperp}[1][\s]{\delta\closesymbol\fourier{f}_{{#1}\vkperp}}

\newcommand{\kperprhoesq}{\kperp^2\rhoe^2}
\newcommand{\kperprhoesqq}{\kperp^4\rhoe^4}
\newcommand{\kperpdesq}{\kperp^2\de^2}

\newcommand{\vRs}[1][\s]{{\vec{R}_{#1}}}
\newcommand{\vRe}{{\vec{R}_e}}
\newcommand{\vRi}{{\vec{R}_i}}
\newcommand{\hs}[1][\s]{h_{#1}}
\newcommand{\he}{\hs[e]}
\newcommand{\hi}{\hs[i]}
\newcommand{\hsk}[1][\s]{\fourier{h}_{{#1}\vk}}
\newcommand{\hek}{\hsk[e]}

\newcommand{\hskperp}[1][\s]{\fourier{h}_{{#1}\vkperp}}

\newcommand{\es}{\varepsilon_\s}
\newcommand{\mus}{\mu_\s}

\newcommand{\ve}{\vec{v}_{E}}
\newcommand{\vchi}{\vec{v}_{\chi}}
\newcommand{\vchiRs}[1][\s]{\avgRs[#1]{\vec{v}_{\chi}}}
\newcommand{\vds}[1][\s]{\vec{v}_{\text{d}#1}}

\newcommand{\vdi}{\vds[i]}

\newcommand{\Bo}{B_0}

\newcommand{\phipot}{\phi}  
\newcommand{\phipotk}{\fourier{\phipot}_\vk}
\newcommand{\phipotkperp}{\fourier{\phipot}_\vkperp}

\newcommand{\Apar}{A_\parallel}

\newcommand{\dApar}{\delta \closesymbol  A_\parallel}
\newcommand{\dApark}{\delta \closesymbol \fourier{A}_{\parallel\vk}}
\newcommand{\dAparkperp}{\delta \closesymbol \fourier{A}_{\parallel\vkperp}}

\newcommand{\dBpar}{\delta \closesymbol  B_\parallel}
\newcommand{\dBpark}{\delta \closesymbol  \fourier{B}_{\parallel\vk}}
\newcommand{\dBparkperp}{\delta \closesymbol  \fourier{B}_{\parallel\vkperp}}
\newcommand{\dBx}{\delta \closesymbol  B_x}

\newcommand{\dEpar}{\delta \closesymbol  E_\parallel}
\newcommand{\upars}[1][\s]{u_{\parallel{#1}}}

\newcommand{\upare}{\upars[e]}

\newcommand{\uparsk}[1][\s]{u_{\parallel{#1}\vk}}

\newcommand{\uparek}{\uparsk[e]}

\newcommand{\vE}{\vec{E}}
\newcommand{\vB}{\vec{B}}
\newcommand{\ub}{\uvec{b}}

\newcommand{\vJ}{\vec{J}}
\newcommand{\vA}{\vec{A}}

\newcommand{\vdB}{\delta \closesymbol \vec{B}}
\newcommand{\vdBperp}{\delta \closesymbol  \vec{B}_\perp}
\newcommand{\vdJ}{\delta\vec{J}}
\newcommand{\vdA}{\delta \closesymbol \vec{A}}

\newcommand{\exb}{\(\vE\times\vB\)}
\newcommand{\vrhos}[1][\s]{\vec{\rho}_{#1}}

\newcommand{\avgRs}[2][\s]{\closesymbol\left\langle #2 \right\rangle_{\vRs[#1]}}
\newcommand{\avgRe}[1]{\closesymbol\left\langle #1 \right\rangle_{\vRe}}
\newcommand{\avgRi}[1]{\closesymbol\left\langle #1 \right\rangle_{\vRi}}
\newcommand{\avgRsinline}[2][\s]{\langle #2 \rangle_{\vRs[#1]}}

\newcommand{\avgr}[1]{\closesymbol\left\langle #1 \right\rangle_{\vec{r}}}

\newcommand{\LB}{L_B}
\newcommand{\Ln}[1][]{L_{n_{#1}}}
\newcommand{\LT}[1][]{L_{T_{#1}}}
\newcommand{\LTs}{\LT[\s]}
\newcommand{\LTe}{\LT[e]}
\newcommand{\LTi}{\LT[i]}
\newcommand{\Lns}{\Ln[\s]}
\newcommand{\Lne}{\Ln[e]}

\newcommand{\Omegas}[1][\s]{\Omega_{#1}}

\newcommand{\Omegai}{\Omegas[i]}

\newcommand{\omegasts}[1][\s]{\omega_{*{#1}}}
\newcommand{\omegaste}{\omegasts[e]}

\newcommand{\omegaTs}[1][\s]{\omega_{T{#1}}}
\newcommand{\omegaTe}{\omegaTs[e]}

\newcommand{\omegads}[1][\s]{\omega_{\text{d}#1}}
\newcommand{\omegade}{\omegads[e]}

\newcommand{\omegacurvs}[1][\s]{\omega_{\kappa{#1}}}

\newcommand{\omegagradBs}[1][\s]{\omega_{\grad \closesymbol B{#1}}}

\newcommand{\zetasts}[1][\s]{\zeta_{*#1}}

\newcommand{\zetast}{\zetasts[\relax]}

\newcommand{\zetaTs}[1][\s]{\zeta_{T#1}}

\newcommand{\zetaT}{\zetaTs[\relax]}

\newcommand{\zetads}[1][\s]{\zeta_{\text{d}#1}}
\newcommand{\zetade}{\zetads[e]}

\newcommand{\zetad}{\zetads[\relax]}

\newcommand{\I}{I}
\newcommand{\J}{J}

\newcommand{\Ical}{\mathcal{I}}
\newcommand{\Jcal}{\mathcal{J}}

\newcommand{\M}{\mathcal{M}}
\newcommand{\K}{\mathcal{K}}
\renewcommand{\L}{\mathcal{L}}

\newcommand{\Lphiphi}{L_{\phi\phi}}
\newcommand{\LphiA}{L_{\phi A}}
\newcommand{\LphiB}{L_{\phi B}}

\newcommand{\LAphi}{L_{A\phi}}
\newcommand{\LAA}{L_{A A}}
\newcommand{\LAB}{L_{A B}}

\newcommand{\LBphi}{L_{B\phi}}
\newcommand{\LBA}{L_{B A}}
\newcommand{\LBB}{L_{B B}}

\newcommand{\Iab}{\I_{a,b}}
\newcommand{\Jab}{\J_{a,b}}

\newcommand{\zetap}{\zeta_+}
\newcommand{\zetam}{\zeta_-}
\newcommand{\zetapm}{\zeta_\pm}

\newcommand{\Zp}{\rmZ_+}
\newcommand{\Zm}{\rmZ_-}

\newcommand{\Qs}[1][\s]{Q_{#1}}

\newcommand{\Qpar}{\mathcal{Q}^{\parallel}_\vkperp}
\newcommand{\Qpars}[1][\s]{\mathcal{Q}^{\parallel}_{#1\vkperp}}

\newcommand{\Qperp}{\mathcal{Q}^{\perp}_\vkperp}
\newcommand{\Qperps}[1][\s]{\mathcal{Q}^{\perp}_{#1\vkperp}}

\newcommand{\kappaT}{\kappa_T}
\newcommand{\kappan}{\kappa_n}

\newcommand{\chiavgperp}{\langle\chi\rangle_\vkperp}
\newcommand{\chiavg}{\langle\chi\rangle_\vk}

\newcommand{\Jac}{\mathcal{J}}

\newcommand{\aQ}{a_\vkperp}

\newcommand{\curv}{\mathcal{C}_\vkperp}

\newcommand{\avgbox}[2][\relax]{#1\langle\!#1\langle#2#1\rangle_{\! \perp}#1\rangle_{\! \psi}}

\newcommand{\paravg}[1]{\left\langle#1\right\rangle_
\parallel}
\newcommand{\paravginline}[2][\relax]{#1\langle#2#1\rangle_\parallel}

\title{The gyrokinetic field invariant and electromagnetic temperature-gradient instabilities in `good-curvature' plasmas}
\author{
	P.~G.~Ivanov$^{1,2}$\thanks{Email: plamen.ivanov@physics.ox.ac.uk}
	P.~Luhadiya$^{2,3}$, 
	T.~Adkins$^{4,5,2}$, 
	A.~A.~Schekochihin$^{2, 3}$
}
\affiliation{
	$^1$\'Ecole Polytechnique F\'ed\'erale de Lausanne (EPFL), Swiss Plasma Center (SPC),\\CH-1015 Lausanne, Switzerland
	\\[\affilskip]
	$^2$Rudolf Peierls Centre for Theoretical Physics, University of Oxford, Oxford OX1 3PU, UK
	\\[\affilskip]
	$^3$Merton College, Oxford, OX1 4JD, 
	\\[\affilskip]
	$^4$Princeton Plasma Physics Laboratory, Princeton, NJ 08540, USA
	\\[\affilskip]
	$^5$Department of Physics, University of Otago, Dunedin, 9016, New Zealand
}

\begin{document}

\maketitle

\begin{abstract}
Curvature-driven instabilities are ubiquitous in magnetised fusion plasmas. By analysing the conservation laws of the gyrokinetic system of equations, we demonstrate that the well-known spatial localisation of these instabilities to regions of `bad magnetic curvature' can be explained using the conservation law for a sign-indefinite quadratic quantity that we call the `gyrokinetic field invariant'. Its evolution equation allows us to define the local effective magnetic curvature whose sign demarcates the regions of `good' and `bad' curvature, which, under some additional simplifying assumptions, can be shown to correspond to the inboard (high-field) and outboard (low-field) sides of a tokamak plasma, respectively. We find that, given some reasonable assumptions, electrostatic curvature-driven modes are always localised to the regions of bad magnetic curvature, regardless of the specific character of the instability. More importantly, we also deduce that any mode that is unstable in the region of good magnetic curvature must be electromagnetic in nature. As a concrete example, we present the \textit{magnetic-drift mode}, a novel good-curvature electromagnetic instability, and compare its properties with the well-known electron-temperature-gradient instability. Finally, we discuss the relevance of the magnetic-drift mode for high-\(\beta\) fusion plasmas, and in particular its relationship with microtearing modes.
\end{abstract}

\section{Introduction}
\label{sec:intro}

Designing successful magnetic-confinement-fusion (MCF) devices relies on understanding the transport of energy and particles in hot magnetised plasmas. In most modern fusion experiments, the majority of this transport out of the confined region of the plasma is a result of turbulent fluctuations on scales much smaller than the size of the device (and usually associated with the Larmor radius of one or more of the particle species in the plasma). These fluctuations are continuously excited by small-scale instabilities (often called `microinstabilities') driven by the gradients of temperature and density of the large-scale plasma equilibrium. Understanding the conditions for the triggering of these instabilities and how they can be suppressed is therefore of crucial importance for designing devices that can support the larger gradients necessary for higher fusion performance. 

Some of the early work on microinstabilities considered a simple inhomogeneous plasma in a straight magnetic field that is either uniform \citep{rudakov61,coppi66} or sheared \citep{coppi67, cowley91, newton10}. Such magnetic-field configurations are often called `slabs' and so these instabilities are known as `slab instabilities'. However, the equilibrium magnetic field in MCF devices is never completely straight. There exists a separate class of instabilities, often called `curvature-driven' (or sometimes `toroidal'), that rely on the magnetic drifts resulting from the nonuniform magnetic field \citep{pogutse68, terry82, guzdar83, romanelli89, biglari89}. By definition, these instabilities do not exist in the slab geometry, sheared or otherwise. In toroidal geometry, one often finds that the corresponding eigenmodes are peaked near the outboard midplane, where the local gradients of the magnetic field and the plasma pressure are aligned. In some simple cases, e.g., the curvature-driven ion-temperature-gradient (cITG) mode, this can be explained by ignoring the parallel variation of the equilibrium entirely. With this simplification, it is easy to show that cITG modes are unstable when the aforementioned gradients are aligned, and become stable at the inboard side of the device, where the gradients oppose each other \citep{beer95_thesis}. Therefore, magnetic curvature that is aligned with the plasma-pressure gradient is often referred to as `bad curvature', on account of its ability to excite plasma microinstabilities. The same bad magnetic curvature is also responsible for large-scale magnetohydrodynamic (MHD) instabilities \citep{bateman77,white14_book}, further underscoring its destabilising nature. Analogously, `good curvature' refers to magnetic curvature that points opposite to the pressure gradient.

Combined with the magnetic drifts, trapped particles can also drive linear instabilities. There are various trapped-electron \citep{adam73, adam76, catto78a, cheng81} and trapped-ion \citep{xu91} modes, which are often considered to be important sources of turbulent fluctuations. These instabilities cannot be described using the same local analysis that is applicable to the cITG mode. In axisymmetric geometry, particles are trapped at the outboard, low-field side of the device, which coincides with the bad-curvature region. However, in nonaxisymmetric geometry, the trapped-particle and bad-curvature regions may not necessarily overlap. This, combined with the fact that trapped-particle modes are typically only unstable in the bad-curvature regions, underlies some schemes for optimising stellarator configurations \citep{rodriguez24}.

In this work, we investigate the properties of gyrokinetic (GK) curvature-driven microinstabilities and their localisation to regions of either good or bad magnetic curvature. Our analysis is based on GK conservation laws and is largely agnostic to the physical mechanisms that destabilise the fluctuations. First, we show that the conservation of a quantity we call the \textit{GK field invariant} can be used to classify modes as either slab-like or curvature-driven. Furthermore, the evolution equation of this invariant provides a natural definition of good and bad magnetic curvature. Given a few reasonable assumptions, we show that, in the electrostatic limit, all curvature-driven modes must be localised to regions where the magnetic curvature is bad. Consequently, modes that are driven unstable in good-curvature regions must be electromagnetic. 

In the second half of this paper, we present a novel low-beta, electromagnetic, curvature-driven instability that we call the \textit{magnetic-drift mode} (MDM) and that is triggered only in good-curvature regions. By comparing it with the well-known electrostatic, curvature-driven electron-temperature-gradient (cETG) instability, we discuss some of the distinctive features of the MDM. We argue that it may be related to some exotic species of microtearing modes (MTMs) seen in GK simulations in toroidal geometry \citep{patel21_thesis,jian21}.

The rest of this paper is structured as follows. Its first part, \cref{sec:conservation_laws}, which begins with a short summary of the relevant parts of GK theory and toroidal magnetic geometry~(\cref{sec:gk}), is devoted to the conservation laws of free energy and of the GK field invariant defined in \cref{sec:invariants_defs}. In \cref{sec:curvature_driven_modes}, we discuss the implications of the field invariant's conservation for the stability of curvature-driven modes and define the local magnetic curvature whereby good- and bad-curvature regions can be distinguished. We discuss the local magnetic curvature in both simplified toroidal geometry (\cref{sec:circular_geometry}) and \(Z\)-pinch geometry (\cref{sec:local_geometry}).  Then, in \cref{sec:lowbeta_dk}, we turn to the MDM and its properties. First, in \cref{sec:ordering}, we discuss the low-beta, drift-kinetic ordering in the magnetic geometry of a \(Z\)-pinch, which provides the minimal model for the MDM. In \cref{sec:2d}, we restrict ourselves to the two-dimensional limit, wherein we find that the electrostatic and electromagnetic fluctuations decouple. Here, we discover the MDM and compare it with the cETG mode. In \cref{sec:mdm_3d}, we provide qualitative estimates for the destabilisation threshold of the MDM in toroidal geometry and argue that this mode might have already been observed in GK simulations. Finally, in \cref{sec:discussion}, we summarise and discuss our results.


\section{Gyrokinetic conservation laws}
\label{sec:conservation_laws}

In this section, we recap the gyrokinetic framework before moving on to discuss the conservation laws that will prove relevant for curvature-driven instabilities. For a more comprehensive review of gyrokinetics, see, e.g., \citet{abel13} or \citet{catto19}. 

\subsection{Gyrokinetic equations in a magnetised plasma}
\label{sec:gk}

In this work, we consider fluctuations that satisfy the GK ordering:
\begin{equation}
	\frac{\omega}{\Omegas} \sim \frac{\nu_{\s \s'}}{\Omegas} \sim \frac{\kpar}{\kperp} \sim \frac{\qs \phipot}{\Ts} \sim \frac{\dBpar}{B} \sim \frac{\abs{\vdBperp}}{B} \sim \frac{\rhos}{L} \ll 1,
	\label{eq:gyrokinetic_ordering}
\end{equation}
where \(\omega\) is the characteristic frequency (or growth rate) of the fluctuations, \(\kpar\) and \(\kperp\) are the characteristic fluctuation wavenumbers in the directions respectively parallel and perpendicular to the equilibrium magnetic field \(\vB\), \(L\) is any characteristic length scale associated with the equilibrium, \(\phipot\) is the fluctuating electrostatic potential (the equilibrium electric field is assumed to be zero), \(\dBpar\) and \(\vdBperp\) are the magnetic-field fluctuations respectively parallel and perpendicular to the mean field, \(\s\) labels the particle species (e.g., ions and electrons), and \(\nu_{\s \s'}\) is the collision frequency between species \(\s\) and \(\s'\). For each species \(\s\), we define its charge \(\qs\), mass \(\ms\), thermal speed \(\vths\), equilibrium temperature \(\Ts = \ms \vths^2/2\), gyrofrequency \(\Omegas = \qs B / \ms c\), and gyroradius \(\rhos=\vths / |\Omegas|\). 

Under \cref{eq:gyrokinetic_ordering}, we expand the particle distribution function as
\begin{equation}
	\fs = \Fs + \dfs,
\end{equation}
where \(\dfs / \Fs \sim \rhos / L\) and the perturbed distribution function is further decomposed as
\begin{equation}
	\dfs(\vr,\vv,t) = - \frac{\qs \phipot(\vr,t)}{\Ts} \Fs(\vRs,\es) + \hs(\vRs, \es, \mus,t),
	\label{eq:perturbed_distribution_function}
\end{equation}
where $\vRs = \vr - \ub \times \vvperp/\Omegas \equiv \vr - \vrhos(\vartheta)$ is the guiding-centre position, \(\ub = \vB / B\), \(\vartheta\)~is the gyroangle, \mbox{\(\es = m_s v^2/2\)} is the particle energy, and \mbox{\(\mus = m_s \vperp^2 / 2B\)} is the particle magnetic moment. The equilibrium distribution \(\Fs\) is a Maxwellian (i.e., an exponential distribution in \(\es\)) with (nonuniform) density \(\ns\) and temperature \(\Ts\). Assuming that the collisions between particles are sufficiently rare, and provided that there are no sonic mean flows,  $\hs$ evolves according to the collisionless gyrokinetic equation
\begin{equation}
	\partd{}{t} \left( \hs - \frac{\qs \avgRs{\chi}}{\Ts} \Fs \right) + \left(\vpar \ub + \vds + \avgRs{\vchi} \right) \bcdot \partd{\hs}{\vRs} + \avgRs{\vchi} \bcdot  \partd{\Fs}{\vRs} = 0,
	\label{eq:gyrokinetic_equation}
\end{equation}
where the parallel velocity is \(\vpar = \sigma \sqrt{2(\es - \mus B)/\ms}\) and \(\sigma = \pm 1\) is its sign. In~\cref{eq:gyrokinetic_equation}, $\avgRsinline{\dots}$ denotes the standard gyroaverage (over \(\vartheta\)) at constant \(\vRs\). The GK potential \mbox{$\chi = \phipot - \vv\bcdot \vdA/c$} is defined in terms of the fluctuating electrostatic potential \(\phipot\) and vector potential \(\vdA\).\footnote{Throughout this work, we have assumed the Coulomb gauge \(\grad \bcdot \vA = 0\).} This potential determines the GK drift velocity \(\vchi = (c/B)\ub\times\grad \chi\), which gives rise to the nonlinear advection of \(\hs\), as well as to the linear drive associated with the gradients of the equilibrium distribution. The magnetic drifts in \cref{eq:gyrokinetic_equation}, which will be a central feature of the discussion below, are
\begin{equation}
	\vds = \frac{\ub}{\Omegas} \times \left( \vpar^2 \ub\bcdot \grad\ub + \frac{1}{2}\vperp^2 \grad\log B \right).
	\label{eq:magnetic_drifts}
\end{equation}

Finally, under the GK ordering \cref{eq:gyrokinetic_ordering} and with the additional assumption that all fluctuations are on scales much larger than the plasma Debye length, the electromagnetic fields appearing in the gyrokinetic equation \eqref{eq:gyrokinetic_equation} are determined by Amp\`ere's law
\begin{equation}
	\grad\times\vdB = \frac{4\pi}{c}\vdJ = \frac{4\pi}{c} \sum_\s \qs \intv \avgr{\vv\hs}
	\label{eq:amperes_law}
\end{equation}
and the quasineutrality condition
\begin{equation}
	0 = \sum_{\s} \qs \dns = \sum_\s \qs \left[ -\frac{\qs \phipot}{\Ts} \ns + \intv\!\avgr{\hs}\right].
	\label{eq:quasineutrality}
\end{equation}
\Cref{eq:amperes_law} is more commonly split into its parallel and perpendicular parts:
\begin{align}
	\grad_\perp^2 \dApar &= - \frac{4\pi}{c} \sum_\s \qs \intv \vpar \avgr{\hs},
	\label{eq:parallel_amperes_law} \\
	\grad_\perp^2 \dBpar& = - \frac{4\pi}{B} \grad_\perp \grad_\perp : \sum_\s \ms \intv \avgr{\vvperp \vvperp \hs},
	\label{eq:perpendicular_amperes_law}
\end{align}
in which form the latter obviously expresses perpendicular pressure balance.

\subsection{Axisymmetric magnetic geometry}
\label{sec:geo}

We assume that the equilibrium is axisymmetric and that there exist well-defined flux surfaces labelled by the poloidal flux \(\psi\). In this case, it can be shown \citep[see, e.g.,][]{abel13} that the plasma equilibrium is, to lowest order in \cref{eq:gyrokinetic_ordering}, a function only of \(\psi\), i.e., \(\ns = \ns(\psi)\), \(\Ts = \Ts(\psi)\), etc. The equilibrium magnetic field is given by \(\vB = \grad\alpha\times\grad\psi\), where the Clebsch angle is \(\alpha =  \varphi - q(\psi)\theta\), \(\varphi\) is the toroidal angle (the symmetry angle of the axisymmetric equilibrium), \(\theta\) is the straight-field-line poloidal angle, and \(q\) is the safety factor  \citep{kruskal58, d’haeseleer91_book}. We take \(\theta=0\) to correspond to the outboard midplane of the device. 

In what follows, we shall be working in a field-line-following coordinate system \((x, y, z)\), where we use \(z\) as the coordinate along the equilibrium magnetic field, while \(x \propto \psi\) and \(y \propto \alpha\) are the radial and poloidal coordinates, respectively, with suitable constant normalisations so that they have units of length. In such a coordinate system, \mbox{\(\grad x \times \grad y = \vB / \Bo\)}, where \(\Bo\) is some constant normalising magnetic field that depends on the choice of \(x\) and \(y\). An example of such a choice for \(x\), \(y\), and \(z\) is given in~\cref{sec:circular_geometry}.

Finally, let us note that even though the results in this section are derived and presented in axisymmetric geometry, they are readily generalisable to non-axisymmetric geometries.

\subsection{Free energy and the gyrokinetic field invariant}
\label{sec:invariants_defs}

The GK system of equations conserves the free energy
\begin{equation}
	W = \sum_\s \avgbox[\bigg]{\intv \frac{\Ts\dfs^2}{2\Fs}} + \avgbox[\bigg]{\frac{\abs{\vdB}^2}{8\pi}},
	\label{eq:free_energy_deltaf}
\end{equation}
where \(\langle . \rangle_\perp\) and \(\langle . \rangle_\psi\) denote an intermediate-scale perpendicular average and the flux-surface average, respectively [see \cref{eq:perpaverage_def} and \cref{eq:fsa_def} for their precise definitions]. Physically, their combination is an average over a finite volume around a given flux surface whose radial size is large compared to the fluctuation wavelength \(\sim \rhos\) but small compared to the scale of equilibrium variation \(L\). It can be shown (see \cref{appendix:conservation_laws}) that the (collisionless)\footnote{Collisions enter as a negative-definite term on the right-hand side of \cref{eq:free_energy_injection}.} time evolution of the free energy is given by
\begin{align}
	\der{W}{t} = \sum_\s \Ts\left(\frac{1}{\Lns} - \frac{3}{2}\frac{1}{\LTs}\right) \Gamma_\s + \sum_\s \frac{\Qs}{\LTs},
	\label{eq:free_energy_injection}
\end{align}
where the right-hand side of \cref{eq:free_energy_injection} depends on the radial particle and heat (more precisely, energy) fluxes for each species, denoted by \(\Gamma_\s\) and \(Q_\s\), respectively:
\begin{align}
	\Gamma_s &= \avgbox[\bigg]{\intv\avgr{\hs \vchi} \bcdot \grad x}, \label{eq:Gammas_def} \\
	\Qs &= \avgbox[\bigg]{\intv\frac{1}{2}\ms v^2\avgr{\hs \vchi} \bcdot \grad x} \label{eq:Qs_def},
\end{align}
and the density and temperature gradient length scales are defined as
\begin{equation}
	\frac{1}{\Lns} \equiv -\der{\ln \ns}{x}, \quad \frac{1}{\LTs} \equiv -\der{\ln \Ts}{x}.
	\label{eq:equilibrium_gradients}
\end{equation}


Note that \cref{eq:free_energy_injection} is a local conservation law valid at any given flux surface. Thus, to lowest order in \cref{eq:gyrokinetic_ordering}, free energy is injected and dissipated at every flux surface separately, i.e., there is no overall transport of free energy across flux surfaces \citep{abel13}. The right-hand side of \cref{eq:free_energy_injection} shows that free energy is injected into fluctuations by the density and heat fluxes along the gradients of the plasma equilibrium. The magnetic drifts are absent from \cref{eq:free_energy_injection}. This is the well-known result that GK fluctuations cannot directly access the energy stored in the equilibrium magnetic field \citep{abel13}. Thus, even though the magnetic drifts are, by definition, required for any curvature-driven instability, we cannot use the conservation of free energy by itself to determine their role in said instabilities. 

To elucidate the role of the magnetic drifts, we construct another nonlinearly conserved quantity, which we call \textit{the GK field invariant}, defined as
\begin{equation}
	Y \equiv \sum_\s \avgbox[\Bigg]{\intv \frac{\Ts}{2\Fs} \avgr{\left(\hs - \frac{\qs\avgRs{\chi}}{\Ts}\Fs\right)^2}} - W,
	\label{eq:Y_definition}
\end{equation}
where \(\avgr{.}\) is the standard gyroaverage at fixed \(\vr\). The field invariant can be shown to satisfy (see \cref{appendix:Y_injection})
\begin{equation}
	\der{Y}{t} = \sum_\s \avgbox[\bigg]{\intv \avgr{\qs\avgRs{\chi}\vpar\ub\bcdot\grad\hs + \qs \avgRs{\chi} \vds \bcdot \gradperp \hs}}.
	\label{eq:Y_injection}
\end{equation}
It is related to the general two-dimensional invariants of GK \citep{schekochihin09, plunk10} and can be thought of as a generalisation of the so-called `electrostatic invariant' \citep{schekochihin09, plunk10, helander13, plunk23}. Note that even though the nonlinear terms do not contribute to the evolution of \(Y\), it still evolves collisionlessly in a uniform plasma and a uniform magnetic field due to the presence of the parallel streaming term on the right-hand side of \cref{eq:Y_injection}. In contrast, \(W\) is a `true' invariant for which \(\rmd W / \rmd t\) vanishes exactly in the absence of equilibrium gradients and collisions. Thus, the relevance of the field invariant to the study of nonlinear turbulent saturation is likely limited to two-dimensional or nearly two-dimensional situations wherein the parallel-streaming term can be ignored \citep{plunk10}.

Nevertheless, the field invariant can be a powerful tool in the study of linear instabilities.\footnote{Indeed, this was already realised by \citet{helander13} for electrostatic trapped-electron modes in stellarators.} The (collisionless) time evolution of \(Y\), given by \cref{eq:Y_injection}, contains two contributions: one from parallel streaming and one from the magnetic drifts. We will shortly argue that the former is the physically important source of \(Y\) for slab instabilities. The latter will turn out to determine the instability of curvature-driven modes. In \cref{appendix:Y_injection}, we show that it can be written as
\begin{align}
	&\sum_\s \avgbox[\bigg]{\intv \qs \avgr{\avgRs{\chi} \vds\bcdot\gradperp\hs}} \nonumber \\
	&= \sum_\s \avgbox[\bigg]{\intv \left(\ms\vpar^2 + \frac{1}{2}\ms\vperp^2 \right) \avgr{\hs \vchi}\bcdot \gradperp\ln B} \nonumber \\
	&\quad+ \beta \der{\ln p}{x} \sum_\s \avgbox[\bigg]{\intv \frac{1}{2}\ms\vpar^2 \avgr{\hs \vchi} \bcdot \grad x},
	\label{eq:Y_magnetic_drift_injection}
\end{align}
where \mbox{\(\beta = 8\pi p / B^2\)} is the plasma beta and \(p = \sum_\s \ns\Ts\) is the equilibrium presssure. \Cref{eq:Y_magnetic_drift_injection} suggests that the injection of \(Y\) due to the magnetic drifts has a form very similar to that of \(W\), viz., it is proportional to the turbulent heat flux across the equilibrium magnetic field, but also depends directly on the gradient of the magnetic field. As the vector \(\gradperp\ln B\) is, in general, not aligned with \(\grad x\), it is not straightforward to relate the injection terms of \(W\) and \(Y\), i.e., the right-hand sides of \cref{eq:free_energy_injection} and \cref{eq:Y_magnetic_drift_injection}. In \cref{sec:curvature_driven_modes}, we deal with this by making some additional assumptions about the nature of the instabilities and the magnetic geometry.

\subsubsection{Local limit}
\label{sec:local}

To make further progress, we restrict ourselves to what is commonly known as the `local \(\delta\!f\) gyrokinetics'. Here `local' means that we are integrating \cref{eq:gyrokinetic_equation} in a domain of perpendicular size that is infinitesimal in comparison with the length scales associated with the plasma equilibrium \(\Fs\) and the equilibrium magnetic field \(\vB\), and so the (perpendicular) gradients associated with the equilibrium are taken to be constant; `\(\delta\!f\)' means that we only consider the evolution of small-amplitude fluctuations over times that are short compared to the transport time scale (over which the equilibrium evolves), with the equilibrium therefore assumed constant in time.

In the local approximation, any fluctuating quantity \(g(\vr)\) can be written as
\begin{equation}
	g(\vr) = \sum_\vkperp g_\vkperp(z) \rme^{\rmi k_x x + \rmi k_y y},
	\label{eq:eikonal}
\end{equation}
where the field-line-following coordinate system \((x, y, z)\) is defined at the end of \cref{sec:gk}. \Cref{eq:eikonal} can be thought of as an ensemble of modes in the ballooning representation \citep{connor78}. Alternatively, and perhaps more practically for numerical simulations, fluctuations have the form \cref{eq:eikonal} when solving \cref{eq:gyrokinetic_equation} in a field-line-following flux tube \citep{beer95} and imposing periodic boundary conditions in the plane perpendicular to the magnetic field.\footnote{We are not, however, interested in the practical numerical implementation of solving \cref{eq:gyrokinetic_equation} in a flux tube. Thus, we are allowed to let our flux tube be `true' to the ballooning representation and hence be infinite along the magnetic-field lines.} Further details on the validity of \cref{eq:eikonal} and the relationship between the ballooning transformation and the local GK formulation can be found in \citet{beer95} and \cref{appendix:ballooning}. For the purposes of the following discussion, we consider modes in a periodic flux tube with an infinite extent along the field lines. In this case, the free energy \cref{eq:free_energy_deltaf} and field invariant \cref{eq:Y_definition} can be expressed as
\begin{equation}
	W = \sum_\s \sum_\vkperp \paravg{\intv \frac{\Ts\abs{\dfskperp}^2}{2\Fs}} + \sum_\vkperp \paravg{\frac{\abs{\dBparkperp}^2 + \kperp^2\abs{\dAparkperp}^2}{8\pi}}
	\label{eq:W_fields_expression}
\end{equation}
and
\begin{align}
	Y = \sum_\vkperp\paravginline[\Bigg]{ &-\sum_{\s} \frac{\qs^2\ns}{2\Ts}\left(1-\rmGamma_{0\s} + \frac{\rmGamma_{1\s}}{2}\right)\abs{\phipotkperp}^2 + \left(\frac{\kperp^2}{8\pi} + \sum_\s \frac{\rmGamma_{0\s}}{8\pi\ds^2}\right)\abs{\dAparkperp}^2 \nonumber \\
	&+\frac{\abs{\dBparkperp}^2}{8\pi} + \sum_\s\ns\Ts  \rmGamma_{1\s} \abs{\frac{\dBparkperp}{B} + \frac{\qs\phipotkperp}{2\Ts}}^2},
	\label{eq:Y_fields_expression}
\end{align}
respectively, where \(\paravginline{.}\) denotes the parallel average [defined in \cref{eq:paravg_def_flux_tube}]. It is defined in such a way that the combination of the sum over \(\vkperp\) and the parallel average is equivalent to the combination of the perpendicular and flux-surface average (see \cref{appendix:ballooning} for details). In \cref{eq:Y_fields_expression}, \mbox{\(\ds = \rhos/\sqrt{\betas}\)} and \(\betas = 8\pi\ns\Ts/B^2\) are the skin depth (or inertial length) and plasma beta of species \(\s\), respectively, and
\begin{align}
	\rmGamma_{0\s} = \rmI_0(\alpha_s)\rme^{-\alpha_\s}, \quad \rmGamma_{1\s} = \left[\rmI_0(\alpha_s) - \rmI_1(\alpha_s)\right] \rme^{-\alpha_\s},
\end{align}
where \(\alpha_\s = \kperp^2\rhos^2/2\), \(\kperp^2 = (k_x \grad x + k_y \grad y)^2\), and \(\rmI_0\) and \(\rmI_1\) are the modified Bessel functions of the first kind. 

When written as \cref{eq:W_fields_expression} and \cref{eq:Y_fields_expression}, it is evident that the field invariant \(Y\) differs from the free energy \(W\) in two crucial ways. First, the distribution function \(\dfs\) does not enter explicitly into the former, which contains contributions only from the electromagnetic fields, justifying the name `field' invariant. Secondly, unlike the free energy, which is always positive, \(Y\) is not sign definite, as is evident in \cref{eq:Y_fields_expression}. Since \(1 - \rmGamma_{0\s}\) and \(\rmGamma_{1\s}\) are both positive, the electrostatic part of the invariant, viz., the part proportional to \(\absinline{\phipotk}^2\) in \cref{eq:Y_fields_expression}, is negative definite, while the rest is positive definite. This will be crucial for the arguments presented in \cref{sec:curvature_driven_modes}. 

In general, the physical meaning of \(Y\) is not transparent. However, in certain limits, it can be related to other, better-known conserved quantities. For instance, in the electrostatic limit, viz., \(\dApark = 0\) and \(\dBpark = 0\), \cref{eq:Y_fields_expression} becomes the usual electrostatic~invariant~\citep{schekochihin09, plunk10, helander13}:
\begin{equation}
	\lim_{\absinline{\vdB}\to0} Y = -\sum_{\vkperp, \s} \frac{\qs^2\ns}{2\Ts}\paravg{\left(1-\rmGamma_{0\s}\right)\abs{\phipotkperp}^2} \approx \avgbox[\bigg]{-\frac{1}{2}\varrho \abs{\ve}^2},
	\label{eq:es_invariant}
\end{equation}
where the last (approximate) equality holds in the long-wavelength limit (\(\kperp\rhos \ll 1\) for all species \(\s\)). In this limit, \(Y\) is simply (the negative of) the total kinetic energy associated with the \exb{} velocity of the particles \(\ve = (c/B)\ub\times\grad\phipot\) and its total density \(\varrho \equiv \sum_\s \ms\ns\). Restoring the contributions from the magnetic fluctuations, we find
\begin{equation}
	Y = \avgbox[\bigg]{-\frac{1}{2}\varrho \abs{\ve}^2 + \frac{\abs{\vdB}^2}{8\pi} +  \frac{\abs{\dApar}^2}{8\pi\de^2}},
	\label{eq:Y_large_scales}
\end{equation}
where we have assumed a `typical' plasma of multiple ion species and one electron species of significantly lower mass. The last term in \cref{eq:Y_large_scales}, which dominates when \mbox{ \(\kperp\ds \ll 1\)}, can be recognised as anastrophy (or `\(\Apar^2\)-stuff') --- a well-known two-dimensional invariant of MHD \citep{fyfe76,pouquet78,schekochihin22}. Thus, in the MHD limit, the field invariant can be though of as the anastrophy with an additional correction due to kinetic physics.

Using \cref{eq:eikonal}, the free-energy conservation law \cref{eq:free_energy_injection} can be written as
\begin{align}
	\der{W}{t} &= \sum_\s \Ts\left(\frac{1}{\Lns} - \frac{3}{2}\frac{1}{\LTs}\right) \Gamma_\s + \sum_{\s,\:\vkperp} \frac{1}{\LTs} \paravg{\Qpars + \Qperps}, \label{eq:W_curv_driven_injection_final}
\end{align}
where we have defined the local (in \(z\)) radial fluxes of energy associated with parallel and perpendicular particle motion:
\begin{align}
	\Qpars(z) &\equiv \im \left[\frac{c k_y}{\Bo} \intv \frac{1}{2}\ms\vpar^2 \hskperp^*\closesymbol \chiavgperp\right], \\
	\Qperps(z) &\equiv \im \left[\frac{c k_y}{\Bo} \intv \frac{1}{2}\ms\vperp^2 \hskperp^*\closesymbol\chiavgperp\right],
\end{align}
where \(\chiavgperp\) is the Fourier-transformed gyroaveraged GK potential [see \cref{eq:chi_fourier}]. These fluxes are related to the total heat flux by \(\Qs = \sum_\vkperp\paravginline{\Qpars + \Qperps}\).

Similarly, for the field invariant, we find
\begin{align}
	\der{Y}{t} &= \left.\der{Y}{t}\right\rvert_\text{slab} + \left.\der{Y}{t}\right\rvert_\text{curv},
	\label{eq:Y_injection_singlemode}
\end{align}
where we have separated the `slab' contribution to \cref{eq:Y_injection} arising from the parallel streaming,
\begin{equation}
	\left.\der{Y}{t}\right\rvert_\text{slab} = \sum_{\s,\: \vkperp}\paravg{\intv\qs\vpar\chiavgperp \ub\bcdot \grad \hskperp^*},
	\label{eq:Y_injection_singlemode_slab}
\end{equation}
and the `curvature' one due to the magnetic drifts,
\begin{equation}
	\left.\der{Y}{t}\right\rvert_\text{curv} = 2\sum_\vkperp\paravg{\curv \Qpar}.
	\label{eq:Y_injection_singlemode_curv}
\end{equation}
Here the \textit{local magnetic curvature} is defined as
\begin{equation}
	\curv(z) \equiv (2+\aQ)\frac{\Bo}{B} (\ub\times\gradperp\ln B) \bcdot \left( \grad y + \frac{k_x}{k_y}\grad x\right) + \frac{\beta}{2}\der{\ln p}{x},
	\label{eq:curv_def}
\end{equation}
where \(\aQ(z) \equiv (\Qperp - 2\Qpar)/2\Qpar\) is the heat-flux anisotropy parameter, defined as the ratio of the total parallel \(\Qpar \equiv \sum_\s \Qpars\) and perpendicular \(\Qperp \equiv \sum_\s \Qperps\) heat fluxes. For isotropic fluctuations, in particular, \(\aQ = 0\).

\Cref{eq:Y_injection_singlemode} provides a natural distinction between slab and curvature-driven modes. By definition, the former are those that survive in a straight uniform magnetic field, where the magnetic drifts vanish, and so does \cref{eq:Y_injection_singlemode_curv}. Thus, the parallel-streaming term \cref{eq:Y_injection_singlemode_slab} is the physically important source of \(Y\) for slab instabilities even if both \cref{eq:Y_injection_singlemode_slab} and \cref{eq:Y_injection_singlemode_curv} happen to be nonzero in the presence of magnetic drifts. In contrast, curvature-driven modes are those whose dominant \(Y\)-injection term is \cref{eq:Y_injection_singlemode_curv}. These modes are destabilised by the combination of the thermal-equilibrium gradients and the magnetic geometry. 

\subsection{Curvature-driven temperature-gradient instabilities}
\label{sec:curvature_driven_modes}

Let us consider a single unstable mode, i.e., a fluctuation with given \(k_x\) and \(k_y\), with amplitude proportional to \(\rme^{-\rmi\omega t}\), where \(\im(\omega) > 0\) is the growth rate. As both \(W\) and \(Y\) are quadratic in the fluctuation amplitude, this implies that the free energy and the field invariant associated with a single mode satisfy
\begin{equation}
	\der{W}{t} = 2\im(\omega) W, \quad \der{Y}{t} = 2\im(\omega) Y.
	\label{eq:W_and_Y_evo_for_single_mode}
\end{equation} 
We now make three simplifying assumptions:
\begin{enumerate}[(i)]
	\item We assume that the injection of free energy of the mode is dominated by the heat rather than the particle fluxes, so we can ignore the latter in \cref{eq:W_curv_driven_injection_final}. We refer to such fluctuations as `temperature-gradient' modes. While this might seem very restrictive, there are, in fact, many instabilities that satisfy this condition. One example are modes for which there is a main species \(\s\), with all other species \(s'\) assumed adiabatic \citep[or `modified' adiabatic in the case of ion-scale turbulence: see][]{hammett93}. For such instabilities, the nonadiabatic distribution functions \(h_{s'}\) for the species \(s'\neq s\) are small, in which case the particle flux vanishes.
	
	\item We assume that the unstable mode is curvature-driven and so the slab injection term in \cref{eq:Y_injection_singlemode} can be ignored. The ratio of \cref{eq:Y_injection_singlemode_slab} and \cref{eq:Y_injection_singlemode_curv} for a curvature-driven mode can be estimated as
	\begin{align}
		\frac{\rmd Y / \rmd t \vert_\text{slab}}{\rmd Y / \rmd t \vert_\text{curv}} \sim \frac{\kpar \vths}{\omegads},
	\end{align}
	where \(\omegads = \vkperp \bcdot \vds\) is the magnetic-drift frequency and \(\kpar \sim (\ub \bcdot \grad \hs) / \hs\) is the effective parallel wavenumber. Therefore, the assumption of neglecting \cref{eq:Y_injection_singlemode_slab} in favour of \cref{eq:Y_injection_singlemode_curv} in \cref{eq:Y_injection_singlemode} can be made asymptotic in the limit \(\kpar \vths \ll \omegads\), i.e., in the limit of long parallel length scales.\footnote{While this is one of the simplest orderings for a curvature-driven mode, it is certainly not the only one. For example, \citet{helander13} argues that, for trapped-electron modes, the injection of \(Y\) is dominated by magnetic curvature even though the ordering of the parallel streaming and magnetic drifts is more complicated.}
	
	\item We assume that the heat flux is dominated by a single, `main' species \(s\). We make this choice purely for the purposes of simplifying the discussion below. Our arguments hold even if multiple species contribute to the injection of \(W\) and \(Y\) if the temperature gradients of the species that contribute the majority of the heat flux are aligned.
\end{enumerate}

Without loss of generality, we can choose the flux-surface label \(x\) so that \(\LTs > 0\). Therefore, due to \cref{eq:W_and_Y_evo_for_single_mode} and the positive-definiteness of \(W\), the heat flux \mbox{\(\Qs = \paravginline{\Qpars + \Qperps}\)} appearing in \cref{eq:W_curv_driven_injection_final} must be positive for any unstable mode with \(\im\:\omega > 0\). This is intuitively clear: temperature-gradient-driven instabilities predominantly transport heat down the temperature gradient, from the hot to the cold regions of the equilibrium. 

Assuming that \(\Qpars \sim \Qperps\) (or at least that they have the same sign), \cref{eq:Y_injection_singlemode}, \cref{eq:Y_injection_singlemode_curv}, and \cref{eq:W_and_Y_evo_for_single_mode} imply that
\begin{align}
	 \im(\omega) \approx \frac{Y}{\paravg{\curv \Qpar}} > 0.
	\label{eq:Y_over_CQpar}
\end{align}
Since the heat flux \(\Qpar(z)\) is positive and the local magnetic curvature \(\curv(z)\) changes sign along the field line (we provide a simple example of this in \cref{sec:circular_geometry}), in order for~\cref{eq:Y_over_CQpar} to be satisfied, the heat flux must be larger at those locations along the field line where \(\curv(z)\) has the same sign as \(Y\). Therefore, if \(Y > 0\), the heat flux, and thus the mode amplitude, must peak at the places where \(\curv(z) > 0\): we call these the regions of good curvature. Analogously, if \(Y < 0\), then the mode amplitude must peak in the bad-curvature regions where \(\curv(z) < 0\). As we saw previously, for electrostatic modes, \(Y\) becomes the well-known electrostatic invariant \cref{eq:es_invariant}, which is negative definite. Thus, we conclude that electrostatic curvature-driven temperature-gradient modes must be localised to regions of bad curvature, regardless of the precise details of the physical mechanism of their instability. Analogously, any curvature-driven temperature-gradient mode that is unstable in a good-curvature region must be predominantly electromagnetic in the sense that the positive-definite \(\dAparkperp\) and \(\dBparkperp\) contributions to \cref{eq:Y_fields_expression} must be larger than the negative-definite electrostatic ones, in order to ensure \(Y > 0\). While such instabilities have been seen before in numerical simulations \citep{patel21_thesis, jian21}, we shall provide in \cref{sec:lowbeta_dk} the first known example of a simple, solvable model of a curvature-driven electromagnetic instability that exists only in good-curvature regions.

Shortly, we shall see that, in simple geometry, \cref{eq:curv_def} corroborates the expectation that the inboard, high-field side of the plasma has predominantly good curvature, while the outboard, low-field one has mostly bad curvature. However, the correlation between inboard--outboard side and good--bad curvature is not perfect. Indeed, as shown by \citet{parisi20},\footnote{The quantities \(\omega_{\kappa e}\) and \(\omega_{\grad B e}\) used by \citet{parisi20} to quantify the local magnetic curvature and its effect on the ETG instability are closely related to \cref{eq:curv_def}. Indeed, their sum is proportional to \(\curv\) for \(\aQ=0\).} modes driven by bad curvature can be found far away from the outboard midplane.

Before we consider a concrete example of a good-curvature electromagnetic instability, let us briefly discuss the meaning of \(\curv\) in two different simplified magnetic geometries and show that it recovers the intuitive picture of good and bad curvature.

\subsubsection{Large-aspect-ratio circular flux surfaces}
\label{sec:circular_geometry}

\begin{figure}
	\centering
	\begingroup\import{figs}{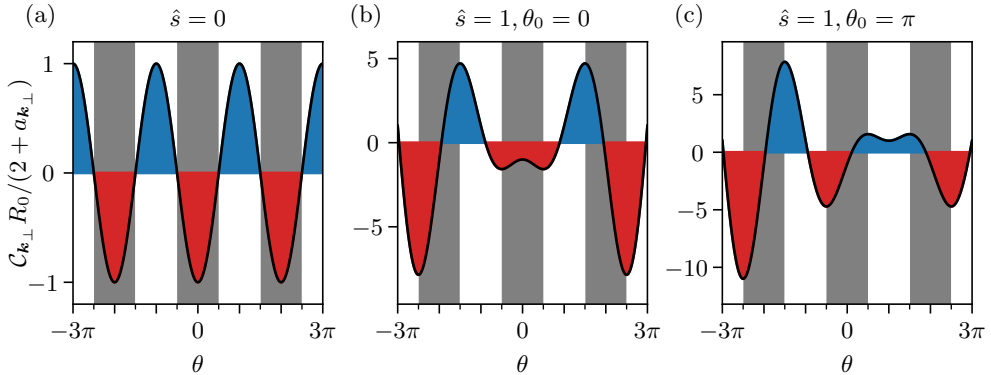}\endgroup
	\caption{Panels (a)--(c) visualise the circular-flux-surface local curvature \cref{eq:curv_circular_geometry} as a function of \(\theta\) for three different values of \(\hat{s}\) and \(\theta_0\), as indicated in the title of each panel. The regions of good and bad local curvature are shaded in blue and red, respectively. The grey shaded regions correspond to the outboard side of the device, viz., \mbox{\(-\pi/2 + 2\pi n < \theta < \pi/2 + 2\pi n\)} for some \(n \in \integers\).}
	\label{fig:C_circ}
\end{figure}

Using the standard choice of field-line-following coordinates made by \citet{beer95}, viz.,
\begin{equation}
	x \equiv \frac{q_0}{\Bo r_0}\left(\psi - \psi_0\right), \quad y \equiv -\frac{r_0}{q_0}(\alpha - \alpha_0), \quad z \equiv \theta,
	\label{eq:xyz_def}
\end{equation}
the local curvature in large-aspect-ratio, circular-flux-surface, low-beta configurations is
\begin{equation}
	\curv = -\frac{2+\aQ}{R_0} \left[\cos\theta + \hat{s}(\theta-\theta_0)\sin\theta\right].
	\label{eq:curv_circular_geometry}
\end{equation}
In the above, \(\psi_0\) and \(\alpha_0\) are the coordinates of the field line at \(x=y=0\), \(q_0 = q(\psi_0)\) is the safety factor, \(R_0\) and \(r_0\) are the major and minor radius of its flux surface, respectively, \(\Bo\) is any suitable normalising magnetic field, \(\hat{s} = (r_0/q_0)\rmd q / \rmd r\) is the magnetic shear at \(r=r_0\), and \(\theta_0 = -k_x/\hat{s}k_y\) is the poloidal angle at which the radial projection of the wavenumber vanishes, viz., \(\vkperp(\theta_0)\bcdot\grad x = 0\). 

In \cref{fig:C_circ}, we plot \cref{eq:curv_circular_geometry} for different values of \(\hat{s}\) and \(\theta_0\). \Cref{fig:C_circ}(a) shows the simplest case of zero magnetic shear (\(\hat{s} = 0\)), wherein the local magnetic curvature 
\begin{equation}
	\curv \approx -\frac{2+\aQ}{R_0} \cos\theta
	\label{eq:curv_circular_low_beta}
\end{equation}
is independent of \(\theta_0\). In this case, the regions of good and bad curvature coincide perfectly with the inboard and outboard sides of the device, respectively. Panels (b) and (c) of \cref{fig:C_circ} demonstrate the inherent symmetry of \cref{eq:curv_circular_geometry} whereby the local effective magnetic curvature \(\curv\) changes from bad to good (or vice versa) as a function of \(\theta - \theta_0\) if the mode is `reflected' to the diametrically opposite side of the flux surface. More concretely, \(\curv \mapsto -\curv\) under the mapping \(\theta \mapsto \theta + \pi\) and \(\theta_0 \mapsto \theta_0 + \pi\), which is evident from~\cref{eq:curv_circular_geometry}.

Note that the definition of good and bad curvature as \(\curv > 0\) or \(\curv < 0\) is mode-specific because \cref{eq:curv_circular_low_beta} [and, indeed, \cref{eq:curv_def}] depends explicitly on the structure of the mode via \(\aQ\). Indeed, \(\aQ < -2\) reverses the relationship between the sign of \(\curv\) and the outboard/inboard location in the toroidal geometry. However, this is a rather exotic scenario wherein the parallel and perpendicular heat fluxes have opposite signs. As far as the authors are aware, no such instabilities are known to exist in fusion plasmas.

\subsubsection{\(Z\)-pinch}
\label{sec:local_geometry}

An even simpler description of curvature-driven modes can be obtained by ignoring entirely the variation of field along \(z\). Sometimes called the `local kinetic approximation' \citep{terry82,romanelli89,zocco18}, this is equivalent to the magnetic geometry of a \(Z\)-pinch if we also ignore the radial magnetic drifts. This \(Z\)-pinch geometry is the minimal model for curvature-driven instabilities in the absence of trapped-particle effects and has been a popular setting for simple models of both linear and nonlinear physics \citep{ricci06, kobayashi12, kobayashi15, ivanov20, ivanov22, adkins22, hallenbert22, adkins23, ivanov23, hoffmann23}.

In this geometry, we take \((x, y, z)\) to be local Cartesian coordinates, where \(z\) is now the distance along the field lines, and define the gradient scale length of the magnetic field \(B = B(x)\):
\begin{equation}
	\frac{1}{\LB} \equiv -\der{\ln B}{x}.
	\label{eq:LB_def}
\end{equation}
Hence, we obtain
\begin{equation}
	\curv = -\frac{2+\aQ}{\LB}.
\end{equation}
Therefore, in the \(Z\)-pinch geometry, \(\LB < 0\) and \(\LB > 0\) correspond to good and bad curvature, respectively. As discussed before, we expect to find electrostatic curvature-driven instabilities only in the case \(\LB > 0\) (note that we are still assuming \(\LTs > 0\)). In the next section, we present an example of an electromagnetic instability in the \(Z\)-pinch geometry that exists exclusively in good-curvature regions, viz., \(\LB < 0\).

\section{Low-beta, drift-kinetic fluctuations in the local kinetic limit}
\label{sec:lowbeta_dk}

\subsection{Ordering and equations}
\label{sec:ordering}

Our goal is to simplify the GK equation \cref{eq:gyrokinetic_equation} in order to distil a minimum model for the good-curvature electromagnetic instability discussed in \cref{sec:mdm}. We limit ourselves to the low-beta, zero-magnetic-shear, \(Z\)-pinch geometry, and consider fluctuations that obey the ordering
\begin{equation}
	\frac{\me}{\mi}\ll\betae \sim \kperp^2\rhoe^2 \ll \kperpdesq \sim 1 \ll \kperp^2 \rhoi^2
	\label{eq:lowbeta_kperp_ordering}
\end{equation}
and evolve on time scales
\begin{equation}
	\omega \sim \kpar \vthe \sim \omegaste \sim \omegaTe \sim \omegade
	\label{eq:lowbeta_freq_ordering}
\end{equation}
in a plasma of two species, electrons and ions. The drift frequencies in \cref{eq:lowbeta_freq_ordering} are
\begin{align}
	\omegaste = \frac{k_y c \Te}{e B \Lne}, \quad \omegaTe = \frac{k_y c \Te}{e B \LTe}, \quad \omegads = \frac{k_y c \Te}{e B \LB}.
	\label{eq:drift_frequency}
\end{align}
In \cref{appendix:drift_kinetic_derivation}, we show that, under the above assumptions, the fluctuations obey the electron drift-kinetic equation [see \cref{eq:electron_drift_kinetics_appendix}]. Under the low-beta assumption, \(\dBpar\) is small and this drift-kinetic equation is closed by quasineutrality \cref{eq:quasineutrality} and the parallel component of Amp\`ere's law \cref{eq:parallel_amperes_law} [see \cref{eq:quasineutrality_mdm_appendix} and \cref{eq:parallel_amperes_mdm_appendix}, respectively]. In \cref{appendix:lowbeta_conservation}, we derive the forms of the free-energy and field-invariant conservation laws under the orderings \cref{eq:lowbeta_kperp_ordering,eq:lowbeta_freq_ordering}. As expected, the free energy is a positive-definite quantity, injected by the radial heat flux. In contrast, the field invariant contains negative-definite electrostatic and positive-definite electromagnetic contributions. It is injected by the slab term, proportional to \(\kpar\), and by the curvature term, proportional to the magnetic-field gradient \cref{eq:LB_def} and the radial heat flux.

In \cref{appendix:analytical_disp}, we outline the derivation of the dispersion relation for the low-beta, drift-kinetic fluctuations. In the three-dimensional case (viz., \(\kpar \neq 0\)), the dispersion relation is given by unwieldy and mostly unenlightening expressions that can be found in \cref{appendix:analytical_disp}. We now focus on the two-dimensional case, which represents a minimal model for (at least one example of) the good-curvature-driven instabilities.

\subsection{Two-dimensional fluctuations}
\label{sec:2d}

In the two-dimensional limit, viz., \(\kpar= 0\), the electron drift-kinetic equation decouples into its even and an odd parts in \(\vpar\) [see \cref{eq:gk_2d_even_appendix} and \cref{eq:gk_2d_odd_appendix}]. The former, together with quasineutrality, describes purely electrostatic (\(\dApar = 0\)) modes. Their dispersion relation is found to be
\begin{align}
	&-1-\tau^{-1}+\frac{\omega-\omegaste}{2\omegade}\rmZ\left(\sqrt{\frac{\omega}{2\omegade}}\right)^2 \nonumber \\ 
	&- \frac{\omegaTe}{2\omegade}\left[2\sqrt{\frac{\omega}{2\omegade}}\rmZ\left(\sqrt{\frac{\omega}{2\omegade}}\right) + \left(\frac{\omega}{\omegade}-1\right)\rmZ\left(\sqrt{\frac{\omega}{2\omegade}}\right)^2\right] = 0,
	\label{eq:es_etg_2d}
\end{align}
where \(\tau \equiv e\Ti / \qi \Te\) and \(\rmZ(\zeta)\) is the usual plasma dispersion function \citep{faddeeva54, fried61} given by
\begin{equation}
	\rmZ(\zeta) = \frac{1}{\sqrt{\pi}} \intinf{u} \frac{\rme^{-u^2}}{u - \zeta}
	\label{eq:Z_def}
\end{equation}
for \(\im(\zeta) > 0\) and analytically continued to the entire complex plane. \Cref{eq:es_etg_2d} is the same as the dispersion relation derived by \citet{biglari89} and \citet{zocco18} for the electrostatic cITG instability, up to replacing \(\tau \mapsto \tau^{-1}\) and \(i \mapsto e\), which accounts for the fact that we are dealing with cETG instead. The solutions to \cref{eq:es_etg_2d} contain the familiar cETG modes that require bad magnetic curvature to be unstable. 

The odd part of the electron drift-kinetic equation and parallel Amp\`ere's law also form a closed system, this time of purely electromagnetic (\(\phipot = 0\)) modes. Physically, these are fluctuations with no density or temperature perturbations but a nonzero parallel current. Their dispersion relation is
\begin{align}
	&-\kperpdesq + \frac{\omega - \omegaste}{\omegade} \left[2 + 2\sqrt{\frac{\omega}{2\omegade}}\rmZ\left(\sqrt{\frac{\omega}{2\omegade}}\right) + \frac{1}{2}\rmZ\left(\sqrt{\frac{\omega}{2\omegade}}\right)^2 \right] \nonumber \\
	&-\frac{\omega\omegaTe}{\omegade^2}\left[1 + \sqrt{\frac{\omega}{2\omegade}}\rmZ\left(\sqrt{\frac{\omega}{2\omegade}}\right) +\frac{1}{2}\rmZ\left(\sqrt{\frac{\omega}{2\omegade}}\right)^2 \right]=0.
	\label{eq:dispersion_MDM2D}
\end{align}

The appearance of a square root in \cref{eq:es_etg_2d} and \cref{eq:dispersion_MDM2D} means that additional care must be taken due to the presence of a branch cut in the complex plane. This is a well known problem \citep{kim94, kuroda98, sugama99, helander11, mishchenko18} stemming from the presence of the magnetic drifts. The square-root branch must be chosen so that it agrees with the principal branch of the square-root function in the upper half of the complex plane. Its branch cut can be anywhere in the lower half of the complex \(\omega\) plane. More details on how to handle the branch cuts and the analytic continuation of the dispersion relation in a \(Z\)-pinch can be found in \citet{ivanov23}.

To avoid the complications arising from the square root, we have chosen to fix the sign of the magnetic-field gradient to \(\omegade > 0\). This is necessary in order to put \(\omegade\) under the square root in \cref{eq:es_etg_2d} and \cref{eq:dispersion_MDM2D}. With this choice, `bad curvature' corresponds to \(\omegaTe > 0\), `good curvature' to \(\omegaTe < 0\). This differs from our discussion in \cref{sec:curvature_driven_modes} and~\cref{sec:local_geometry} where we fixed the sign of the temperature-gradient length scale \(\LTs\) but varied the local sign of the magnetic gradient. Note that we will only investigate equilibria in which the temperature and density gradients are aligned, i.e., \(\LTs\) and \(\Lns\) will always have the same relative sign.

Before we investigate the novel, good-curvature electromagnetic instability that is about to emerge, let us recap some of the properties of the more widely known electrostatic cETG instability. This will be useful later on in order to compare and contrast the electrostatic and electromagnetic curvature-driven instabilities.

\subsubsection{Two-dimensional cETG instability}
\label{sec:cETG}

\begin{figure}
	\begingroup\import{figs/fig2}{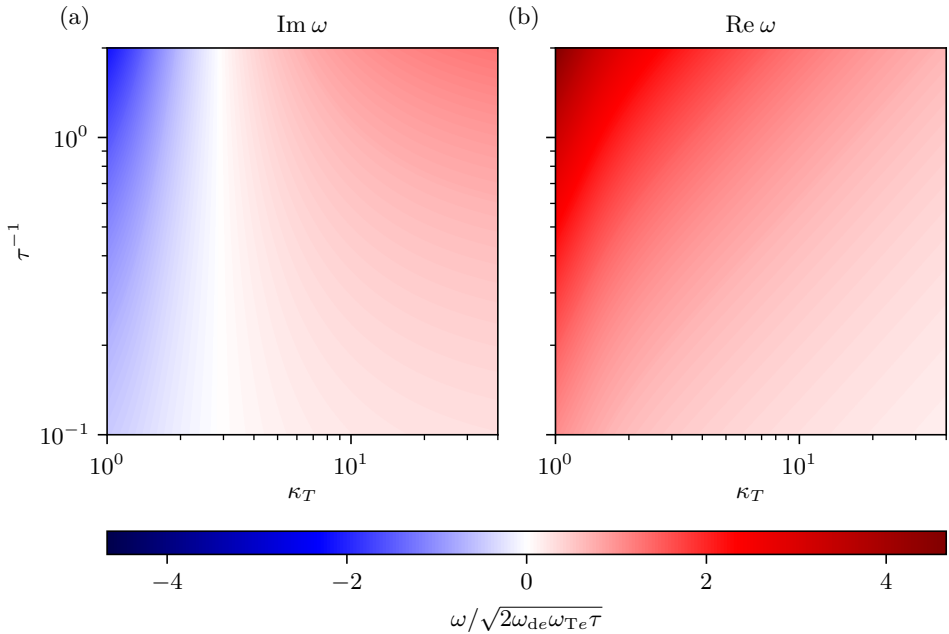}\endgroup
	\caption{Growth rate (a) and frequency (b) of the cETG mode as a function of the normalised temperature gradient \(\kappaT\) and inverse temperature ratio \(\tau^{-1}\) at zero density gradient (\(\kappan = 0\)). These are the solutions of \cref{eq:es_etg_2d}. The growth rate and frequency have been normalised by the fluid growth rate \cref{eq:cETG_fluid_dispersion}. Only the unstable, i.e., \(\im\:\omega>0\), region is shown.}
	\label{fig:etg_omega}
\end{figure}

In the two-dimensional limit, the electrostatic dispersion relation \cref{eq:es_etg_2d} has three independent parameters, which are the temperature ratio \(\tau\) and the normalised equilibrium gradients
\begin{equation}
	\kappaT \equiv \frac{\omegaTe}{\omegade} = \frac{\LB}{\LTe}, \quad \kappan \equiv \frac{\omegaste}{\omegade} = \frac{\LB}{\Lne}.
	\label{eq:kappaT_and_kappan_defs}
\end{equation}
In this notation, \(\kappaT > 0\) corresponds to bad magnetic curvature. Note that if we normalise \(\omega\) to any of the diamagnetic frequencies (or, indeed, a combination of them), then the dispersion relation for the normalised frequency is independent of the perpendicular wavenumber \(\vkperp\). Equivalently, any linear solution to the (electrostatic) electron drift-kinetic equation \cref{eq:gk_2d_even_appendix} obeys \(\omega \propto k_y\). This is a reflection of the scale invariance of electrostatic drift kinetics \citep{adkins23}, which is only broken at small scales by the finite-Larmor-radius effects (viz., the Bessel functions) that we neglected in \cref{sec:ordering}. \Cref{fig:etg_omega} shows the growth rate and frequency of the unstable cETG mode. The instability exists only when \(\kappaT > \kappa_{T\text{crit}}\), where \(\kappa_{T\text{crit}} > 0\) is some positive critical temperature gradient that depends on \(\tau\) and \(\kappan\). Additionally, the (real) frequency of the cETG mode satisfies \(\re(\omega)/\omegaTe > 0\), i.e., the phase velocity of the mode is in the electron diamagnetic direction.

A key characteristic of the cETG instability is that it has a physically transparent `fluid' limit. Mathematically, this limit corresponds to \(\omega \gg \omegade\), wherein we can use the asymptotic expansion
\begin{equation}
	\rmZ(\zeta) \sim -\frac{1}{\zeta} -\frac{1}{2\zeta^3} - \frac{3}{4\zeta^5} + \order{\zeta^{-7}},
	\label{eq:Z_largezeta_expansion}
\end{equation}
valid for \(\abs{\zeta} \gg 1\) and \(\im\:\zeta > 0\). The electrostatic dispersion relation \cref{eq:es_etg_2d} then simplifies to
\begin{equation}
	\omega^2 = -2\omegaTe\omegade\tau.
	\label{eq:cETG_fluid_dispersion}
\end{equation}
This is the well-known fluid cETG dispersion relation \citep[e.g.,][]{adkins22} that yields unstable solutions only if the curvature is bad, i.e., if \(\omegaTe\) and \(\omegade\) have the same sign (or, equivalently, \(\kappaT > 0\)). Evidently, the `fluid' assumption \(\omega \gg \omegade\) is consistent with \cref{eq:cETG_fluid_dispersion} only in the strongly driven regime \(\omegaTe \gg \omegade\). 

The `fluidisation' of ETG in the limit \mbox{\(\omega \gg \omegade\)} can be justified rigorously by performing a systematic decomposition of the drift-kinetic equation into an infinite hierarchy of fluid moments. One popular way of doing this is the Hermite-Laguerre decomposition \citep{smith97_thesis, watanabe04, zocco11, zocco15, loureiro16, mandell18, adkins22, frei22, frei23, mischenko24_thesis}. In the limit where the mode frequency is much larger that the `kinetic' frequencies \(\kpar \vthe\) and \(\omegade\), the moment hierarchy is naturally truncated to the six lowest-order fluid moments listed in Appendix A.5.4 of \citet{adkins22}. Reducing this to the two-dimensional limit (\(\kpar = 0\)), we find a closed system of fluid equations describing the perturbations of the electron density \(\dne\), parallel temperature \(\dTpare\), and perpendicular temperature \(\dTperpe\) \citep{adkins22}:
\begin{align}
	&\partd{}{t}\frac{\dnek}{\ne} + \rmi \omegade\left(\frac{\dTparek}{\Te} + \frac{\dTperpek}{\Te}\right) = 0, \label{eq:dne_etg} \\
	&\partd{}{t}\frac{\dTparek}{\Te}  + \rmi \omegaTe \frac{e\phipotk}{\Te} = 0, \label{eq:dTpare_etg} \\
	&\partd{}{t}\frac{\dTperpek}{\Te} + \rmi \omegaTe \frac{e\phipotk}{\Te} = 0, \label{eq:dTperpe_etg}
\end{align}
where the density and electrostatic potential are related by \(\dnek/\ne = -e\phipotk / \tau \Te\). The system \crefrange{eq:dne_etg}{eq:dTperpe_etg} is the minimal fluid model for the strongly driven, two-dimensional, electrostatic collisionless cETG instability. It is straightforward to confirm that its dispersion relation is \cref{eq:cETG_fluid_dispersion}. The fluidisation of cETG is discussed further in \cref{appendix:fluid_limit}.

\subsubsection{Two-dimensional magnetic-drift mode}
\label{sec:mdm}


\begin{figure}
	\begingroup\import{figs/fig3}{fig3.pgf}\endgroup
	\caption{Growth rate (a) and frequency (b) of the MDM as a function of \(\kperpdesq\) and \(\kappaT\). These are the solutions of \cref{eq:dispersion_MDM2D}. The solid black lines are the stability boundaries, on which \(\im\:\omega=0\). Their asymptotic slopes at large \(\kappaT\) are given by \cref{eq:upper_stability_line_slope} and \cref{eq:lower_stability_line_slope}. On the horizontal dotted grey line, the MDM root of \cref{eq:dispersion_MDM2D} is \(\omega = 0\). The value of \(\kperpdesq\) there is given by \cref{eq:kperp2de2_min}. The vertical dashed-dotted grey lines denote the ends of the solid black stability boundaries and are given by the limits of the black dashed line~\cref{eq:dashed_line_limits}. In the hatched region, there does not exist a unique root that is continuously connected to the unstable MDM solution, thus we have omitted that part of the plot [see also the discussion after \cref{eq:lower_stability_line_slope}]. }
	\label{fig:mdm_omega}
\end{figure}

The electromagnetic dispersion relation \cref{eq:dispersion_MDM2D} has three independent parameters, which are \(\kperpdesq\) and the normalised gradients \cref{eq:kappaT_and_kappan_defs}. Note that, in contrast with the electrostatic case, the perpendicular wavenumber enters explicitly as a parameter because the presence of the electron skin depth \(\de\) breaks the scale invariance of the equations. \Cref{fig:mdm_omega} show the growth rate and frequency of the unstable solutions of \cref{eq:dispersion_MDM2D} as a function of \(\kperpdesq\) and \(\kappaT\). The instability is present only when \(\kappaT < 0\), i.e., when the magnetic curvature is good. This is the MDM. There are three stability boundaries, where the growth rate of the MDM vanishes. The details of their derivation can be found in \cref{appendix:mdm_stability}.

The first of these (the dotted grey line in \cref{fig:mdm_omega}) lies at \(\kperpdesq = k_{\perp\text{,min}}^2\de^2\), where
\begin{equation}
	 k_{\perp\text{,min}}^2\de^2 = -\frac{4-\pi}{2}\kappan
	\label{eq:kperp2de2_min}
\end{equation}
and is bounded by
\begin{equation}
	-\frac{4-\pi}{\pi-2} + \kappan \leq \kappaT \leq \frac{2}{3}\kappan.
	\label{eq:dashed_line_limits}
\end{equation}
On this line, \(\omega = 0\), i.e., both the frequency and growth rate of the MDM vanish. Note that, as promised above, we only consider equilibria where the temperature and density gradients are aligned. In particular, if \(\kappaT\) is negative, then so is \(\kappan\).

The upper stability boundary (the solid black line in \cref{fig:mdm_omega}) asymptotes to
\begin{equation}
	\kperpdesq \approx -0.61\kappaT
	\label{eq:upper_stability_line_slope}
\end{equation}
at large \(\kappaT\). To the left of it, the MDM solution of \cref{eq:dispersion_MDM2D}, having crossed the positive real line in the \(\omega\) plane, is a damped mode. The lower stability boundary (the other solid black line in \cref{fig:mdm_omega}) asymptotes to
\begin{equation}
	\kperpdesq \approx -0.09\kappaT
	\label{eq:lower_stability_line_slope}
\end{equation}
at large \(\kappaT\). On this boundary, the MDM solution and its complex conjugate collapse into a repeated root of \cref{eq:dispersion_MDM2D}, which, to the right of the boundary, splits into two purely real solutions. There is an ambiguity as to which of these should be thought of as the continuation of the MDM, so we have omitted that part of the plot. A similar issue occurs below \(k_{\perp\text{,min}}^2\de^2\), and so we have left that part of the plot empty as well. There are no unstable solutions in either of these regions, whether continuously connected to the MDM solution or not. We shall continue the discussion of the MDM solution in the complex plane in \cref{sec:magnetic_drift_wave}.

A characteristic feature of the MDM is that the sign of its real frequency is not given by either \(\omegaTe\) or \(\omegade\). The mode can propagate in either the electron (\(\re \: \omega/\omegaTe > 0\)) or the ion (\(\re \: \omega/\omegaTe < 0\)) diamagnetic direction, depending on the size, but not the sign, of the temperature gradient. In contrast, the electrostatic cETG instability can be understood as a destabilised drift wave whose phase velocity is always in the electron diamagnetic direction. This variability of the direction of poloidal propagation of the MDM is not entirely surprising. Indeed, a defining property of the good magnetic curvature is that the magnetic drifts oppose the diamagnetic drifts that arise from the equilibrium gradients. Therefore, both poloidal directions can be associated with one of the electron drifts. In the strongly driven limit \(\abs{\kappaT} \to \infty\), the maximum growth rate is attained at
\begin{equation}
	\kperpdesq \approx -0.14\kappaT,
\end{equation}
where the complex frequency is given by
\begin{equation}
	\frac{\omega}{\omegade} \approx 0.12 + 1.88\rmi.
\end{equation}
Thus, the phase velocity of the fastest-growing, strongly driven MDM is in the direction of the electron magnetic drifts, which is opposite to that of the electron diamagnetic flows.

There is, however, an important physical distinction between the magnetic drifts, caused by the magnetic-field gradients, and the diamagnetic ones, resulting from the inhomogeneous plasma equilibrium. The magnetic drifts are `real' particle drifts, while the diamagnetic ones are only `apparent' drifts, in the sense that they are not associated with the motion of individual particles. Thus, resonant effects (like Landau damping) are possible only with the magnetic drifts. However, the MDM is not a resonant instability. This is manifestly true as the mode can propagate in either direction relative to the poloidal particle motion (given by the magnetic drift). We discuss this further in \cref{sec:magnetic_drift_wave}, where we show that the MDM root of the dispersion relation \cref{eq:dispersion_MDM2D} is continuously connected to a non-resonant, undamped magnetic drift wave propagating against the magnetic drifts. 

Finally, unlike cETG, the MDM is not a fluid instability since its complex frequency always satisfies \(\omega \lesssim \omegade\). Consequently, the MDM does not have a fluid counterpart and cannot be captured by a low-order truncation of the infinite hierarchy of fluid moments of electron drift-kinetic equation. This is discussed further in \cref{appendix:fluid_limit}.

\subsubsection{Physical ingredients and complex-plane behaviour of the magnetic-drift mode}
\label{sec:magnetic_drift_wave}

\begin{figure}
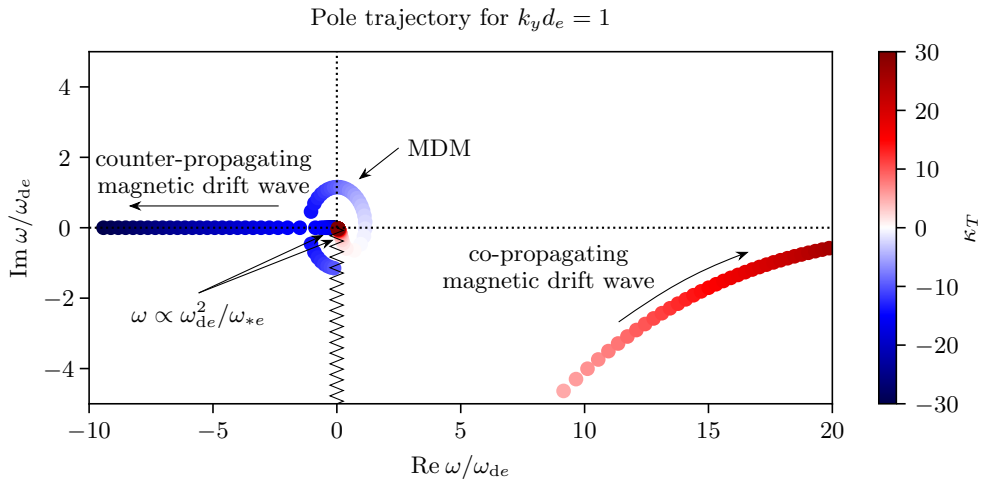

		\begin{tikzonpgf}{figs}{fig4.pgf}
			
		\draw[decorate,decoration={zigzag,segment length=4pt}] (0.338, 0.535) -- (0.338, 0.17);	
		
		\draw[-{Stealth[scale=1]}] (0.28, 0.58) -- node[midway, above, align=center] {counter-propagating\\magnetic drift wave} (0.13, 0.58);
		
		\draw[-{Stealth}] (0.62, 0.34) to[out=35, in=200] (0.75, 0.485);
		
		\def\textloc{(0.55, 0.45)};
		\node[align=center] at \textloc {co-propagating\\magnetic drift wave};
		
		\def\mdmtextloc{(0.4, 0.7)};
		\node[anchor=west] at \mdmtextloc {MDM};
		\draw[-{Stealth}] \mdmtextloc -- (0.36, 0.635);
		
		\def\inertialtextloc{(0.2, 0.4)};
		\node[anchor=north] at \inertialtextloc {\(\omega \propto \omegade^2/\omegaste\)};
		\draw[-{Stealth}] \inertialtextloc -- (0.325, 0.52);
		\draw[-{Stealth}] \inertialtextloc -- (0.335, 0.51);
	\end{tikzonpgf}
	\caption{A visualisation of the motion of the MDM solution to \cref{eq:dispersion_MDM2D} in the complex \(\omega\) plane as a function of \(\kappaT\). The colours indicate the value of \(\kappaT\), as indicated in the colour bar on the right. The zigzagging black line is the branch cut of the square root in \cref{eq:dispersion_MDM2D}.}
	\label{fig:pole_trajectory}
\end{figure}

Since the MDM is neither resonant nor `fluidisable', it is difficult to come up with a convincing physical picture of the feedback mechanism that underpins this instability. Let us explore how the root of the dispersion relation \cref{eq:dispersion_MDM2D} that corresponds to the MDM connects to other known low-beta modes. In \cref{fig:pole_trajectory}, we show the complex-plane trajectory of the roots of \cref{eq:dispersion_MDM2D} as functions of the temperature gradient \(\kappaT\).\footnote{Of course, there are infinitely many roots of \cref{eq:dispersion_MDM2D} below the real line. Here, we have focused on the part of the complex plane that contains the physically important, i.e., the least damped, Landau modes.} 

At large temperature gradients, viz., \(\abs{\kappaT} \gg 1\), there are two weakly damped (or undamped) solutions. The first one satisfies \(\omega \sim \omegaTe \gg \omegade\) for \(\abs{\kappaT} \gg 1\) and can be identified as the magnetic drift wave \citep{adkins22, chandran24}. As \(\omega \gg \omegade\), this wave is a `fluid' mode, and, as expected, can be described using only low-order moments of the distribution function. Setting \(\LB^{-1} = 0\), we take the \(\vpar\) moment of the electron drift-kinetic equation \cref{eq:gk_2d_odd_appendix} to find the electron-momentum equation
\begin{equation}
	\ne\me\partd{\upare}{t} = -\frac{\vdBperp}{B}\bcdot\grad \pe -e\ne \dEpar.
	\label{eq:upar_magnetic_drift_wave}
\end{equation}
Physically, \cref{eq:upar_magnetic_drift_wave} expresses the change in parallel momentum of the electrons as a result of the inductive parallel electric field (note that since \(\kpar=0\), \(\dEpar\) has no electrostatic part)
\begin{equation}
	\dEpar = -\frac{1}{c}\partd{\dApar}{t}
\end{equation}
and the equilibrium pressure gradient along the perturbed field line
\begin{equation}
	-\frac{\vdBperp}{B} \bcdot \grad \pe = \left(\frac{\pe}{\Lne} + \frac{\pe}{\LTe}\right)\frac{\dBx}{B} \propto \left(\omegaste + \omegaTe\right)\partd{\dApar}{y}.
\end{equation}
Combining \cref{eq:upar_magnetic_drift_wave} and the parallel Amp\`ere's law \cref{eq:parallel_amperes_mdm_appendix}, we obtain a purely oscillating mode with frequency
\begin{equation}
	\omega = \frac{\omegaste + \omegaTe}{1 + \kperpdesq}.
	\label{eq:magnetic_drift_wave_freq}
\end{equation}
With the inclusion of finite but small magnetic drifts, the fate of the magnetic drift wave is determined by the relative sign of \(\omegade\) and the diamagnetic frequencies \(\omegaste\) and \(\omegaTe\). If the magnetic drifts are aligned with the direction of propagation, i.e., if \(\kappaT > 0\), then the wave experiences a form of Landau damping due to the resonance between the poloidal drift of the electrons and the phase velocity of the wave.\footnote{Mathematically, this can be seen from the form of the denominator in \cref{eq:Iab_def_appendix}. As the two-dimensional case corresponds to dropping the \(u\) factor, it is clear that resonance is possible only if the signs of \(\zeta\propto\omega\) and \(\zetad\propto\omegade\) are the same.} In contrast, if the magnetic drifts are opposite to the direction of propagation of the magnetic drift wave, i.e., if \(\kappaT < 0\), no resonance is possible and the wave remains undamped. The difference in behaviour of co- and counter-propagating drift waves is evident in \cref{fig:pole_trajectory}. There, the dark red points show the behaviour of the solution as \(\kappaT \to +\infty\), where the solution converges to a damped, co-propagating magnetic drift wave. In contrast, as \(\kappaT \to -\infty\) (dark blue points), the solution is an undamped, counter-propagating magnetic drift wave.

Apart from the magnetic drift wave \cref{eq:magnetic_drift_wave_freq}, a small but finite magnetic-drift frequency \(\abs{\omegade} \ll \abs{\omegaTe}\) gives rise to another wave whose frequency satisfies
\begin{equation}
	\omega \sim \frac{(\kperp^2 - k_{\perp\text{,min}}^2)\de^2\omegade^2}{\omegaTe}.
	\label{eq:inertial_magnetic_drift_wave_scaling}
\end{equation}
This mode is purely oscillatory if \(\kappaT < 0\) and damped if \(\kappaT > 0\) due to the resonance with the magnetic drifts. Unlike the magnetic drift wave, the wave \cref{eq:inertial_magnetic_drift_wave_scaling} cannot be described in terms of low-order moments as \(\omega \ll \omegade\). As \(\kappaT\) is increased from \(-\infty\) towards zero, the magnetic drift waves \cref{eq:magnetic_drift_wave_freq} and \cref{eq:inertial_magnetic_drift_wave_scaling} are both undamped and counter-propagating (relative to the magnetic drift) if \(\kperpdesq > k_{\perp\text{,min}}^2\de^2\). The MDM appears when their frequencies meet at some finite, negative \(\kappaT\). This value of \(\kappaT\) lies on the lower stability boundary in \cref{fig:mdm_omega}. As \(\kappaT\) increases, the MDM solution changes from counter- (\(\omega < 0\)) to co-propagating (\(\omega > 0\)), before crossing the real line again when \(\kappaT\) crosses the upper stability boundary in \cref{fig:mdm_omega} at some \(\kappaT < 0\). Increasing \(\kappaT\) further to~\(+\infty\) connects the MDM solution to the co-propagating mode satisfying \cref{eq:inertial_magnetic_drift_wave_scaling}, which is resonantly damped. As \(\kappaT \to +\infty\), there is also a co-propagating magnetic drift wave. This wave, unlike its counter-propagating part, is not connected to the MDM solution and experiences finite damping due to a resonance with the magnetic drifts. \Cref{fig:pole_trajectory} shows the motion of the modes in the complex \(\omega\) plane as a function of \(\kappaT\).

Finally, we can gain more insight into the MDM by artificially removing terms from the equations. While unphysical, this can help us understand how the feedback mechanism works and what the minimal set of ingredients for this instability is. We find that removing the \(\grad B\) drifts, viz., the magnetic drifts proportional to \(\vperp^2\), does very little to change the behaviour of the instability. Even in their absence, we find a mode with the same qualitative behaviour as the MDM. However, removing the curvature drifts eliminates the instability completely. This is in stark contrast with the curvature-driven ETG instability discussed in \cref{sec:cETG}, which is qualitatively unchanged if either (but not both!) of the drifts is removed. In the strongly driven regime, its growth rate simply reduces by a factor of two if one of the drifts is not present. We outline these calculations in \cref{appendix:nogradB_or_curvature}. Then, in \cref{appendix:delta_equilibrium}, we abandon the Maxwellian equilibrium entirely and consider a greatly simplified \(\Fe\) made up of two infinitely narrow (in \(\vpar\)) colliding beams that allows us to obtain qualitatively the same mode. That model suggests that the stabilisation at \(\re\:\omega > 0\), viz., the upper stability boundary in \cref{fig:mdm_omega}, is a consequence of the Landau-like damping of the co-propagating MDM. 

\subsection{Magnetic-drift modes in three dimensions}
\label{sec:mdm_3d}

\begin{figure}
	\begingroup\import{figs/fig5}{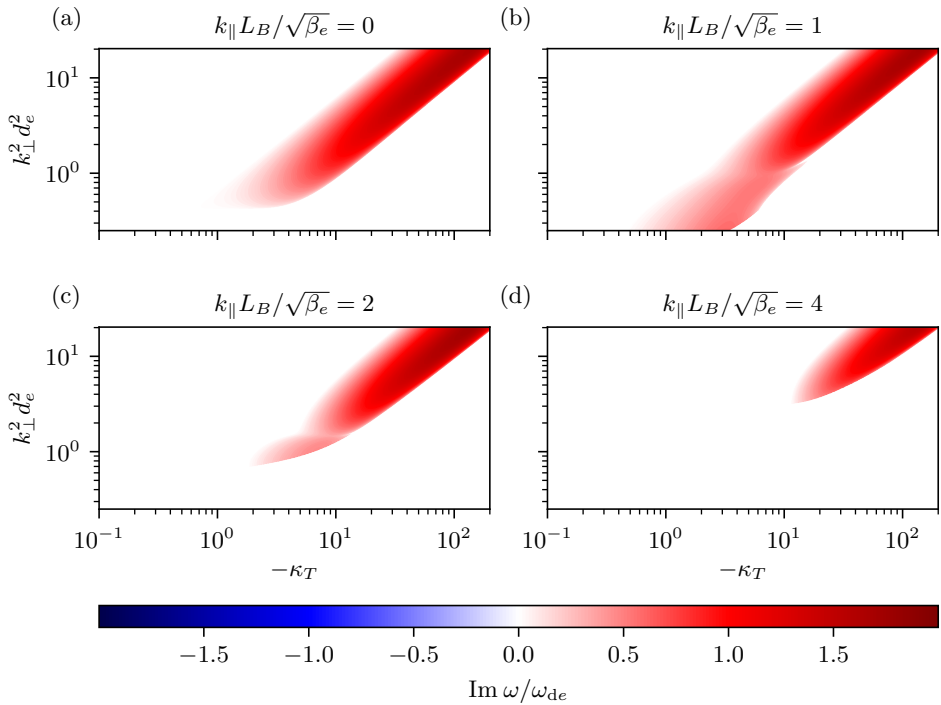}\endgroup
	\caption{MDM growth rate for \(\kappan = -1\) (as in \cref{fig:mdm_omega}) for four different values of \(\kpar\) as indicated in the title of each panel. Note that \cref{eq:lowbeta_kperp_ordering} and \cref{eq:lowbeta_freq_ordering} imply that the parallel wavenumber always satisfies \(\kpar \LB \sim \sqrt{\betae}\).}
	\label{fig:mdm_3d}
\end{figure}

Restoring finite \(\kpar\) requires solving the three-dimensional dispersion relation \cref{eq:disp_appendix}. \Cref{fig:mdm_3d} is a visualisation of such a solution for three-dimensional fluctuations in the \(Z\)-pinch geometry as a function of \(\kperpdesq\) and \(\kappaT\). We see that finite \(\kpar\) stabilises the MDM at large scales and low temperature gradients, but the instability survives at large gradients.

We can make the following heuristic argument about the size of \(\kpar\) that will be sufficient to affect the MDM. Taking the strongly driven case \(\omegaTe \gg \omegade\) (or, equivalently, \(\kappaT \gg 1\)), we expect that three-dimensional effects will not introduce any qualitative changes to the mode if
\begin{equation}
	\kpar \vthe \lesssim \omegade \quad \Rightarrow \quad \kpar \LB \lesssim k_y\rhoe.
	\label{eq:kparvthe_vs_omegade}
\end{equation}
By \cref{eq:upper_stability_line_slope} and \cref{eq:lower_stability_line_slope}, the MDM exists at scales \(\kperpdesq \sim \kappaT\), so \cref{eq:kparvthe_vs_omegade} implies
\begin{equation}
\kpar \lesssim \sqrt{\frac{\betae}{\LB\LTe}}.
\end{equation}
Since the MDM is destabilised only in the good-curvature region, the parallel extent of the mode is limited by the connection length between the inboard and outboard side of the device, viz., 
\begin{equation}
	\kpar \gtrsim \frac{1}{qR},
\end{equation}
where \(q\) is the safety factor and \(R\sim\LB\) is the major radius. Therefore, we expect the MDM to be destabilised in toroidal geometry if
\begin{equation}
	\frac{\betae\LB}{\LTe} \gtrsim \frac{1}{q^2},
	\label{eq:MDM_toroidal_condition}
\end{equation}
which shows that, for a fixed geometry, steeper temperature gradients and/or higher values of \(\betae\) will trigger the MDM. In \cref{eq:MDM_toroidal_condition}, we neglect order-unity factors that will undoubtedly impact the true condition for MDM destabilisation in toroidal geometry. However, we expect this simple scaling estimate to capture the trend of the MDM threshold as a function of \(\betae\) and of the temperature gradient.

In addition to the threshold for the MDM, we can make a rough estimate of the expected growth rate and compare it to that of other known instabilities. Recall that both the real and imaginary parts of the frequency of the MDM satisfy \(\omega \sim \omegade\). Combining this with the requirement that \(\kperpdesq \sim \kappaT\), the growth rate of the MDM can be estimated as
\begin{equation}
	\gamma_\text{MDM} \sim \frac{\rhoe \vthe k_y }{\LB} \sim \frac{\vthe}{\sqrt{\LB\LTe}}\sqrt{\betae}.
	\label{eq:mdm_growth_estimate}
\end{equation}
To put this in the context of more common GK instabilities, let us compare it with the usual estimate for the growth rate of the cITG instability, viz.,
\begin{equation}
	\gamma_\text{ITG} \sim \frac{\vthi}{\sqrt{\LB\LTi}}.
	\label{eq:itg_growth_estimate}
\end{equation}
Assuming similar temperature gradients, viz., \(\LTi \sim \LTe\), we find
\begin{equation}
	\frac{\gamma_\text{MDM}}{\gamma_\text{ITG}} \sim \sqrt{\frac{\betae\mi}{\me}},
	\label{eq:mdm_over_itg_estimate}
\end{equation}
and so the MDM growth rate is (potentially) larger than the cITG one in any device where \(\betae \gg \me/\mi\).\footnote{Indeed, our derivation of the MDM is based on this ordering, see \cref{eq:lowbeta_kperp_ordering}.} Of course, \crefrange{eq:mdm_growth_estimate}{eq:mdm_over_itg_estimate} are very rough estimates that are true only within potentially important factors of order unity. They also fail to consider the possible mitigation of the instabilities due either to neglected linear effects, like toroidal geometry and magnetic shear, or to nonlinear effects, e.g., suppression of fluctuations as a result of multiscale interactions \citep{waltz07, candy07, maeyama15, maeyama17, hardman19, hardman20}. It is also worth noting that, in electron-scale simulations of GK turbulence, even if present, the MDM is likely to be subdominant to the more common electrostatic ETG instabilities, whose growth rate,
\begin{equation}
	\gamma_\text{ETG}\sim\frac{\vthe}{\sqrt{\LB\LTe}},
\end{equation}
is a factor of \(\betae^{-1/2}\) larger than \cref{eq:mdm_growth_estimate}. While this could make it harder to identify the MDM in linear simulations, its effects may still be crucial for the nonlinear turbulent state because it can drive turbulence in the good-curvature regions where the ETG modes are stable.

It is likely that what we call the MDM here is, in fact, related to other, already observed good-curvature electromagnetic GK instabilities. The intermediate-wavelength microtearing mode (iMTM) reported by \citet{patel21_thesis} is an electromagnetic, good-curvature mode with a phase speed in the ion diamagnetic direction. Additionally, \citet{jian21} have performed linear and nonlinear GK simulations of high-\(\beta\) DIII-D equilibria, wherein they have identified a mode that they call `slab MTM'. Linearly, this mode is peaked at the good-curvature side of the device. Nonlinearly, it drives turbulent transport predominantly at the inboard side. Furthermore, they have identified it as a slab mode due to its strong dependence on \(q\), being destabilised at large values of \(q\), consistent with the analysis leading to \cref{eq:MDM_toroidal_condition}. Both of these examples match the properties of the MDM, at least qualitatively.

\section{Discussion}
\label{sec:discussion}

By considering the conservation laws of the free energy~\cref{eq:free_energy_injection} and the field invariant~\cref{eq:Y_injection} in GK, we have shown that the localisation of curvature-driven modes to regions of either good or bad magnetic curvature is a necessary condition for instability. In particular, these conservation laws, together with some reasonable assumptions stated in~\cref{sec:curvature_driven_modes}, allow us to prove that all electrostatic curvature-driven instabilities are localised to the so-called bad-curvature regions of the plasma, where the local magnetic curvature~\cref{eq:curv_def} is negative. Consequently, any instabilities in the good-magnetic-curvature region must be electromagnetic. It is important to note that, modulo the assumptions stated in~\cref{sec:curvature_driven_modes}, these results are independent of the precise nature of the instability and are thus equally valid for all curvature-driven modes. 

The MDM, whose dispersion relation in a low-beta \(Z\)-pinch is given by~\cref{eq:dispersion_MDM2D}, is, as far as the authors are aware, the first example of a simple, analytically derivable, good-curvature instability. In the simple model wherein we have derived it, the MDM exists at perpendicular length scales comparable to the electron skin depth \(\de\) and has a growth rate that is \(\sqrt{\betae}\) smaller than that of the cETG instability and much larger than that of the cITG instability in plasmas with \(\betae \gg \me/\mi\). The fact that it is localised to the good-curvature regions of the equilibrium, where electrostatic modes like cETG and cITG are stable, means that the MDM can be the dominant one there. As discussed in \cref{sec:magnetic_drift_wave}, the MDM is neither resonant nor fluid. Physically, it arises as a result of the \(\vpar^2\)-dependent magnetic-curvature drifts and the parallel streaming of electrons along perturbed field lines. Even though simplified, the model wherein we derived the MDM is a particular asymptotic limit of GK. Therefore, we expect that, under the right conditions (some of which are outlined in \cref{sec:mdm_3d}), the MDM exists in toroidal geometry, too. Indeed, it is likely that is related to some exotic varieties of MTM seen in recent high-\(\beta\) GK simulations \citep{patel21_thesis,jian21}.

Given that, both theoretically and experimentally, the vast majority of fusion-plasma research has concentrated on regimes where electrostatic instabilities dominate, it is unsurprising that the community has adopted the term `bad curvature' for the locations where these modes are unstable. However, the pursuit of high-\(\beta\) fusion devices, e.g., STEP \citep{tholerus24}, has opened a Pandora's box of electromagnetic instabilities and turbulence that challenge conventional intuition and suggest that the performance of such devices could be worse than expected if electromagnetic instabilities are not kept in check \citep{kennedy23, giacomin24}. The unstable MDM is likely only one of a class of good-curvature electromagnetic instabilities that are yet to be discovered and understood and whose role in the turbulent transport of high-\(\beta\) plasmas is currently unknown. For example, good-curvature instabilities driven by trapped particles in nonaxisymmetric plasmas could drive turbulence in stellarator equilibria otherwise thought to be (relatively) immune to trapped-particle instabilities \citep{rodriguez24}. It is thus clear that the study of the good-curvature instabilities, like the MDM, is a fertile field for future work.

\section*{Acknowledgements}
We would like to thank the organisers of the 7\textsuperscript{th} Vlasovia (2024) workshop in Florence, where an early version of this work was first presented. We would also like to thank the organisers of the 15\textsuperscript{th} Plasma Kinetics Working Meeting (2024) and its hosts at the Wolfgang Pauli Institute in Vienna.

\section*{Funding}
This work was supported by the Engineering and Physical Sciences Research Council (EPSRC) [EP/R034737/1]. PGI and AAS were also supported in part by the Simons Foundation via a Simons Investigator award to AAS. TA's contributions were supported in part by the U.S. Department of Energy under contract number DE-AC02-09CH11466. The United States Government retains a non-exclusive, paid-up, irrevocable, world-wide license to publish or reproduce the published form of this manuscript, or allow others to do so, for United States Government purposes. TA acknowledges the support of the Royal Society Te Ap\=arangi, through Marsden-Fund grant MFP-UOO2221.

\section*{Declaration of interests}
The authors report no conflict of interest.

\appendix

\section{Spatial averaging}
\label{appendix:spatial_average}

The average \(\avgbox{g(\vr)}\) that appears in the GK conservation laws is defined as a combination of a perpendicular and a flux-surface average, viz.,
\begin{equation}
	\avgbox{g(\vr)} = \left\langle\left\langle g(\vr) \right\rangle_\perp \right\rangle_\psi,
	\label{eq:avgbox_def}
\end{equation}
where
\begin{equation}
	\left\langle g(\vr) \right\rangle_\perp = \int_{S_\lambda} \rmdint[2]{\vr} g(\vr) \Bigg/ \int_{S_\lambda} \rmdint[2]{\vr}
	\label{eq:perpaverage_def}
\end{equation}
is an average in the plane perpendicular to \(\vB\) over an area \(S_\lambda\) with linear size \(\lambda\) that is an intermediate length between the microscales associated with the gyroradii and the macroscale of the equilibrium, viz.,
\begin{equation}
	\rhos \ll \lambda \ll L.
\end{equation}
By assumption in deriving the GK equation, if \(g(\vr)\) is a fluctuating quantity with perpendicular spatial scales
\begin{equation}
	\frac{g}{\abs{\gradperp g}} \sim \rhos,
\end{equation}
then the perpendicular average of \(g\) is an equilibrium-like quantity with
\begin{equation}
	\frac{\left\langle g \right\rangle_\perp}{\abs{\gradperp \left\langle g \right\rangle_\perp}} \sim L.
\end{equation}
In other words, the perpendicular average is a kind of coarse-graining that eliminates microscales. As a consequence, averaging a gradient of a fluctuating quantity reduces it by a factor of \(\rhos/L\), viz.,
\begin{equation}
	\frac{\abs{\left\langle \gradperp g \right\rangle_\perp}}{\abs{\gradperp g}} \sim \frac{\rhos}{L} \ll 1.
	\label{eq:perpaverage_kills_gradperp}
\end{equation}

The flux-surface average in \cref{eq:avgbox_def} is defined as
\begin{equation}
	\left\langle g(\vr) \right\rangle_\psi \equiv \lim_{\Delta \psi \to 0} \left[\int_{\Delta V(\psi)} \rmdint[3]{\vr} g(\vr) \Bigg/ \int_{\Delta V(\psi)} \rmdint[3]{\vr}\right],
	\label{eq:fsa_def}
\end{equation}
where \(\Delta V(\psi) = V(\psi+\Delta \psi) - V(\psi)\) and \(V(\psi)\) is the volume of the flux surface labelled by \(\psi\). The only property of the flux-surface average that we will need is that, for any \(g\),
\begin{equation}
	\left\langle \grad \bcdot(\vB g) \right\rangle_\psi = 0,
	\label{eq:flux_surface_average_kills_parallel_grad}
\end{equation}
which is straightforward to derive using \cref{eq:fsa_def}, the divergence theorem, and the fact that \(\vB\bcdot\vec{n} = 0\) where \(\vec{n}\) is the normal vector to the flux surface \citep{abel13}.

\subsection{Ballooning modes and local gyrokinetics}
\label{appendix:ballooning}

Following \citet{connor78}, we can represent any fluctuating quantity \(g(\vr)\) associated with a linear mode that satisfies \cref{eq:gyrokinetic_ordering} as
\begin{equation}
	g(\vr) = \sum_{p = -\infty}^{+\infty} \hat{g}(\psi, \theta + 2\pi p) \rme^{\rmi S(\psi, \theta + 2\pi p, \varphi)} + \text{c.c.},
	\label{eq:ballooning_rep}
\end{equation}
where the fast (viz., on the scale of \(\rhos\)) variation across the magnetic field lines is captured entirely by the eikonal \(S\) and \(\hat{g}\) varies slowly (viz., on the equilibrium length scale \(L\)) as a function of the flux-surface label \(x\). The eikonal is expressed as
\begin{equation}
	S(\psi, \theta, \varphi) = n(q\theta - \varphi) + S_\psi(\psi),
\end{equation}
where \(\varphi\), \(q\), and \(\theta\) are defined in \cref{sec:geo}, \(n\) is the toroidal wavenumber, \(q(\psi)\) is the safety factor, and \(\psi\) is the flux-surface label (typically, the poloidal flux). Since the perpendicular average \cref{eq:perpaverage_def} eliminates modes with fast perpendicular variation,
\begin{equation}
	\left\langle f(\vr) g (\vr) \right\rangle_\perp = 2\re\sum_{p = -\infty}^{+\infty} \hat{f}(\psi, \theta + 2\pi p)\hat{g}^*(\psi, \theta + 2\pi p)
	\label{eq:ballooning_product_avg}
\end{equation}
for any two quantities \(f(\vr)\) and \(g(\vr)\) expressed as \cref{eq:ballooning_rep}.\footnote{Strictly speaking, we have assumed that \(q\) is irrational. Here we do not discuss the complications that arise at rational flux surfaces.} Flux-surface-averaging \cref{eq:ballooning_product_avg}, we find
\begin{align}
	\avgbox{f(\vr)g(\vr)} &= 2\re \sum_{p = -\infty}^{+\infty} \lim_{\Delta \psi \to 0} \frac{1}{\Delta V(\psi)} \int_{\Delta V(\psi)} \rmdint[3]{\vr} \hat{f}(\psi, \theta + 2\pi p)\hat{g}^*(\psi, \theta + 2\pi p) \nonumber \\
	&= 2\re \sum_{p = -\infty}^{+\infty} \lim_{\Delta \psi \to 0} \frac{1}{\Delta V(\psi)} \int_{\Delta V(\psi)} \frac{\rmd \varphi \rmd \theta \rmd \psi}{\Jac} \hat{f}(\psi, \theta + 2\pi p)\hat{g}^*(\psi, \theta + 2\pi p) \nonumber \\
	&= 2\re \sum_{p = -\infty}^{+\infty} \lim_{\Delta \psi \to 0} \frac{\Delta \psi}{\Delta V(\psi)} \int_0^{2\pi} \rmd \varphi \int_0^{2\pi} \frac{\rmd \theta}{\Jac} \hat{f}(\psi, \theta + 2\pi p)\hat{g}^*(\psi, \theta + 2\pi p) \nonumber \\
	&= 2\re \paravg{\hat{f}(\psi, \theta)\hat{g}^*(\psi, \theta)},
\end{align}
where we have defined the parallel average of any function \(\hat{f}(\psi, \theta)\) to be
\begin{equation}
	\paravg{\hat{f}(\psi, \theta)} \equiv \int_{-\infty}^{+\infty} \frac{\rmd \theta}{\Jac} \hat{f}(\psi, \theta) \Bigg / \int_{0}^{2\pi} \frac{\rmd \theta}{\Jac}.
	\label{eq:ballooning_avgbox}
\end{equation} 
In the above, \(\Jac = (\grad \varphi \times \grad \theta)\bcdot \grad \psi\) is the Jacobian, and we have used the fact that, in axisymmetric geometry, \(\Jac\) is independent of \(\varphi\) at fixed \(\theta\) and \(\psi\).

Alternatively, in a GK flux tube with local coordinates \((x, y, z)\) and periodic boundary conditions across the field lines, we can Fourier transform in \(x\) and \(y\) as in \cref{eq:eikonal}. In such a flux tube, we define \(\avgbox{.}\) as the average over the entire flux tube
\begin{equation}
	\avgbox{g(\vr)} = \frac{1}{V_1} \intr g(\vr)
	\label{eq:flux_tube_avgbox_def}
\end{equation}
where \(V_1\) is an appropriate normalising volume. Then, using \cref{eq:eikonal} and \cref{eq:flux_tube_avgbox_def}, we find that, for two flux-tube quantities \(f(\vr)\) and \(g(\vr)\),
\begin{equation}
	\avgbox{f(\vr)g(\vr)} = \frac{S_\perp}{V_1}\sum_\vkperp \int_{-\infty}^{+\infty} \frac{\rmd z}{\Jac} f_\vkperp g_\vkperp^*,
\end{equation}
where \(S_\perp = \int \rmd x \rmd y\) is the perpendicular surface area of the flux tube and \mbox{\(\Jac = (\grad x \times \grad y)\bcdot \grad z\)} is the Jacobian of the flux-tube coordinate system. Note that \(\Jac\) is a function only of \(z\) since it is an equilibrium quantity whose length scale of variation \(L\) is much larger than the perpendicular extent of the flux tube (which is comparable to~\(\rhos\)). The above is equivalent to summing a spectrum of ballooning modes and averaging as per~\cref{eq:ballooning_avgbox} if we define the parallel average in the flux tube as
\begin{equation}
	\paravg{g} \equiv \frac{S_\perp}{V_1} \int_{-\infty}^{+\infty} \frac{\rmd z}{\Jac} g.
	\label{eq:paravg_def_flux_tube}
\end{equation}
Note that the choice of \(x \propto \psi\) and \(y \propto \alpha\) that we made in \cref{sec:gk}, together with \(z = \theta\) implies \mbox{\(\Jac \propto (\grad \varphi \times \grad \theta)\bcdot \grad \psi\)}. In this case, the normalising volume is
\begin{equation}
	V_1 = \int_{\theta \in (0, 2\pi)} \rmdint[3]{\vr} = S_\perp \int_{0}^{2\pi} \frac{\rmd \theta}{\Jac},
\end{equation}
i.e., it is the volume of the flux tube per \(2\pi\) period in \(\theta\), and \cref{eq:paravg_def_flux_tube} becomes
\begin{equation}
	\paravg{g} = \int_{-\infty}^{+\infty} \frac{\rmd \theta}{\Jac} g \Bigg / \int_{0}^{2\pi} \frac{\rmd \theta}{\Jac}.
	\label{eq:paravg_def_flux_tube_theta}
\end{equation}
Thus, \cref{eq:ballooning_avgbox} and \cref{eq:paravg_def_flux_tube} do indeed represent the same average.

More details on the correspondence between the ballooning representation and the local flux tubes can be found in \citet{beer95}.

\section{Gyrokinetic conservations laws}
\label{appendix:conservation_laws}

\subsection{Alternative expression for \(W\)}

Let us show that the free energy \cref{eq:free_energy_deltaf} can also be expressed as
\begin{equation}
	W = \sum_\s\avgbox[\Bigg]{\intv\avgr{\frac{\Ts\hs^2}{2\Fs}}} + \avgbox[\Bigg]{\frac{\dBpar^2}{8\pi} + \frac{\abs{\gradperp\dApar}^2}{8\pi} - \sum_\s \frac{\qs^2\ns}{2\Ts}\phipot^2}.
	\label{eq:free_energy_gk}
\end{equation}
This is a standard result \citep[see, e.g.,][]{abel13}, but we prove it here for completeness. Using \cref{eq:perturbed_distribution_function}, we find
\begin{equation}
	\sum_\s \avgbox[\Bigg]{\intv \frac{\Ts\dfs^2}{2\Fs}} = \sum_\s \avgbox[\Bigg]{\intv \avgr{\frac{\Ts\hs^2}{2\Fs} - \qs \phipot \hs + \frac{\qs^2\phipot^2}{2\Ts}\Fs}},
	\label{eq:W_expression_step1}
\end{equation}
where the overall gyroaverage \(\avgr{.}\) signifies that the \(\rmdint[3]{\vv}\) integral is performed with \(\vr\) held constant. By \cref{eq:quasineutrality}, we have
\begin{equation}
	\sum_\s \avgbox[\bigg]{\intv \avgr{\qs \phipot \hs}} = \sum_\s \avgbox[\bigg]{\intv \qs\phipot\avgr{\hs}} = \sum_\s \avgbox[\bigg]{\frac{\qs^2\ns}{\Ts}\phipot^2},
	\label{eq:W_expression_step2}
\end{equation}
and so \cref{eq:W_expression_step1} becomes
\begin{equation}
	\sum_\s \avgbox[\Bigg]{\intv \frac{\Ts\dfs^2}{2\Fs}} = \sum_\s \avgbox[\Bigg]{\intv \avgr{\frac{\Ts\hs^2}{2\Fs}}} - \sum_\s\avgbox[\Bigg]{\frac{\qs^2\ns}{2\Ts}\phipot^2}.
	\label{eq:W_expression_step3}
\end{equation}
Finally, using
\begin{equation}
	\abs{\vdB}^2 = \abs{\dBpar\ub - \ub \times \gradperp\dApar}^2 = \dBpar^2 + \abs{\gradperp\dApar}^2,
\end{equation}
we obtain \cref{eq:free_energy_gk}. Note that, in the local limit, \cref{eq:free_energy_gk} can be written as
\begin{align}
	W \! = \! \sum_\s\avgbox[\Bigg]{\intv \! \avgr{\frac{\Ts\hs^2}{2\Fs}}} \! \! +\!\sum_\vkperp\!\paravg{\frac{\abs{\dBparkperp}^2}{8\pi} + \frac{\kperp^2\abs{\dAparkperp}^2}{8\pi} - \sum_\s \frac{\qs^2\ns}{\Ts}\abs{\phipotkperp}^2}\!.
	\label{eq:W_fourier}
\end{align}

\subsection{Injection terms for \(W\)}
\label{appendix:W_injection}
Here, we derive \cref{eq:free_energy_injection} and give explicit formulae for the particle and heat fluxes. To begin, using \cref{eq:gyrokinetic_equation}, we find
\begin{align}
	&\der{}{t} \sum_\s \avgbox[\Bigg]{\intv \avgr{\frac{\Ts\hs^2}{2\Fs}}} \nonumber \\
	&= \! \sum_\s \avgbox[\Bigg]{\intv \! \avgr{\frac{\Ts\hs}{\Fs} \left[ \partd{}{t}\frac{\qs\avgRs{\chi}}{\Ts}\Fs \! - \vpar\ub\bcdot\grad\hs \! -  \vds\!\bcdot\grad\hs \! -  \vchiRs\!\!\bcdot\grad\Fs\right]}}\!,
	 \label{eq:W_injection_step1}
\end{align}
where the \(\hs\vds\bcdot\grad\hs\) term vanishes by \cref{eq:perpaverage_kills_gradperp} because it is an exact derivative, viz.,
\begin{equation}
	\hs\vds\bcdot\gradperp\hs = \frac{1}{2}\gradperp \bcdot \left(\vds \hs^2\right)
	\label{eq:magn_drift_are_div}
\end{equation} 
where the equality is correct to lowest order in \cref{eq:gyrokinetic_ordering} because \(\vds\) is an equilibrium-like quantity whose perpendicular derivatives are small compared to those of \(\hs\). Writing the velocity-space integral in terms of \(\es, \mus\) and \(\vartheta\) using
\begin{equation}
	\intv = \sum_\sigma \int \frac{\rmd\es\rmd\mus\rmd\vartheta}{\ms^2\abs{\vpar}}B,
	\label{eq:velocity_jacobian}
\end{equation}
we find that the parallel streaming term in \cref{eq:W_injection_step1} also vanishes:
\begin{align}
	&\avgbox[\bigg]{\intv \avgr{\frac{\Ts\hs}{\Fs} \vpar \ub\bcdot\grad\hs}} \propto \avgbox[\bigg]{\intv \vpar \ub \bcdot \grad \left(\frac{\Ts}{\Fs}\avgr{\hs^2}\right)} \nonumber \\
	&= \avgbox[\bigg]{\sum_\sigma\int \frac{\rmd \es \rmd \mus\rmd\vartheta}{\ms^2\abs{\vpar}}\vpar\vB\bcdot\grad \left(\frac{\Ts}{\Fs}\avgr{\hs^2}\right)} \nonumber \\
	&=\avgbox[\bigg]{\grad\bcdot\sum_\sigma\int \sigma \frac{\rmd \es \rmd \mus\rmd\vartheta}{\ms^2} \vB \frac{\Ts}{\Fs}\avgr{\hs^2}} = 0,
\end{align}
where we used \(\grad \bcdot \vB = 0\), \cref{eq:flux_surface_average_kills_parallel_grad}, and the fact that \(\Ts\) and \(\Fs\) vary only in the perpendicular direction to lowest order in \cref{eq:gyrokinetic_ordering}.

To handle the first term in the square bracket in \cref{eq:W_injection_step1}, we will need the following identity:
\begin{align}
	&\avgbox[\bigg]{\intv \avgr{\hs \avgRs{\chi}}} \nonumber \\ 
	&= \!\avgbox[\bigg]{\sum_\sigma \! \int \! \frac{\rmd\es\rmd\mus\rmd\vartheta}{\ms^2\abs{\vpar}}B \hs\Big(\vr - \vrhos(\vartheta), \es, \mus\Big) \! \! \int \! \frac{\rmd \vartheta'}{2\pi}\chi\Big(\vr - \vrhos(\vartheta) + \vrhos(\vartheta'), \es, \mus, \vartheta'\Big)\! } \nonumber \\
	&= \!\avgbox[\bigg]{\sum_\sigma \! \int \! \frac{\rmd\es\rmd\mus\rmd\vartheta}{\ms^2\abs{\vpar}}B \int \frac{\rmd \vartheta'}{2\pi} \hs\Big(\vr' - \vrhos(\vartheta'), \es, \mus\Big) \chi(\vr', \es, \mus, \vartheta')} \nonumber \\
	&= \!\avgbox[\bigg]{\sum_\sigma \! \int \! \frac{\rmd\es\rmd\mus\rmd\vartheta'}{\ms^2\abs{\vpar}}B \hs\Big(\vr' - \vrhos(\vartheta'), \es, \mus\Big) \chi(\vr', \es, \mus, \vartheta')} \nonumber \\
	&= \!\avgbox[\bigg]{\intv \avgr{\hs \chi}},
	\label{eq:avgRs_chi_times_hs_integral}
\end{align}
where, to obtain the final line, we have relabelled \(\vr' \mapsto \vr\), \(\vartheta' \mapsto \vartheta\) and used \cref{eq:velocity_jacobian}; the change of variables \(\vr' = \vr - \vrhos(\vartheta) + \vrhos(\vartheta')\) does not change the perpendicular spatial average in \(\avgbox{.}\) to lowest order as \(\abs{\vrhos}\sim\rhos \ll L\). Using \cref{eq:avgRs_chi_times_hs_integral}, we find
\begin{align}
	&\sum_\s \qs \avgbox[\Bigg]{\avgr{\intv \hs\partd{\avgRs{\chi}}{t}}} &\nonumber \\
	&= \avgbox[\bigg]{\partd{\phi}{t} \sum_\s \qs \intv \avgr{\hs}} - \avgbox[\bigg]{\partd{\vdA}{t} \bcdot \sum_\s \frac{\qs}{c} \intv \avgr{\vv\hs}} \nonumber \\
	&= \der{}{t}\sum_\s \avgbox[\bigg]{\frac{\qs^2\ns}{2\Ts}\phi^2} - \avgbox[\bigg]{\partd{\vdA}{t}\bcdot\frac{\grad\times\vdB}{4\pi}} \nonumber \\
	&= \der{}{t}\sum_\s \avgbox[\bigg]{\frac{\qs^2\ns}{2\Ts}\phi^2} - \der{}{t}\avgbox[\bigg]{ \frac{\abs{\vdB}^2}{8\pi}},
	\label{eq:W_injection_step2}
\end{align}
where we used \cref{eq:amperes_law} and \cref{eq:quasineutrality}. Therefore, adding together \cref{eq:W_expression_step3} and the second integral in \cref{eq:free_energy_deltaf} and taking a time derivative, we find
\begin{equation}
	\der{W}{t} = -\sum_\s \avgbox[\Bigg]{\intv \avgr{\frac{\Ts}{\Fs}\hs\vchiRs\bcdot\grad\Fs}}.
	\label{eq:W_time_derivative_appendix}
\end{equation}

To calculate the right-hand side of \cref{eq:W_time_derivative_appendix}, we use \cref{eq:equilibrium_gradients} to write, for a Maxwellian \(\Fs\),
\begin{align}
	\grad \Fs =  - \grad x  \left[\frac{1}{\Lns} + \frac{1}{\LTs} \left( \frac{v^2}{\vths^2} - \frac{3}{2} \right) \right] \Fs.
	\label{eq:equilibrium_gradient}
\end{align}
Finally, using \cref{eq:avgRs_chi_times_hs_integral}, we find that the time evolution of \(W\) is given by \crefrange{eq:free_energy_injection}{eq:Qs_def}.

\subsection{Expression for \(Y\)}

Let us now derive the expression \cref{eq:Y_fields_expression} for the GK field invariant \(Y\). Expanding the the first term in the integrand in \cref{eq:Y_definition}, we find
\begin{equation}
	\sum_\s \avgbox[\Bigg]{\intv\avgr{ \frac{\Ts \hs^2}{2\Fs} - \qs\avgRs{\chi}\hs + \frac{\qs^2\left(\avgRs{\chi}\right)^2}{2\Ts}\Fs}}.
	\label{eq:Y_expression_step1}
\end{equation}
The second term in \cref{eq:Y_expression_step1} is handled just like \cref{eq:W_injection_step2} to find
\begin{align}
	&\sum_\s \avgbox[\bigg]{\intv \avgr{\qs\avgRs{\chi}\hs}} = \avgbox[\Bigg]{\sum_\s\frac{\qs^2\ns}{\Ts}\phipot^2 - \frac{\abs{\gradperp\dApar}^2}{4\pi} - \frac{\dBpar^2}{4\pi}}.
	\label{eq:Y_expression_step2}
\end{align}
For the last term in the integrand in \cref{eq:Y_expression_step1}, we use the Fourier-space representation of the GK potential
\begin{equation}
	\avgRs{\chi} = \sum_\vk \chiavgperp \rme^{\rmi \vk\bcdot\vRs},
\end{equation}
where
\begin{align}
	\chiavgperp \equiv \rmJ_0(b_\s)\left(\phipotkperp - \frac{\vpar\dAparkperp}{c}\right) + \frac{2\rmJ_1(b_\s)}{b_\s}\frac{\Ts}{\qs}\frac{\vperp^2}{\vths^2}\frac{\dBparkperp}{\Bo}.
	\label{eq:chi_fourier}
\end{align}
Then,
\begin{align}
	&\sum_\s \avgbox[\Bigg]{\intv \avgr{\frac{\qs^2\left(\avgRs{\chi}\right)^2}{2\Ts}\Fs}} \nonumber \\ 
	&= \sum_{\vkperp, \s}  \paravginline[\Bigg]{\intv\Bigg[ \frac{\qs^2}{2\Ts}\abs{\phipotkperp}^2\rmJ_0^2 + \frac{\qs^2\vpar^2}{2c^2\Ts}\abs{\dAparkperp}^2\rmJ_0^2 + \frac{\Ts\vperp^4}{2\vths^4}\abs{\frac{\dBparkperp}{B}}^2\frac{4\rmJ_1^2}{b_\s^2} \nonumber \\
	&\qquad \qquad \qquad \qquad  + \frac{\qs\vperp^2}{2\vths^2}\left(\phipotkperp\frac{\dBparkperp^*}{B} +\phipotkperp^*\frac{\dBparkperp}{B}\right) \frac{2\rmJ_1\rmJ_0}{b_\s} \Bigg]\Fs}.
	\label{eq:chi_squared_nasty}
\end{align}
Using the identity
\begin{equation}
	\int_0^{+\infty} \rmdint{x}x\rmJ_n(px)\rmJ_m(qx)\rme^{-ax^2} = \frac{1}{2a}\rmI_n\left(\frac{pq}{2a}\right)\rme^{-(p^2+q^2)/2a}\delta_{mn},
\end{equation}
valid for any \(m, n \in \integers\), one can show that
\begin{align}
	&\intv \rmJ_0^2 \Fs = \ns \rmGamma_{0\s}, \label{eq:first_gamma_eq} \\
	&\intv \frac{2\rmJ_1\rmJ_0}{b_\s}\frac{\vperp^2}{\vths^2}\Fs = \ns \rmGamma_{1\s}, \\
	&\intv \rmJ_0^2\frac{\vpar^2}{\vths^2} \Fs = \frac{1}{2}\ns\rmGamma_{1\s}, \\
	&\intv \frac{4\rmJ_1^2}{b_\s^2}\frac{\vperp^4}{\vths^4}\Fs = 2\ns\Fs. \label{eq:last_gamma_eq}
\end{align}
Substituting \crefrange{eq:first_gamma_eq}{eq:last_gamma_eq} into \cref{eq:chi_squared_nasty}, we find
\begin{align}
	&\sum_\s \avgbox[\Bigg]{\intv \avgr{\frac{\qs^2\left(\avgRs{\chi}\right)^2}{2\Ts}\Fs}} \nonumber \\
	&=\sum_{\vkperp, \s}\paravginline[\Bigg]{ \frac{\qs^2\ns\rmGamma_{0\s}}{2\Ts}\abs{\phipotkperp}^2 + \frac{\rmGamma_{0\s}\abs{\dAparkperp}^2}{8\pi\ds^2} + \frac{\betas\rmGamma_{1\s}\abs{\dBparkperp}^2}{8\pi}  \nonumber \\
	&\qquad\qquad + \frac{\qs\ns\rmGamma_{1\s}}{2}\left(\phipotkperp\frac{\dBparkperp^*}{B} + \phipotkperp^*\frac{\dBparkperp}{B}\right)  }.
	\label{eq:Y_expression_step3}
\end{align}
Finally, combining \cref{eq:W_fourier}, \cref{eq:Y_expression_step1}, \cref{eq:Y_expression_step2}, and \cref{eq:Y_expression_step3}, we arrive at
\begin{align}
	&Y = \sum_\vkperp\paravginline[\Bigg]{-\sum_\s \frac{\qs^2\ns}{2\Ts}\left(1-\rmGamma_{0\s}\right)\abs{\phipotkperp}^2 + \frac{\kperp^2\abs{\dAparkperp}^2}{8\pi} + \sum_\s \frac{\rmGamma_{0\s}\abs{\dAparkperp}^2}{8\pi\ds^2}\nonumber \\
	&\qquad+ \frac{\abs{\dBparkperp}^2}{8\pi} + \sum_\s \frac{\betas\rmGamma_{1\s}\abs{\dBparkperp}^2}{8\pi} + \frac{1}{2}\sum_\s\qs\ns\rmGamma_{1\s}\left(\phipotkperp\frac{\dBparkperp^*}{B} + \phipotkperp^*\frac{\dBparkperp}{B}\right)}.
\end{align}
Combining the last two terms in the square bracket using 
\begin{equation}
	\frac{\betas\abs{\dBparkperp}^2}{8\pi} + \frac{1}{2}\qs\ns\left(\phipotkperp\frac{\dBparkperp^*}{B} + \phipotkperp^*\frac{\dBparkperp}{B}\right) = \ns\Ts\abs{\frac{\dBparkperp}{B} + \frac{\qs\phipotkperp}{2\Ts}}^2 - \frac{\qs^2\ns}{4\Ts}\abs{\phipotkperp}^2\!,
\end{equation}
we obtain \cref{eq:Y_fields_expression}.

\subsection{Injection terms for \(Y\)}
\label{appendix:Y_injection}

Taking a time derivative of the integral in \cref{eq:Y_definition} and substituting \cref{eq:gyrokinetic_equation}, we obtain
\begin{align}
	&\der{}{t}\sum_\s \avgbox[\Bigg]{\intv \frac{\Ts}{2\Fs} \avgr{\left(\hs - \frac{\qs\avgRs{\chi}}{\Ts}\Fs\right)^2}} = \nonumber \\
	&=\!\sum_\s\avgbox[\Bigg]{\intv \! \frac{\Ts}{2\Fs} \avgr{\!\left(\hs \!-\! \frac{\qs\avgRs{\chi}}{\Ts}\Fs\right) \!\! \left(\!-\vpar\ub\bcdot\grad\hs \! - \vds\!\bcdot\!\grad\hs\!-\vchiRs\!\!\bcdot\!\grad\Fs\right)\!}\!} \nonumber \\
	&=\!\sum_\s\avgbox[\Bigg]{\intv \bigg\langle\qs\avgRs{\chi}\vpar\ub\bcdot\grad\hs + \qs \avgRs{\chi}\vds\bcdot\grad\hs - \frac{\Ts}{\Fs}\hs\vchiRs\bcdot\grad\Fs \nonumber \\ 
	&\qquad\qquad\qquad\qquad +\qs \avgRs{\chi} \vchiRs\bcdot\grad\Fs\bigg\rangle_\vr},
	\label{eq:Y_injection_step1}
\end{align}
where the terms proportional to \(\hs \vds\bcdot\grad\hs\) and \(\hs\vpar\ub\bcdot\grad\hs\) have vanished for the same reason they did in \cref{eq:W_injection_step1}. Additionally,
\begin{equation}
	\avgRs{\chi} \vchiRs\bcdot\grad\Fs = \avgRs{\chi} \frac{c}{\Bo} \left(\ub \times \gradperp \avgRs{\chi}\right)\bcdot \grad\Fs = -\frac{1}{2}\gradperp \bcdot \left(\avgRe{\chi}^2 \ub \times \grad\Fs\right)
\end{equation}
also vanishes to lowest order after integration [cf. \cref{eq:magn_drift_are_div}]. Finally, subtracting \cref{eq:W_time_derivative_appendix} from~\cref{eq:Y_injection_step1}, we obtain~\cref{eq:Y_injection}.

Using \cref{eq:magnetic_drifts} and \cref{eq:avgRs_chi_times_hs_integral}, the magnetic-drift injection term becomes
\begin{align}
	&\sum_\s \qs \avgbox[\bigg]{\intv \avgr{\avgRs{\chi} \vds\bcdot\gradperp\hs}} \nonumber \\
	&= \sum_\s \avgbox[\bigg]{\intv \avgr{\hs \left[\ms\vpar^2 \vchi\bcdot (\ub\bcdot\grad\ub) + \frac{1}{2}\ms\vperp^2 \vchi\bcdot \gradperp\ln\Bo \right]}}.
	\label{eq:Y_magnetic_drift_injection_before_force_balance}
\end{align}
Using \(\grad\times\vB = (4\pi/c)\vJ\) and \(\vJ \times \vB = c\grad p\), where \(\vJ\) is the equilibrium current and \(p = \sum_\s \ns\Ts\) is the equilibrium presssure, \cref{eq:Y_magnetic_drift_injection_before_force_balance} is simplified to \cref{eq:Y_magnetic_drift_injection}.

\section{Derivation of the low-beta drift-kinetic model}
\label{appendix:drift_kinetic_derivation}

Our goal is to simplify the GK equation \cref{eq:gyrokinetic_equation} in order to distil a minimum model for the good-curvature, electromagnetic instability discussed in \cref{sec:mdm}. First, we limit ourselves to the low-beta, zero-magnetic-shear \(Z\)-pinch geometry, wherein \cref{eq:magnetic_drifts} simplifies to
\begin{equation}
	\vds = -\frac{1}{\Omegas\LB} \left( \vpar^2 + \frac{1}{2}\vperp^2 \right)\uvec{y},
	\label{eq:magnetic_drifts_zpinch}
\end{equation}
where
\begin{equation}
	\LB^{-1} \equiv -\uvec{y} \bcdot (\ub\times\grad \ln B)
	\label{eq:R_and_LB_defs}
\end{equation}
is the (inverse) length scale associated with the gradient of the magnetic-field strength. The main benefit of the \(Z\)-pinch geometry is that \cref{eq:gyrokinetic_equation} becomes homogeneous along the field lines. We are thus free to impose periodic boundary conditions in all three spatial dimensions and Fourier transform the distribution functions\footnote{Geometrically, in a `true' \(Z\)-pinch, the field lines are closed circles with radius \(R\). This implies that the parallel wavenumber is quantised as \(\kpar = n/R\), where \(n \in \integers\). Here, however, we will not necessarily limit ourselves to these values of \(\kpar\). Physically, this can be achieved by deforming the \(Z\)-pinch into a screw pinch. Practically speaking, our setup is that of a shearless, triply periodic flux tube \citep{beer95} with magnetic geometry that is artificially constant along the field lines.}
\begin{equation}
	\hs = \sum_\vk \hsk \rme^{\rmi \vk\bcdot\vRs}
\end{equation}
and fields
\begin{equation}
	\phipot = \sum_\vk \phipotk \rme^{\rmi \vk\bcdot\vr}, \quad \dApar = \sum_\vk \dApark \rme^{\rmi \vk\bcdot\vr}, \quad \dBpar = \sum_\vk \dBpark \rme^{\rmi \vk\bcdot\vr}.
\end{equation}
Like in \cref{eq:chi_fourier}, the Fourier-transformed gyroaveraged GK potential is
\begin{align}
	\avgRs{\chi} = \sum_\vk \chiavg \rme^{\rmi \vk\bcdot\vRs} = \sum_\vk \left[\rmJ_0(b_\s)\left(\phipotk - \frac{\vpar\dApark}{c}\right) + \frac{2\rmJ_1(b_\s)}{b_\s}\frac{\Ts}{\qs}\frac{\vperp^2}{\vths^2}\frac{\dBpark}{\Bo}\right] \rme^{\rmi \vk\bcdot\vRs}.
	\label{eq:chi_fourier_3D}
\end{align}
The GK equation \cref{eq:gyrokinetic_equation} can then be written as
\begin{align}
	&\partd{}{t} \left( \hsk - \frac{\qs \chiavg}{\Ts}\Fs \right) + \rmi \kpar \vpar \hsk + \rmi\omegads \left(\frac{2\vpar^2}{\vths^2} + \frac{\vperp^2}{\vths^2} \right) \hsk \nonumber \\
	&\quad -\rmi \left[ \omegasts + \omegaTs \left( \frac{v^2}{\vths^2} - \frac{3}{2} \right) \right] \frac{\qs \avgRs{\chi_\vk}}{\Ts}\Fs = 0, 
	\label{eq:gk_fourier_space}
\end{align}
where the drift frequencies are
\begin{equation}
	\omegasts = - \frac{k_y c \Ts}{\qs \Bo \Lns}, \quad \omegaTs = -\frac{k_y c \Ts}{\qs B \LTs}, \quad \omegads = -\frac{k_y c \Ts}{\qs B \LB}.
	\label{eq:drift_frequency_appendix}
\end{equation}
The field equations \crefrange{eq:quasineutrality}{eq:perpendicular_amperes_law} become
\begin{align}
	\sum_\s \frac{\qs^2 \ns}{\Ts} \phipotk & =  \sum_\s \qs   \intv  \rmJ_0(b_\s) \hsk, \label{eq:quasineutrality_fourier} \\
	k_\perp^2 \dApark & = \frac{4\pi}{c} \sum_\s \qs \intv \vpar \rmJ_0(b_\s)\hsk , \label{eq:parallel_amperes_fourier} \\
	\frac{\dBpark}{\Bo} &= - \frac{1}{2} \sum_\s \frac{\betas}{\ns} \intv \frac{v_\perp^2}{\vths^2} \frac{2 \rmJ_1 (b_\s)}{b_\s} \hsk.
	\label{eq:perpendicular_amperes_law_fourier}
\end{align}

By restricting ourselves only to fluctuations that obey the orderings \cref{eq:lowbeta_kperp_ordering} and \cref{eq:lowbeta_freq_ordering}, we are able to make several simplifications. First, \(\hi\)'s contributions to Maxwell's equations \crefrange{eq:quasineutrality_fourier}{eq:perpendicular_amperes_law_fourier} can be neglected because \(\rmJ_0(b_i)\sim\rmJ_1(b_i)\sim (\kperp\rhoi)^{-1/2} \ll 1\) in the limit \(\kperp\rhoi \gg 1\), i.e., averaging over the large Larmor orbits of the ions results in an adiabatic ion response. Secondly, \(\kperp\rhoe\ll1\) implies that the electrons become drift kinetic. This simplifies the Bessel functions in \cref{eq:chi_fourier_3D} and \crefrange{eq:quasineutrality_fourier}{eq:perpendicular_amperes_law_fourier} to \mbox{\(\rmJ_0(b_e) \approx 2\rmJ_1(b_e)/b_e \approx 1\)}. Additionally, \(\betae \ll 1\) implies that the parallel magnetic-field fluctuations are small. Indeed, using \cref{eq:parallel_amperes_fourier} and \cref{eq:perpendicular_amperes_law_fourier}, we find
\begin{equation}
	\frac{\dBpar}{\abs{\vdBperp}} \sim \frac{\dBpark}{\kperp\dApark} \sim \frac{\me\vthe^2}{\Bo} \frac{c\kperp}{e\vthe} \sim \kperp\rhoe \sim \sqrt{\betae} \ll 1.
\end{equation}

Putting everything together, under the ordering \cref{eq:lowbeta_kperp_ordering}, the low-beta dynamics are described by the electron drift-kinetic equation
\begin{align}
	&\partd{}{t} \left(\hek + \frac{e\phipotk}{\Te} \Fe - \frac{2\vpar}{\vthe}\frac{\dApark}{\rhoe B} \Fe \right) + \rmi\kpar\vpar\hek + \rmi\omegade\left(\frac{2\vpar^2}{\vthe^2} + \frac{\vperp^2}{\vthe^2}\right)\hek \nonumber \\
	&\quad + \rmi\left[\omegaste + \omegaTe \left( \frac{v^2}{\vths^2} - \frac{3}{2} \right) \right] \left(\frac{e\phipotk}{\Te} - \frac{2\vpar}{\vthe}\frac{\dApark}{\rhoe B}\right) \Fe = 0
	\label{eq:electron_drift_kinetics_appendix}
\end{align}
and the field equations
\begin{align}
	-(1 + \tau^{-1})\frac{e\phipotk}{\Te} &=\frac{1}{\ne}\intv \hek , \label{eq:quasineutrality_mdm_appendix} \\
	-\kperpdesq \frac{\dApark}{\rhoe B} &= \frac{1}{\ne}\intv \frac{\vpar}{\vthe}\hek = \frac{\uparek}{\vthe}, \label{eq:parallel_amperes_mdm_appendix}
\end{align}
where \(\upare\) is the perturbed parallel electron flow. In the two-dimensional limit, viz., \(\kpar= 0\), \cref{eq:electron_drift_kinetics_appendix} decouples into its even part (in \(\vpar\))
\begin{align}
	&\partd{}{t} \left(\hek^\text{(even)} + \frac{e\phipotk}{\Te} \Fe\right) + \rmi \omegade\left(\frac{2\vpar^2}{\vthe^2} + \frac{\vperp^2}{\vthe^2}\right)\hek^\text{(even)} \nonumber \\
	&\quad + \rmi\left[\omegaste + \omegaTe \left( \frac{v^2}{\vths^2} - \frac{3}{2} \right) \right] \frac{e\phipotk}{\Te}\Fe = 0, 	\label{eq:gk_2d_even_appendix}
\end{align}
closed by \cref{eq:quasineutrality_mdm_appendix}, and its odd part
\begin{align}
	&\partd{}{t} \left(\hek^\text{(odd)} - \frac{2\vpar}{\vthe}\frac{\dApark}{\rhoe B} \Fe \right) + \rmi \omegade\left(\frac{2\vpar^2}{\vthe^2} + \frac{\vperp^2}{\vthe^2}\right)\hek^\text{(odd)} \nonumber \\
	&\quad - \rmi\left[\omegaste + \omegaTe \left( \frac{v^2}{\vths^2} - \frac{3}{2} \right) \right] \frac{2\vpar}{\vthe}\frac{\dApark}{\rhoe B} \Fe = 0
	\label{eq:gk_2d_odd_appendix}
\end{align}
closed by \cref{eq:parallel_amperes_mdm_appendix}.

\subsection{Low-beta conservation laws}
\label{appendix:lowbeta_conservation}

Here, we use the orderings \cref{eq:lowbeta_kperp_ordering,eq:lowbeta_freq_ordering} to derive the asymptotic forms of the conservation laws \cref{eq:free_energy_injection} and \cref{eq:Y_injection}.

First, the \(\hi\) contribution to the free energy \cref{eq:free_energy_gk} is independently conserved and thus can be ignored. To see this, we use \cref{eq:gyrokinetic_equation} to find
\begin{align}
	&\der{}{t} \intrV \intv \avgr{\frac{\Ti\hi^2}{2\Fi}} \nonumber \\ &= \intrV \intv \avgr{\frac{\Ti\hi}{\Fi}\left(\frac{\qi \avgRi{\chi}}{\Fi} - \vpar \ub\bcdot\grad\hi - \vdi \bcdot \grad \hi - \vchiRs[i]\bcdot\grad\Fi\right)} \nonumber \\
	&= \intrV \intv \avgr{\frac{\Ti\hi}{\Fi}\left(\frac{\qi \avgRi{\chi}}{\Fi} - \vchiRs[i]\bcdot\grad\Fi\right)},
	\label{eq:lowbeta_hi_evolution}
\end{align}
where the parallel-streaming and magnetic-drift terms vanish for the same reason as they do in the derivation of the free energy's evolution equation given in \cref{appendix:W_injection}. For the remaining terms, we take advantage of the fact that, under \cref{eq:lowbeta_kperp_ordering}, the ion gyroaverage of the GK potential is small, viz.,
\begin{equation}
	\avgRi{\chi_\vk} \sim \frac{\chi_\vk}{\sqrt{\kperp\rhoi}} \ll \chi_\vk,
\end{equation}
where we used \cref{eq:chi_fourier} and the asymptotic expansions
\begin{equation}
	\abs{\rmJ_0(x)} \sim \abs{\rmJ_1(x)} \sim \frac{1}{\sqrt{x}},
	\label{eq:ion_bessels_are_small}
\end{equation}
valid for \(x \gg 1\). Therefore, the last line of \cref{eq:lowbeta_hi_evolution} vanishes to lowest order. 

Neglecting the \(\hi\) contributions to \cref{eq:free_energy_gk}, taking advantage of \(\kperp\rhoe \ll 1\), dropping the parallel magnetic fluctuations, and using
\begin{align}
	\rmGamma_{0i} = \order{\frac{1}{\kperp\rhoi}}, \quad
	\rmGamma_{0e} = 1 - \frac{\kperprhoesq}{2} + \order{\kperprhoesqq},
\end{align}
\cref{eq:W_curv_driven_injection_final} and \cref{eq:Y_injection_singlemode_curv} become
\begin{align}
	&\der{W}{t} = \frac{Q^{\parallel}_e + Q^{\perp}_e}{\LTe}, \quad \der{Y}{t} = \left.\der{Y}{t}\right\rvert_\text{slab} + \left.\der{Y}{t}\right\rvert_\text{drift}, \label{eq:W_evolution_lowbeta}
\end{align}
where
\begin{align}
	&\left.\der{Y}{t}\right\rvert_\text{slab} = \ne\Te\vthe\intrV\! \left(\frac{\upare}{\vthe}\partd{}{z}\frac{e\phipot}{\Te} + \frac{\dApar}{\rhoe B}\:\partd{}{z}\frac{\dppare}{\ne\Te}\right), \label{eq:Y_evolution_lowbeta_slab} \\
	&\left.\der{Y}{t}\right\rvert_\text{drift} = -\frac{2Q^{\parallel}_e + Q^{\perp}_e}{\LB}\label{eq:Y_evolution_lowbeta_drifts}.
\end{align}
The free energy and field invariant are given by
\begin{align}
	W &= \intrV\intv\frac{\Te\he^2}{2\Fe} + \intrV\left( \frac{\abs{\gradperp\dApar}^2}{8\pi} - \sum_\s \frac{\qs^2\ns}{2\Ts}\phipot^2\right),
	\label{eq:W_lowbeta} \\
	Y &= \intrV\left(\frac{\dApar^2}{8\pi\de^2} + \frac{\abs{\gradperp\dApar}^2}{8\pi} -\frac{\qi^2\ni}{2\Ti}\phipot^2\right),\label{eq:Y_lowbeta}
\end{align}
respectively. In the above, \(\dppare \equiv \intv \me\vpar^2\he\) is the perturbed parallel electron pressure and
\begin{align}
	Q^{\parallel}_e &= \rhoe\vthe\Te\intrV \intv \frac{\vpar^2}{\vthe^2}\left(-\frac{1}{2}\partd{}{y}\frac{e\phipot}{\Te} + \frac{\vpar}{\vthe}\partd{}{y}\frac{\dApar}{\rhoe B}\right)\he, \label{eq:Qpare} \\
	Q^{\perp}_e &= \rhoe\vthe\Te\intrV \intv \frac{\vperp^2}{\vthe^2}\left(-\frac{1}{2}\partd{}{y}\frac{e\phipot}{\Te} + \frac{\vpar}{\vthe}\partd{}{y}\frac{\dApar}{\rhoe B}\right)\he \label{eq:Qperpe}
\end{align}
are the radial fluxes of energy associated with parallel and perpendicular electron motion, respectively. The first terms in the parentheses in \cref{eq:Qpare} and \cref{eq:Qperpe} represent advection of parallel and perpendicular temperature, respectively, by the radial component of the \exb{} drift
\begin{equation}
	\ve \bcdot \grad x = -\frac{c}{\Bo}\partd{\phipot}{y},
	\label{eq:ExB_flow_normalised}
\end{equation}
the second ones are the radial projection of transport along the exact field lines \citep[sometimes called `flutter' transport;][]{callen77, manheimer78}, whose radial perturbation is
\begin{equation}
	\dBx = \partd{\dApar}{y}.
	\label{eq:dBx_normalised}
\end{equation}

\section{Derivation of the low-beta dispersion relation}
\label{appendix:analytical_disp}

The derivation of the dispersion relation is a direct application of the results of \citet{ivanov23}. The linear GK dispersion relation for a two-species, \(Z\)-pinch  plasma is
\begin{equation}
	\begin{vmatrix}
		\Lphiphi & \LphiA & \LphiB \\
		\LAphi & \LAA & \LAB \\
		\LBphi & \LBA & \LBB
	\end{vmatrix} = 0,
	\label{eq:disp_GK}
\end{equation}
where
\begin{align}
	L_{\phi \phi} & = - \sum_\s \frac{\qs^2 \ns \Ts[r]}{q_r^2 \ns[r] \Ts} \left\{ 1+ \left[\zeta_\s - \zeta_{*\s} + \eta_\s \zeta_{*\s} \left( \partial_a + \partial_b + \frac{3}{2} \right) \right] \left.\Ical_{a,b}^{(\s)} \right|_{a=b=1}\right\} ,\label{eq:L_phiphi} \\
	L_{\phi A} & =  2\sum_\s \frac{\qs^2 \ns \vths \Ts[r]}{q_r^2 \ns[r] \vths[r] \Ts} \left[\zeta_\s - \zeta_{*\s} + \eta_\s \zeta_{*\s} \left( \partial_a + \partial_b + \frac{3}{2} \right) \right] \left.\Jcal_{a,b}^{(\s)}\right|_{a=b=1}, \label{eq:L_phia} \\
	L_{\phi B} & = \sum_\s \frac{\qs \ns}{q_r \ns[r]} \left[\zeta_\s - \zeta_{*\s} + \eta_\s \zeta_{*\s} \left( \partial_a + \partial_b + \frac{3}{2} \right) \right] \partial_b \left.\K_{a,b}^{(\s)}\right|_{a=b=1}, \label{eq:L_phib} \\
	L_{A \phi} & = - \sum_\s \frac{\qs^2 \ns \vths \Ts[r]}{q_r^2 \ns[r] \vths[r] \Ts} \left[\zeta_\s - \zeta_{*\s} + \eta_\s \zeta_{*\s} \left( \partial_a + \partial_b + \frac{3}{2} \right) \right] \left.\Jcal_{a,b}^{(\s)}\right|_{a=b=1} ,\label{eq:L_aphi}\\
	L_{AA} & = - \frac{B_0^2 ( k_\perp \rhos[r])^2}{8 \pi \ns[r] \Ts[r]} - 2 \sum_\s \frac{\qs^2 \ns m_r}{q_r^2 \ns[r] m_\s} \left[\zeta_\s - \zeta_{*\s} + \eta_\s \zeta_{*\s} \left( \partial_a + \partial_b + \frac{3}{2} \right) \right] \partial_a \left.\Ical_{a,b}^{(\s)}\right|_{a=b=1}, \label{eq:L_aa} \\
	L_{AB} & = \sum_s \frac{\qs \ns \vths}{q_r \ns[r] \vths[r]} \left[\zeta_\s - \zeta_{*\s} + \eta_\s \zeta_{*\s} \left( \partial_a + \partial_b + \frac{3}{2} \right) \right] \partial_b \left.\L_{a,b}^{(\s)}\right|_{a=b=1}, \label{eq:L_ab} \\
	L_{B\phi} & = - \sum_\s  \frac{\beta_\s}{2} \frac{\qs \Ts[r]}{q_r \Ts} \left[\zeta_\s - \zeta_{*\s} + \eta_\s \zeta_{*\s} \left( \partial_a + \partial_b + \frac{3}{2} \right) \right] \partial_b \left.\K_{a,b}^{(\s)}\right|_{a=b=1}, \label{eq:L_bphi} \\
	L_{B A} & = \sum_\s  \beta_\s \frac{\qs \Ts[r] \vths}{q_r \Ts \vths[r]} \left[\zeta_\s - \zeta_{*\s} + \eta_\s \zeta_{*\s} \left( \partial_a + \partial_b + \frac{3}{2} \right) \right] \partial_b \left.\L_{a,b}^{(\s)} \right|_{a=b=1},\label{eq:L_ba} \\
	L_{BB} & = - 1 +  \sum_\s \frac{\beta_\s}{2} \left[\zeta_\s - \zeta_{*\s} + \eta_\s \zeta_{*\s} \left( \partial_a + \partial_b + \frac{3}{2} \right) \right] \partial_b^2 \left.\M_{a,b}^{(\s)}\right|_{a=b=1}, \label{eq:L_bb}
\end{align}
where have normalised our expressions using an arbitrary reference species \(r\). We have also defined the following integrals
\begin{align}
	\Ical_{a,b}^{(\s)} & = \frac{1}{\sqrt{\pi}} \int_{-\infty}^\infty \rmd u \int_0^\infty \rmd \mu \: \frac{ \rme^{-a u^2 - b \mu}}{u - \zeta_\s + \left(2u^2 \zeta_{\kappa \s} + \mu  \zeta_{Bs} \right)} \rmJ_0^2(b_\s),  \label{eq:iab_flr} \\
	\Jcal_{a,b}^{(\s)} & = \frac{1}{\sqrt{\pi}} \int_{-\infty}^\infty \rmd u \int_0^\infty \rmd \mu \: \frac{ u \rme^{-a u^2 - b \mu}}{u - \zeta_\s + \left(2u^2 \zeta_{\kappa \s} + \mu  \zeta_{Bs} \right)} \rmJ_0^2(b_\s),  \label{eq:jab_flr} \\
	\K_{a,b}^{(\s)} & = \frac{1}{\sqrt{\pi}} \int_{-\infty}^\infty \rmd u \int_0^\infty \rmd \mu \: \frac{ \rme^{-a u^2 - b \mu}}{u - \zeta_\s + \left(2u^2 \zeta_{\kappa \s} + \mu  \zeta_{Bs} \right)} \frac{2 \rmJ_0 (b_\s) \rmJ_1 (b_\s)}{b_\s},  \label{eq:kab_flr} \\
	\L_{a,b}^{(\s)} & = \frac{1}{\sqrt{\pi}} \int_{-\infty}^\infty \rmd u \int_0^\infty \rmd \mu \: \frac{ u \rme^{-a u^2 - b \mu}}{u - \zeta_\s + \left(2u^2 \zeta_{\kappa \s} + \mu  \zeta_{Bs} \right)} \frac{2 \rmJ_0 (b_\s) \rmJ_1 (b_\s)}{b_\s},  \label{eq:lab_flr} \\
	\M_{a,b}^{(\s)} & = \frac{1}{\sqrt{\pi}} \int_{-\infty}^\infty \rmd u \int_0^\infty \rmd \mu \: \frac{  \rme^{-a u^2 - b \mu}}{u - \zeta_\s + \left(2u^2 \zeta_{\kappa \s} + \mu  \zeta_{Bs} \right)} \left[\frac{2 \rmJ_1 (b_\s)}{b_\s} \right]^2, \label{eq:mab_flr} 
\end{align}
and normalised frequencies
\begin{align}
	\zeta_\s = \frac{\rmi p}{|k_\parallel| \vths}, \quad \zeta_{*\s} = \frac{\omega_{*s}}{|k_\parallel| \vths}, \quad \zeta_{\kappa \s} = \frac{\omega_{\kappa s}}{|k_\parallel| \vths}, \quad \zeta_{\grad \closesymbol B \s} = \frac{\omega_{\grad \closesymbol B \s}}{|k_\parallel| \vths},
	\label{eq:normalised_frequencies_s}
\end{align}
where \(\omegasts\) is defined in \cref{eq:drift_frequency_appendix} and the frequencies associated with the curvature and \(\grad B\) drifts are, respectively,
\begin{align}
	\omegacurvs = \frac{\vths^2 }{2\Omega_s} \vec{k}_\perp \bcdot \left[\vec{b}_0 \times (\vec{b}_0 \bcdot \grad )\vec{b}_0 \right], \quad \omegagradBs = \frac{\vths^2}{2\Omega_s} \vec{k}_\perp \bcdot \left(\vec{b}_0 \times \grad \log B \right).
	\label{eq:magnetic_drift_frequencies}
\end{align}
Using the orderings \cref{eq:lowbeta_kperp_ordering} and \cref{eq:lowbeta_freq_ordering}, we can make several simplifications to \mbox{\crefrange{eq:L_phiphi}{eq:L_bb}}. First, \(\kperp\rhoi \gg 1\) implies that the ion integrals \crefrange{eq:iab_flr}{eq:mab_flr} are small since \mbox{\(b_i = \kperp\vperp/\Omegai \sim \kperp\rhoi\)} and \(\rmJ_0(b_i) \to 0\), \(\rmJ_1(b_i) \to 0\) as \(b_i \to \infty\). Secondly, as the electrons are drift kinetic, viz., \(b_e \sim \kperp\rhoe \ll 1\), we find that \(\rmJ_0(b_e) \to 1\) and \(2\rmJ_1(b_e)/b_e \to 1\). Finally, the equilibrium pressure balance implies that, in the low-beta limit, \(\omegacurvs = \omegagradBs = \omegads\), where the latter is defined in \cref{eq:drift_frequency_appendix}.

Combining all of this, the electron integrals \crefrange{eq:iab_flr}{eq:mab_flr} simplify as
\begin{align}
	&\Ical_{a,b}^{(e)} = \K_{a,b}^{(e)} = \M_{a,b}^{(e)} = \Iab, \\
	&\Jcal_{a,b}^{(e)} = \L_{a,b}^{(e)} = \Jab,
\end{align}
where \(\Iab\) and \(\Jab\) are given by
\begin{align}
	\Iab &= \frac{1}{\sqrt{\pi}}\intinf{u}\inthalfinf{\mu} \frac{\rme^{-au^2 - b\mu}}{u - \zeta + \zetad (2u^2 + \mu)}, \label{eq:Iab_def_appendix} \\
	\Jab &= \frac{1}{\sqrt{\pi}}\intinf{u}\inthalfinf{\mu} \frac{u\rme^{-au^2 - b\mu}}{u - \zeta + \zetad (2u^2 + \mu)}. \label{eq:Jab_def_appendix} \\
\end{align}
We have normalised the frequencies as per \cref{eq:normalised_frequencies_s} with \(\s = e\), the normalised magnetic-drift frequency is \(\zetade \equiv \omegade / \absinline{\kpar}\vthe\), and the \(e\) subscript has been dropped. Finally, as~\(\betae \ll 1\), \crefrange{eq:L_bphi}{eq:L_bb} imply \(\LBphi \to 0\), \(\LBA \to 0\), and \(\LBB \to -1\), so \cref{eq:disp_GK} becomes
\begin{equation}
	0 = \begin{vmatrix}
		\Lphiphi & \LphiA & \LphiB \\
		\LAphi & \LAA & \LAB \\
		\LBphi & \LBA & \LBB
	\end{vmatrix}
	= \begin{vmatrix}
		\Lphiphi & \LphiA & \LphiB \\
		\LAphi & \LAA & \LAB \\
		0 & 0 & -1
	\end{vmatrix}
	= -(\Lphiphi \LAA - \LphiA \LAphi),
	\label{eq:disp_appendix}
\end{equation}
where, using electrons as the reference species in \crefrange{eq:L_aa}{eq:L_bb} (i.e., setting \(r=e\)), we have
\begin{align}
	\Lphiphi &= -1-\tau^{-1} - \left[\zeta - \zetast + \zetaT \left(\partial_a + \partial_b + \frac{3}{2}\right)\right]\left.\Iab\right\rvert_{a = b = 1}, \label{eq:Lphiphi_appendix} \\
	\LAA &= -\kperpdesq - 2 \left[\zeta - \zetast + \zetaT \left(\partial_a + \partial_b + \frac{3}{2}\right)\right]\left.\partial_a \Iab\right\rvert_{a = b = 1}, \label{eq:LAA_appendix} \\
	\LphiA &= -2\LAphi = 2 \left[\zeta - \zetast + \zetaT \left(\partial_a + \partial_b + \frac{3}{2}\right)\right]\left.\Jab\right\rvert_{a = b = 1}. \label{eq:LphiA_appendix}
\end{align}
Using the results of \citet{ivanov23}, we find
\begin{align}
	&\left.\Iab\right\rvert_{a = b = 1} = -\frac{1}{2\zetad}\Zp\Zm, \label{eq:Iaa_explicit} \\
	& \left.\partial_a\Iab\right\rvert_{a = b = 1} = -\frac{1}{2\zetad}\left[2 + \frac{1}{\zetad}\left(\Zp - \Zm\right) + \zetap\Zm + \zetam\Zp - \left(\frac{1}{4\zetad^2} - \frac{1}{2}\Zp\Zm\right)\right],\\
	&\left.\Jab\right\rvert_{a = b = 1} = -\frac{1}{2\zetad}\left(\Zp - \Zm - \frac{\Zp\Zm}{2\zetad}\right),\\
	&\left(\partial_a + \partial_b + \frac{3}{2}\right)\left.\Iab\right\rvert_{a = b = 1} = \frac{1}{2\zetad}\left[\zetap\Zm + \zetam\Zp + \left(\frac{\zeta}{\zetad} + \frac{1}{4\zetad^2} -1\right)\Zp\Zm\right], \\
	&\left(\partial_a + \partial_b + \frac{3}{2}\right)\left.\partial_a\Iab\right\rvert_{a = b = 1} = \frac{1}{2\zetad}\bigg[\frac{1}{2\zetad^2} + \frac{\zeta}{\zetad} - \frac{1}{2\zetad}\left(\Zp - \Zm\right) - \frac{1}{4\zetad^2}\left(\zetap\Zm + \zetam\Zp\right), \nonumber \\
	&\quad + \frac{1}{\zetad}\left(\zetap^2\Zp - \zetam^2\Zm\right) + \frac{\zeta}{2\zetad}\left(\zetap\Zp + \zetam\Zm\right) + \frac{1}{4\zetad^2}\left(\frac{3}{2} - \frac{1}{4\zetad^2} - \frac{\zeta}{\zetad} + 2\zeta\zetad\right)\Zp\Zm\bigg], \\
	&\left(\partial_a + \partial_b + \frac{3}{2}\right)\left.\Jab\right\rvert_{a = b = 1} = \frac{1}{2\zetad}\bigg[\frac{1}{2\zetad} - \frac{1}{2}\left(\Zp - \Zm\right) - \frac{1}{2\zetad}\left(\zetap\Zm + \zetam\Zp\right) \nonumber \\
	&\quad + \zetap^2\Zp - \zetam^2\Zm + \frac{1}{2\zetad}\left(1 - \frac{1}{4\zetad^2}-\frac{\zeta}{\zetad}\right)\Zp\Zm\bigg], \label{eq:derivatives_of_Jaa_explicit}
\end{align}
where \(\rmZ_\pm = \rmZ(\zetapm)\) and
\begin{equation}
	\zetapm \equiv \frac{\sqrt{1 + 8\zetad\zeta} \pm 1}{4\zetad}.
	\label{eq:zetapm_def}
\end{equation}
Substituting \crefrange{eq:Iaa_explicit}{eq:derivatives_of_Jaa_explicit} into \crefrange{eq:Lphiphi_appendix}{eq:LphiA_appendix} and then into \cref{eq:disp_appendix}, we obtain a bulky but explicit dispersion relation for the system \crefrange{eq:electron_drift_kinetics_appendix}{eq:parallel_amperes_mdm_appendix} that is straightforward to solve numerically.

\subsection{Two-dimensional limit}

The two-dimensional limit, i.e., \(\kpar \to 0\), corresponds to \(\zeta \sim \zetad \to \infty\), and so \cref{eq:zetapm_def} becomes
\begin{equation}
	\zeta_\pm \approx \sqrt{\frac{\zeta}{2\zetad}} = \sqrt{\frac{\omega}{2\omegade}}.
	\label{eq:zetapm_2d}
\end{equation}
In writing \cref{eq:zetapm_2d}, we have made use of our choice \(\omegade > 0\), which we discussed at the end of~\cref{sec:lowbeta_dk}. The two-dimensional limit is equivalent to dropping the \(u\) term in the denominator of the integrands in \cref{eq:Iab_def_appendix} and \cref{eq:Jab_def_appendix}. As its integrand becomes odd in \(u\), \cref{eq:Jab_def_appendix} vanishes. In this case, \(\LphiA = \LAphi = 0\), so the dispersion relation \cref{eq:disp_appendix} becomes
\begin{equation}
	D_\text{2D} \equiv \Lphiphi \LAA = 0,
	\label{eq:disp_2d}
\end{equation}
where \(\Lphiphi\) and \(\LAA\) are given by
\begin{align}
	\Lphiphi &= -1-\tau^{-1}+\frac{\omega-\omegaste}{2\omegade}\rmZ\left(\sqrt{\frac{\omega}{2\omegade}}\right)^2 \nonumber \\ 
	&\quad- \frac{\omegaTe}{2\omegade}\left[2\sqrt{\frac{\omega}{2\omegade}}\rmZ\left(\sqrt{\frac{\omega}{2\omegade}}\right) + \left(\frac{\omega}{\omegade}-1\right)\rmZ\left(\sqrt{\frac{\omega}{2\omegade}}\right)^2\right],
	\label{eq:Lphiphi_2D}\\
	\LAA&=-\kperpdesq + \frac{\omega - \omegaste}{\omegade} \left[2 + 2\sqrt{\frac{\omega}{2\omegade}}\rmZ\left(\sqrt{\frac{\omega}{2\omegade}}\right) + \frac{1}{2}\rmZ\left(\sqrt{\frac{\omega}{2\omegade}}\right)^2 \right] \nonumber \\
	&\quad-\frac{\omega\omegaTe}{\omegade^2}\left[1 + \sqrt{\frac{\omega}{2\omegade}}\rmZ\left(\sqrt{\frac{\omega}{2\omegade}}\right) +\frac{1}{2}\rmZ\left(\sqrt{\frac{\omega}{2\omegade}}\right)^2 \right]=0.
	\label{eq:LAA_2D}
\end{align}
Thus, the two-dimensional dispersion relation \cref{eq:disp_2d} splits into the electrostatic \cref{eq:es_etg_2d} and the electromagnetic \cref{eq:dispersion_MDM2D} ones, given by \(\Lphiphi = 0\) and \(\LAA = 0\), respectively.

\section{Fluid approximation for the magnetic-drift mode}
\label{appendix:fluid_limit}

To explore the cETG and MDM instabilities using a set of fluid moments, we follow Appendix A.5 of \citet{adkins22} and project the drift-kinetic equation~\cref{eq:electron_drift_kinetics_appendix} onto a set of Hermite (in \(\vpar\)) and Laguerre (in \(\vperp\)) basis functions. After truncating the infinite hierarchy to a total of \(M\) Hermite and \(L\) Laguerre functions, we obtain a set of \(M\times L\) fluid equations, whose normal modes can be found by computing the eigenvalues of an \((M\times L, M\times L)\) matrix. The details of the decomposition can be found in \citet{adkins22}. For the purposes of the discussion here, \(M\) and \(L\) can be thought of as the `resolution' of the fluid approximation in \(\vpar\) and \(\vperp\), respectively.

\begin{figure}
	\centering\begingroup\import{figs}{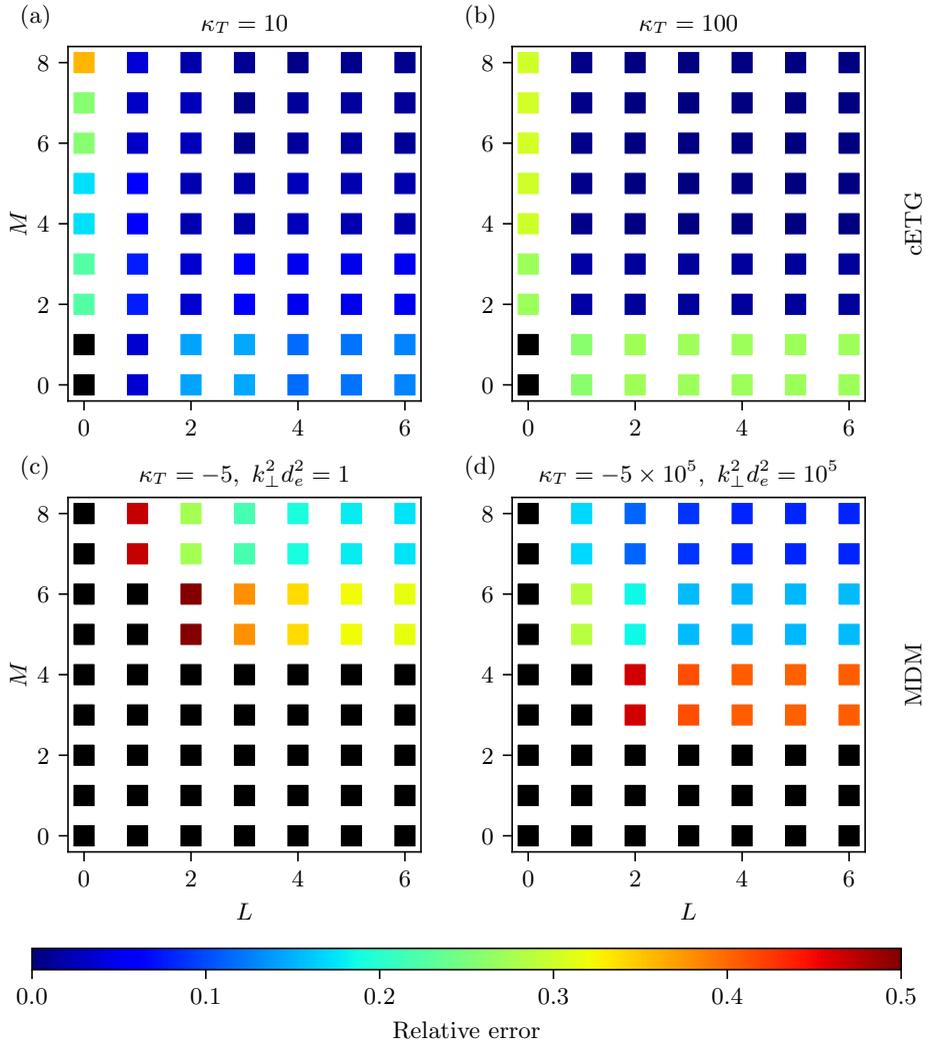}\endgroup
	\caption{Panels (a) and (b) show the relative error between the unstable solution of \cref{eq:es_etg_2d} and that of the truncated Hermite-Laguerre hierarchy [see (A 64) of \citealt{adkins22}] at varying numbers of Hermite and Laguerre moments, denoted by \(M\) and \(L\), respectively, and two different values of \(\kappaT\), as labelled on the panels. Here, we define the relative error between two quantities \(a\) and \(b\) as \(\abs{a-b}/\text{max}\lbrace a,b\rbrace\). Black denotes cases where the truncated fluid hierarchy has no unstable solution. For either \(L  = 0\) or \(M < 2\), as \(\kappaT \to \infty\), the growth rate is underpredicted by a factor of \(\sqrt{2}\) and so the relative error converges to \(1 - 1/\sqrt{2} \approx 0.29\). Panels (c) and (d) show the relative error for the MDM dispersion relation \cref{eq:dispersion_MDM2D} at two different values of \(\kappaT\) and \(\kperpdesq\), as labelled on each panel. }
	\label{fig:combined_LM}
\end{figure}

\begin{figure}
	\centering\begingroup\import{figs}{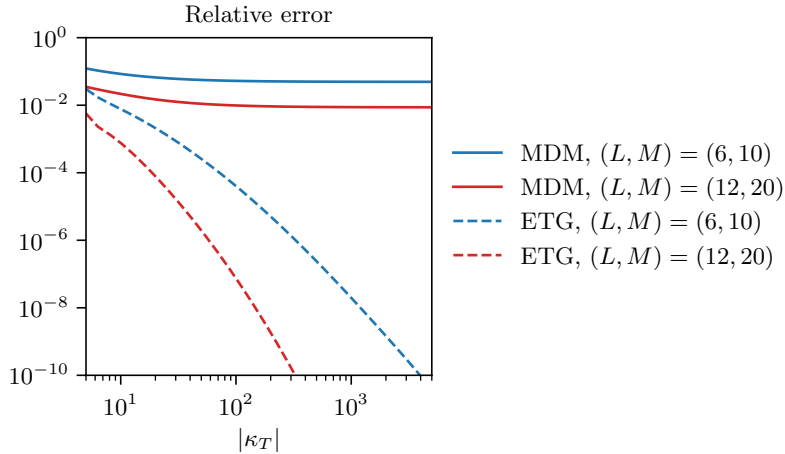}\endgroup
	\caption{Relative error between the unstable solutions obtained by solving the ETG~\cref{eq:es_etg_2d} and the MDM \cref{eq:dispersion_MDM2D} kinetic dispersion relations and those obtained from the truncated Hermite-Laguerre hierarchy at two different values of \((L, M)\) as a function of the temperature gradient \(\kappaT\) (as labelled in the legend). The ETG and MDM solutions are found at \(\kappaT > 0\) and \(\kappaT < 0\), respectively. For both cases, \(\kappan = 0\) and \(\tau = 1\). To ensure that the MDM remains unstable as we increase \(\abs{\kappaT}\), we have set \(\kperpdesq = -\kappaT/5\).}
	\label{fig:mdm_fluid_error}
\end{figure}

In \cref{fig:combined_LM}(a,b), we show the relative error between the complex cETG frequency, obtained by solving \cref{eq:es_etg_2d} directly, and by finding the unstable normal mode of the truncated Hermite-Laguerre fluid hierarchy. As we increase the temperature gradient, the fluid and kinetic solutions converge as long as \(M > 2\) and \(L > 1\), i.e., if the fluctuations of parallel \cref{eq:dTpare_etg} and perpendicular \cref{eq:dTperpe_etg} temperature are included in the fluid truncation. This is the expected behaviour given the discussion in \cref{sec:cETG}. If either \(\dTpare\) or \(\dTperpe\) is not included, it is easy to show that the resulting fluid dispersion relation is \(\omega^2 = -\omegaTe\omegade\tau\), and thus the growth rate is underpredicted by a factor of \(\sqrt{2}\) (see also \cref{appendix:nogradB_or_curvature}).

The convergence of the kinetic and truncated-fluid solutions shows that the cETG instability is `fluidisable', i.e., it can be described (asymptotically) using a set of fluid equations. The strongly driven limit \(\kappaT \to \infty\) is the natural one for the collisionless `fluidisation' of the electron drift-kinetic equation \cref{eq:electron_drift_kinetics_appendix}. To see this, note that the frequencies in \cref{eq:electron_drift_kinetics_appendix} can be split into two groups: the `kinetic frequencies', containing the parallel streaming \(\kpar \vpar\) and the magnetic drifts \(\omegade\), and the `fluid frequencies', \(\omegaste\) and \(\omegaTe\), which are proportional to the gradients of the plasma equilibrium profiles. In the Hermite-Laguerre decomposition, the terms proportional to the fluid frequencies appear only in the equations for the six lowest-order moments \citep{zocco11,adkins22}, while those proportional to the kinetic frequencies are responsible for coupling to higher-order moments.\footnote{In the representation of the kinetic system via the moment hierarchy, this coupling is the manifestation of phase mixing.} Thus, in the strongly driven limit, wherein the fluid frequencies are asymptotically larger than the kinetic ones, if the mode frequency scales with the fluid frequencies [which is the case for the cETG dispersion relation~\cref{eq:cETG_fluid_dispersion}], the infinite hierarchy is asymptotically truncated to (at most) the six lowest-order moments. Note that this truncation is entirely collisionless.

In contrast to cETG, as discussed in \cref{sec:mdm}, the MDM satisfies \(\omega \lesssim \omegade\), and so we do not expect to be able to `fluidise' it. In \cref{fig:combined_LM}(c,d), we show the relative error between the MDM frequency, obtained by solving \cref{eq:dispersion_MDM2D} directly, and by finding the unstable normal mode of the truncated Hermite-Laguerre fluid hierarchy. We observe that, even at very large values of the temperature gradient, the fluid approach does not provide a good approximation to the kinetic solution. To be more precise, for a fixed choice of \(L\) and \(M\), it is impossible to find parameters \(\kappaT\) and \(\kperpdesq\) for which the kinetic and fluid solutions converge to each other. This is further illustrated in \cref{fig:mdm_fluid_error} where we vary \(\kappaT\) for fixed \(L\) and \(M\). We find that as \(\kappaT \to \infty\), the error in the ETG solution decreases to zero. In contrast, as \(\kappaT \to -\infty\), the error in the MDM solution converges to a finite, nonzero value. Thus, we conclude that the MDM cannot be captured by a finite set of fluid equations.\footnote{Of course, if we were to increase \(L\) and \(M\) at fixed \(\kappaT\) and \(\kperpdesq\), the unstable normal mode of the truncated fluid system would approach the kinetic solution. Indeed, \cref{fig:mdm_fluid_error} shows that doubling both \(L\) and \(M\) decreases the error significantly.}

\section{MDM stability boundaries}
\label{appendix:mdm_stability}


Since the complex solutions \(\omega\) to \cref{eq:dispersion_MDM2D} are a smooth function of the parameters \(\kperpdesq\), \(\LB/\LTe\), and \(\LB/\Lne\), the region of instability seen in \cref{fig:mdm_omega} is bounded by lines on which \(\omega \in \reals\), the stability boundaries. Recall that we have assumed that \(\omegade > 0\). Therefore, \(\sqrt{\omega/2\omegade}\), which appears as the argument of the \(\rmZ\) functions in \cref{eq:dispersion_MDM2D}, lies on the positive real or positive imaginary axes if \(\omega > 0\) or \(\omega < 0\), respectively. When \(\omega > 0\), we recover the upper solid black line in \cref{fig:mdm_omega}, while the lower one corresponds to \(\omega < 0\). The dashed black line in \cref{fig:mdm_omega} is a special case of the stability boundary where \(\omega = 0\), not just its imaginary part. Let us try to understand this feature of \cref{fig:mdm_omega} first.

\subsection{Small-\(\omega\) limit}
\label{sec:smallomega}

In the limit \(\omega/\omegade \sim \delta \ll 1\), we can expand the \(\rmZ\) functions in the dispersion relation using 
\begin{equation}
	\rmZ(\zeta) = \rmi\sqrt{\pi} - 2\zeta -\rmi\zeta^2 + \frac{4}{3}\zeta^3 + \order{\zeta^4},
	\label{eq:Z_smallzeta_expansion}
\end{equation}
to obtain from \cref{eq:LAA_2D}
\begin{align}
	\LAA = &-\kperpdesq - \frac{4-\pi}{2}\kappan + \frac{4-\pi - (\pi-2)(\kappan-\kappaT)}{2}\frac{\omega}{\omegade} \nonumber \\ 
	&- \rmi \sqrt{\frac{\pi}{2}}\frac{2\kappan-3\kappaT}{3}\left(\frac{\omega}{\omegade}\right)^{3/2} + \order{\delta^2}.
	\label{eq:dispersion_MDM2D_smallomega}
\end{align}
Keeping only terms to \(\orderinline{\delta^0}\), we find that \(\omega \ll \omegade\) solutions to \(\LAA = 0\) exist only if \(\kperpdesq + (4-\pi)\kappan/2 \sim \delta \ll 1\). Indeed, the horizontal dotted line in \cref{fig:mdm_omega}, which specifies the minimum \(\kperpdesq\) for the instability and on which \(\omega = 0\) is an exact solution, is \(\kperpdesq = k_{\perp\text{,min}}^2\de^2\), where
\begin{equation}
	k_{\perp\text{,min}}^2\de^2 = -\frac{4-\pi}{2}\kappan.
\end{equation}
Note that \cref{eq:kperp2de2_min} implies that \(\omega = 0\) is a solution only if \(\kappan < 0\). Expanding 
\begin{equation}
	\omega = \omega_0 + \omega_1 + \dots
	\label{eq:smallomega_expansion}
\end{equation}
in \(\delta\ll1\), we find
\begin{equation}
	\frac{\omega_0}{\omegade} = \frac{2\kperpdesq + (4-\pi)\kappaT}{4-\pi - (\pi-2)(\kappan - \kappaT)}.
	\label{eq:smallomega_realpart}
\end{equation}
Thus, to lowest order, \(\omega \approx \omega_0 \in \reals\).

To determine the growth rate, we proceed to next order, which, in combination with~\cref{eq:smallomega_realpart}, yields
\begin{align}
	\frac{\omega_1}{\omegade} &= \frac{\rmi\sqrt{2\pi}}{3}\frac{(2\kappan-3\kappaT)\left[2\kperpdesq + (4-\pi)\kappan\right]}{\left[4-\pi-(\pi-2)(\kappan-\kappaT)\right]^2}\sqrt{\frac{\omega_0}{\omegade}}.
	\label{eq:smallomega_omega1}
\end{align}
If \(\omega_0/\omegade < 0\), then \(\omega_1\) is again purely real. In fact, it is not difficult to see that if \(\omega_0/\omegade < 0\), then all terms in the perturbative expansion \cref{eq:smallomega_expansion} are purely real. This happens because all integer powers of \(\omega\) in the expansion \cref{eq:dispersion_MDM2D_smallomega} have real coefficients while the half-integer ones have purely imaginary ones, which is itself a consequence of the pattern of alternating purely imaginary and purely real coefficients in the expansion of the \(\rmZ\) function \cref{eq:Z_smallzeta_expansion}. Therefore, the small-$\omega$ expansion can yield solutions with finite \(\im\:\omega\) only if \(\omega_0/\omegade > 0\).

If this is satisfied, then \cref{eq:smallomega_omega1} has a positive imaginary part if
\begin{equation}
	(2\kappan-3\kappaT)\left[2\kperpdesq + (4-\pi)\kappan\right] > 0.
	\label{eq:smallomega_growth_condition}
\end{equation}
Suppose that
\begin{equation}
	2\kperpdesq + (4-\pi)\kappan < 0
	\label{eq:kappan_cannot_be_positive}
\end{equation}
and
\begin{equation}
	2\kappan-3\kappaT < 0.
	\label{eq:smallomega_needed_for_contradiction}
\end{equation}
Since we have demanded \(\omega_0/\omegade > 0\), we need, from \cref{eq:smallomega_realpart},
\begin{equation}
	\kappaT < -\frac{4-\pi}{\pi-2} + \kappan,
\end{equation}
which, together with \cref{eq:smallomega_needed_for_contradiction}, yields
\begin{equation}
	\frac{\kappan}{3} > \frac{4-\pi}{\pi-2},
\end{equation}
which is a contradiction with the assumption \cref{eq:kappan_cannot_be_positive}, which is impossible when \(\kappan < 0\). Thus, \cref{eq:smallomega_growth_condition} can be fulfilled only if 
\begin{equation}
	2\kperpdesq + (4-\pi)\kappan > 0
	\label{eq:smallomega_first_inequality}
\end{equation}
and
\begin{equation}
	2\kappan-3\kappaT > 0 \implies \kappaT < \frac{2}{3}\kappan.
	\label{eq:dashedline_left}
\end{equation}
Combining \cref{eq:smallomega_first_inequality} and \cref{eq:smallomega_realpart}, we find also that
\begin{equation}
	\kappaT > -\frac{4-\pi}{\pi-2} + \kappan.
	\label{eq:dashedline_right}
\end{equation}
Together, \cref{eq:dashedline_left} and \cref{eq:dashedline_right} give the left and right ends of the horizontal dashed line in \cref{fig:mdm_omega}, respectively. 

\subsection{Finite positive \(\omega\)}
\label{sec:positive_omega}

Now, we turn to the case of marginal stability with the real frequency \(\omega > 0\) and let \(s \equiv \sqrt{\omega/2\omegade}\), so the \(\rmZ\) functions in \cref{eq:dispersion_MDM2D} become
\begin{equation}
	\rmZ(s) = \frac{1}{\sqrt{\pi}} \intinf{u} \frac{\rme^{-u^2}}{u-s} = \re\:\rmZ(s) + \rmi \sqrt{\pi}\rme^{-s^2},
\end{equation}
where, since \(s \in \reals^+\), the real part of \(\rmZ\) is given by the principal value of \cref{eq:Z_def}, viz.,
\begin{equation}
	\re\:\rmZ(s) = \frac{1}{\sqrt{\pi}} \mathcal{P}\intinf{u} \frac{\rme^{-u^2}}{u-s}.
\end{equation}
Taking the imaginary part of the dispersion relation \cref{eq:dispersion_MDM2D}, we find
\begin{equation}
	\frac{\re\:\rmZ(s)}{s} + \frac{2s^2(2 - \kappaT) - 2\kappan}{2s^2(1 - \kappaT) - \kappan}=0.
	\label{eq:positive_omega_imag_part}
\end{equation}
We now take two limits of \cref{eq:positive_omega_imag_part}: \(s \ll 1\) and \(\kappaT \gg 1\). 

When \(s \ll 1\), we can use \cref{eq:Z_smallzeta_expansion}, to expand
\begin{align}
	&\frac{\re\:\rmZ(s)}{s} + \frac{2s^2(2 - \kappaT) - 2\kappan}{2s^2(1 - \kappaT) - \kappan} = 2\left(\frac{2}{3} - \frac{\kappaT}{\kappan}\right)s^2 + \order{s^4},
\end{align}
which then implies that the solution of \cref{eq:positive_omega_imag_part} satisfies \(s \to 0\) only if \(\kappaT \to 2\kappan/3\). This connects the positive-\(\omega\) stability boundary to the left end of the small-\(\omega\) stability boundary given by \cref{eq:dashedline_left}. 

In the strongly driven limit \(\kappaT \gg 1\), \cref{eq:positive_omega_imag_part} gives \(\re\:\rmZ(s)\approx-s\), which has a unique solution at \(s \approx 1.063\). In this limit, the dispersion relation \cref{eq:dispersion_MDM2D} becomes
\begin{equation}
	\kperpdesq \approx -2s^2\kappaT\left\lbrace1+s\re\:\rmZ(s) + \frac{1}{2}\left[\re\:\rmZ(s)\right]^2-\frac{\pi}{2}\rme^{-2s^2}\right\rbrace,
\end{equation}
which, upon substitution of \(s \approx 1.063\), gives us
\begin{equation}
	\kperpdesq \approx -0.61\kappaT.
\end{equation}
This is the asymptotic slope of the top solid line in \cref{fig:mdm_omega}.

\subsection{Finite negative \(\omega\)}
\label{sec:negative_omega}

If \(\omega < 0\), then the argument of the \(\rmZ\) functions in \cref{eq:dispersion_MDM2D} is purely imaginary. Let
\begin{equation}
	\sqrt{\frac{\omega}{2\omegade}} \equiv \rmi v,
\end{equation}
where \(v > 0\). Then, \cref{eq:Z_def} implies \(\rmZ(\rmi v) = \rmi \im \: \rmZ(\rmi v)\), where
\begin{equation}
	\im \:\rmZ(\rmi v) = \frac{v}{\sqrt{\pi}}\intinf{u} \frac{\rme^{-u^2}}{u^2 + v^2}.
\end{equation}
In this case, the dispersion relation \cref{eq:dispersion_MDM2D} is purely real and can be written as
\begin{equation}
	\kperpdesq = -\left(2v^2 + \kappan\right) \left[2 - 2v\:\im\:\rmZ - \frac{1}{2}\left(\im\:\rmZ\right)^2\right] + 2v^2\kappaT \left[1 - v\:\im\:\rmZ - \frac{1}{2}\left(\im\:\rmZ\right)^2\right].
	\label{eq:dispersion_MDM2D_negativeomega}
\end{equation}
Once again, we consider two limits: \(v \ll 1\) and \(\kappaT \gg 1\).

When \(v \ll 1\), we expand \cref{eq:dispersion_MDM2D_negativeomega} to find
\begin{align}
	\kperpdesq = &-\frac{4-\pi}{2}\kappan - \left[4-\pi-(\pi-2)\left(\kappan-\kappaT\right)\right]v^2 \nonumber \\ 
	&+ 2\sqrt{\pi}\left(\kappaT-\frac{2}{3}\kappan\right)v^3 + \order{v^4}.
	\label{eq:dispersion_expansion_negativeomega}
\end{align}
Anticipating that the negative-\(\omega\) stability boundary connects to the right end \cref{eq:dashedline_right} of the small-\(\omega\) one, \cref{eq:dashedline_left} and \cref{eq:dashedline_right} imply that the coefficients of \(v^2\) and \(v^3\) in \cref{eq:dispersion_expansion_negativeomega} are positive and negative, respectively. Thus, the right-hand side of \cref{eq:dispersion_expansion_negativeomega} peaks at finite \(v > 0\), and so \cref{eq:dispersion_expansion_negativeomega} can be solved for any \(\kperpdesq\) up to the value of that peak. Differentiating the right-hand side of \cref{eq:dispersion_expansion_negativeomega} and substituting the value of \(v\) at the local maximum, we find
\begin{equation}
	\kperpdesq = -\frac{4-\pi}{2}\kappan - \frac{\left[4-\pi-(\pi-2)\left(\kappan-\kappaT\right)\right]^3}{3\pi\left(3\kappaT-2\kappan\right)^2}.
\end{equation}
Thus, the start of the bottom solid line in \cref{fig:mdm_omega} is a cubic curve.

In the strongly driven limit \(\kappaT \gg 1\), \cref{eq:dispersion_MDM2D_negativeomega} simplifies to
\begin{equation}
	\frac{\kperpdesq}{\kappaT} \approx 2v^2 \left[1 - v\im\:\rmZ - \frac{1}{2}\left(\im\:\rmZ\right)^2\right].
	\label{eq:dispersion_negativeomega_stronglydriven}
\end{equation}
The right-hand side of \cref{eq:dispersion_negativeomega_stronglydriven} is a positive function with a peak at \(v \approx 0.95\). Substituting this value into \cref{eq:dispersion_negativeomega_stronglydriven}, we find that \cref{eq:dispersion_negativeomega_stronglydriven} has solutions for \(v \in \reals\) only if \mbox{\(\kperpdesq < k_{\perp, c}\de^2\)}, where
\begin{equation}
	k_{\perp, c}\de^2 \approx -0.09\kappaT.
	\label{eq:stability_boundary_negative_omega_stronglydriven}
\end{equation}
This is the asymptotic slope at \(\kappaT \gg 1\) of the lower solid black line in \cref{fig:mdm_omega}.

\section{Reduced models of the magnetic-drift mode}
\label{appendix:reduced_models}

In this appendix, we consider two different simplifications of the two-dimensional MDM dispersion relation \cref{eq:dispersion_MDM2D}. Note that these simplifications are ad hoc in the sense that they cannot be captured by any subsidiary asymptotic expansion of \cref{eq:dispersion_MDM2D}. Nevertheless, they allow us to understand the qualitative ingredients of our instability.

\subsection{\(\grad B\) vs curvature drifts}
\label{appendix:nogradB_or_curvature}

\begin{figure}
	\begingroup\import{figs/fig8}{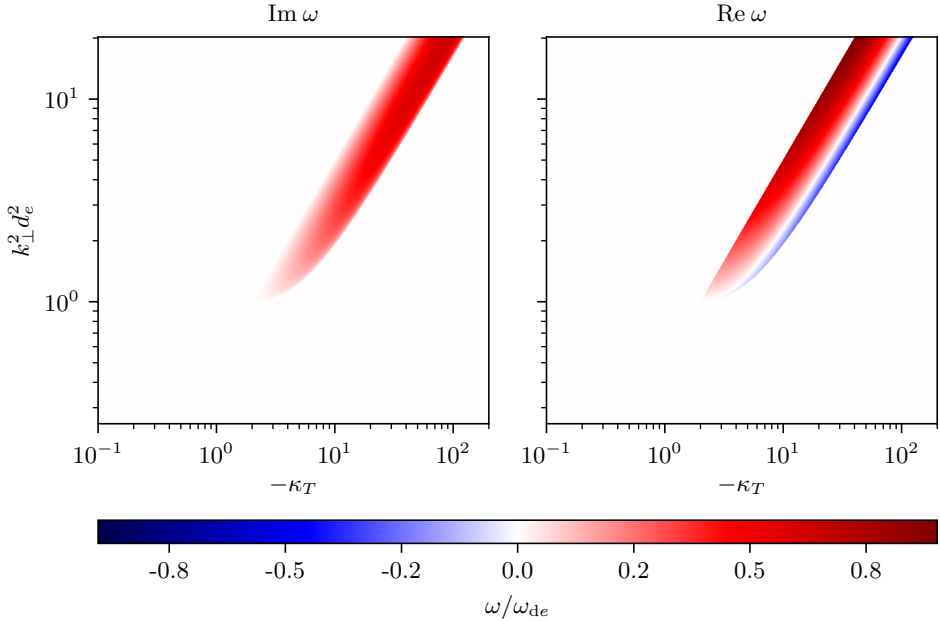}\endgroup
	\caption{(a) Growth rates and (b) frequencies obtained by solving \cref{eq:dispersion_MDM2D_nogradB} numerically for the same parameters as \cref{fig:mdm_omega}. Here, we are plotting only the unstable modes.}
	\label{fig:mdm_omega_nogradB}
\end{figure}

First, let us discuss the consequences of turning off either the \(\grad B\) or curvature drift, starting with the former. Removing the \(\grad B\) drift, i.e., the \(\vperp^2\)-dependent contribution to the magnetic drifts in \cref{eq:electron_drift_kinetics_appendix}, is equivalent to changing the definition of \(\Iab\) in \cref{eq:Iab_def_appendix} to
\begin{equation}
\Iab^{(\text{no \(\grad B\)})} = \frac{1}{\sqrt{\pi}}\intinf{u}\inthalfinf{\mu} \frac{\rme^{-au^2 - b\mu}}{- \zeta + 2\zetad u^2} = \frac{1}{b\sqrt{\pi}}\intinf{u} \frac{\rme^{-au^2}}{- \zeta + 2\zetad u^2},
\label{eq:Iab_nogradB}
\end{equation}
where we have explicitly taken the two-dimensional limit discussed in \cref{sec:2d}. Assuming without loss of generality that \(\zetad > 0\), we can carry out the integration in \cref{eq:Iab_nogradB} as
\begin{align}
	\Iab^{(\text{no \(\grad B\)})} &= \frac{1}{b\sqrt{8\zeta\zetad}}\frac{1}{\sqrt{\pi}}\intinf{u} \rme^{-au^2}\left(\frac{1}{u-\sqrt{\zeta/2\zetad}} - \frac{1}{u+\sqrt{\zeta/2\zetad}}\right) \nonumber \\
	&= \frac{1}{b\sqrt{2\zeta\zetad}}\frac{1}{\sqrt{\pi}}\intinf{u}\frac{ \rme^{-au^2}}{u-\sqrt{\zeta/2\zetad}} \nonumber \\
	&= \frac{1}{b\sqrt{2\zeta\zetad}} \rmZ\left(\sqrt{\frac{a\zeta}{2\zetad}}\right),
	\label{eq:Iab_nogradB_solution}
\end{align}
where we changed variables \(u \mapsto -u\) to go from the first to the second line and then used the definition~\cref{eq:Z_def} of the plasma dispersion function. Substituting \cref{eq:Iab_nogradB_solution} into \cref{eq:LAA_appendix}, we obtain the dispersion relation
\begin{align}
	0 = \LAA^{(\text{no \(\grad B\)})} = &-\kperpdesq + \frac{\omega - \omegaste}{\omegade}\left[1 + \sqrt{\frac{\omega}{2\omegade}}\rmZ\left(\sqrt{\frac{\omega}{2\omegade}}\right)\right] \nonumber \\
	&-\frac{\omega\omegaTe}{2\omegade^2} \left[1 + \sqrt{\frac{\omega}{2\omegade}}\rmZ\left(\sqrt{\frac{\omega}{2\omegade}}\right) \left(1 - \frac{\omegade}{\omega}\right)\right],
	\label{eq:dispersion_MDM2D_nogradB}
\end{align}
where the similarity with \cref{eq:dispersion_MDM2D} is evident. The numerical solution of \cref{eq:dispersion_MDM2D_nogradB} shown in \cref{fig:mdm_omega_nogradB} reveals a good-curvature, viz., \(\kappaT = \omegaTe/\omegade < 0\), instability with the same qualitative behaviour as was seen in \cref{fig:mdm_omega}.

In contrast, let us consider turning off the curvature drift, i.e., the \(\vpar^2\)-dependent contributions to the magnetic drifts in \cref{eq:electron_drift_kinetics_appendix}. Analogously to \cref{eq:Iab_nogradB}, we find that \cref{eq:Iab_def_appendix} becomes
\begin{equation}
	\Iab^{(\text{no curv})} = \frac{1}{\sqrt{\pi}}\intinf{u}\inthalfinf{\mu} \frac{\rme^{-au^2 - b\mu}}{- \zeta + \zetad \mu} = \frac{1}{\zetad\sqrt{a}}\inthalfinf{\mu} \frac{\rme^{-b\mu}}{\mu - \zeta/\zetad}.
	\label{eq:Iab_nocurv}
\end{equation}
Again, we assume that \(\zetad > 0\). Additionally, we know that \cref{eq:Iab_nocurv} is an analytic function in the upper half plane, and so let us look for an alternative integral expression for~\cref{eq:Iab_nocurv} for \(\re\:\zeta < 0\) that we will later analytically continue to \(\re\:\zeta > 0\). For \(\re\:\zeta < 0\) and \(\zetad > 0\), we let \(\mu = -(t-1) \zeta / \zetad\) to find
\begin{equation}
	\Iab^{(\text{no curv})} = \frac{1}{\zetad\sqrt{a}} \exp\left(-\frac{b\zeta}{\zetad}\right) \int_1^{+\infty} \frac{\rmdint{t}}{t}\exp\left(\frac{b\zeta}{\zetad}t\right) = \frac{1}{\zetad\sqrt{a}} \exp\left(-\frac{b\zeta}{\zetad}\right) \rmE_1\left(-\frac{b\zeta}{\zetad}\right),
	\label{eq:Iab_nocurv_solution}
\end{equation}
where \(\rmE_1\) is the exponential integral
\begin{equation}
	\rmE_1(x) \equiv \int_1^{+\infty} \rmdint{t} \frac{\rme^{-tx}}{t}.
	\label{eq:E1}
\end{equation}
The function \(\rmE_1\) is analytic in the complex plane with a branch cut along the negative real axis. Substituting \cref{eq:Iab_nocurv_solution} into \cref{eq:LAA_appendix}, we find the dispersion relation
\begin{align}
	0 = \LAA^{(\text{no curv})} = &-\kperpdesq + \frac{\omega - \omegaste}{\omegade}\rme^{-\omega/\omegade}\rmE_1\left(\frac{\omega}{\omegade}\right) \nonumber \\
	&-\frac{\omegaTe}{\omegade} \left[1 + \frac{\omega}{\omegade}\rme^{-\omega/\omegade}\rmE_1\left(\frac{\omega}{\omegade}\right)\right].
	\label{eq:dispersion_MDM2D_nocurv}
\end{align}

To show that this dispersion relation does not contain an MDM-like instability, we consider the simpler case \(\omegaste = 0\), wherein, using \crefrange{eq:Iab_nocurv}{eq:E1}, we can express the entirety of \cref{eq:dispersion_MDM2D_nocurv} as
\begin{equation}
	\inthalfinf{\mu} \frac{\mu\rme^{-\mu}}{\mu - \omega / \omegade} = \frac{1 + \kperpdesq}{1 - \kappaT}.
	\label{eq:dispersion_MDM2D_nocurv_alt}
\end{equation}
Since the right-hand side of \cref{eq:dispersion_MDM2D_nocurv_alt} is always real, the imaginary part of the left-hand side,
\begin{equation}
	\im \inthalfinf{\mu} \frac{\mu\rme^{-\mu}}{\mu - \omega / \omegade} = \im(\omega / \omegade) \inthalfinf{\mu} \frac{\mu\rme^{-\mu}}{\abs{\mu - \omega / \omegade}^2},
	\label{eq:dispersion_MDM2D_nocurv_alt_imaginarypart}
\end{equation}
must vanish. However, \cref{eq:dispersion_MDM2D_nocurv_alt_imaginarypart} is nonzero as long as \mbox{\(\im(\omega) > 0\)}.\footnote{The integral expression \cref{eq:Iab_nocurv} is valid only for unstable modes, so we cannot use \cref{eq:dispersion_MDM2D_nocurv_alt_imaginarypart} to conclude that there are no damped modes without picking a branch and performing the relevant analytical continuation.} Therefore, we conclude that the MDM is destabilised exclusively by the curvature drifts. 

This behaviour is drastically different from that of the ETG instability discussed in~\cref{sec:cETG}. Indeed, analogously to \cref{eq:dispersion_MDM2D_nogradB} and \cref{eq:dispersion_MDM2D_nocurv}, we can derive the two-dimensional ETG dispersion relations in the absence of \(\grad B\),
\begin{align}
	0 = \Lphiphi^{(\text{no \(\grad B\)})} = &-1 - \tau^{-1} - \left(1 - \frac{\omegaste}{\omega}\right)\sqrt{\frac{\omega}{2\omegade}}\rmZ\left(\sqrt{\frac{\omega}{2\omegade}}\right) \nonumber \\
	& + \frac{\omegaTe}{2\omegade}\left[1 + \left(1 - \frac{\omegade}{\omega}\right)\sqrt{\frac{\omega}{2\omegade}}\rmZ\left(\sqrt{\frac{\omega}{2\omegade}}\right)\right],
	\label{eq:Lphiphi_nogradB}
\end{align}
or of curvature drifts,
\begin{align}
	0 = \Lphiphi^{(\text{no curv})} = &-1 - \tau^{-1} - \frac{\omega-\omegaste}{\omegade}\rme^{-\omega/\omegade}\rmE_1\left(-\frac{\omega}{\omegade}\right) \nonumber \\
	&+ \frac{\omegaTe}{\omegade} \left[1 + \left(\frac{\omega}{\omegade} - 1\right)\rme^{-\omega/\omegade}\rmE_1\left(-\frac{\omega}{\omegade}\right)\right].
	\label{eq:Lphiphi_nocurv}
\end{align}
Expanding in the strongly driven (\mbox{\(\omegade \ll \omega \ll \omegaTe\)}), zero-density-gradient (\(\omegaste = 0\)) limit considered in \cref{sec:cETG}, we find that both \cref{eq:Lphiphi_nogradB} and \cref{eq:Lphiphi_nocurv} yield
\begin{equation}
	\omega^2 = -\omegaTe \omegade\tau.
	\label{eq:cETG_disp_onlyonedrift}
\end{equation}
This dispersion relation can also be obtained directly from \crefrange{eq:dne_etg}{eq:dTperpe_etg} by noticing that dropping the \(\grad B\) drift simplifies the fluid equations to
\begin{align}
	&\partd{}{t}\frac{\dnek}{\ne} + \rmi\omegade\frac{\dTparek}{\Te} = 0, \\
	&\partd{}{t}\frac{\dTparek}{\Te}  + \rmi\omegaTe\frac{e\phipotk}{\Te} = 0,
\end{align}
whereas dropping the curvature drift yields
\begin{align}
	&\partd{}{t}\frac{\dnek}{\ne} + \rmi\omegade\frac{\dTperpek}{\Te} = 0, \\
	&\partd{}{t}\frac{\dTperpek}{\Te} + \rmi\omegaTe\frac{e\phipotk}{\Te} = 0.
\end{align}
In either case, we obtain the dispersion relation \cref{eq:cETG_disp_onlyonedrift}. This confirms that the curvature-driven ETG instability is indifferent to which magnetic drift couples the density and temperature fluctuations. 

\subsection{Colliding beams}
\label{appendix:delta_equilibrium}

Possibly the simplest version of the MDM can be obtained if we make the unphysical assumption that the unperturbed distribution function is
\begin{equation}
	\Fe = \frac{1}{2}\ne \delta(v_x)\delta(v_y) \left[\delta(\vpar - v_0) + \delta(\vpar + v_0)\right],
	\label{eq:Fe_beams} 
\end{equation}
where \(v_0 = v_0(x)\) can be thought of as a spatially varying thermal speed: by definition,
\begin{equation}
	\frac{1}{2}\vthe^2 = \frac{1}{\ne}\intv v^2 \Fe = v_0^2.
\end{equation}
The temperature-gradient length scale \cref{eq:equilibrium_gradients} is then
\begin{equation}
	\frac{1}{\LTe} = -2 \der{\ln v_0}{x}.
\end{equation}
The distribution function \cref{eq:Fe_beams} is made up of two counter-propagating beams aligned with the \(z\) direction, with zero net momentum but radially varying energy. Such a distribution function is entirely unphysical in the sense that it would give rise to high-frequency instabilities that are outside the ordering \cref{eq:gyrokinetic_ordering}. In other words, \cref{eq:Fe_beams} is not a valid GK equilibrium distribution. Nevertheless, we can use it as an ad hoc toy model for the MDM.

\begin{figure}
	\begingroup\import{figs/fig9}{fig9.pgf}\endgroup
	\caption{Same as \cref{fig:mdm_omega,fig:mdm_omega_nogradB} but for the unstable solutions of \cref{eq:disp_beams}.}
	\label{fig:mdm_beams}
\end{figure}

Note that the last term in the odd (in \(\vpar\)) drift-kinetic equation \cref{eq:gk_2d_odd_appendix} comes from the last term in \cref{eq:gyrokinetic_equation}, and thus \cref{eq:gk_2d_odd_appendix} is
\begin{align}
	&\partd{}{t} \left(\hek^\text{(odd)} - \frac{2\vpar}{\vthe}\frac{\dApark}{\rhoe B} \Fe \right) + \rmi \omegade\left(\frac{2\vpar^2}{\vthe^2} + \frac{\vperp^2}{\vthe^2}\right)\hek^\text{(odd)}+ \rmi k_y \frac{\rhoe\vthe}{2} \frac{2\vpar}{\vthe}\frac{\dApark}{\rhoe B} \der{\Fe}{x} = 0.
	\label{eq:gk_2d_odd_appendix_genericFe}
\end{align}
Substituting \cref{eq:Fe_beams} instead of a Maxwellian \(\Fe\) into \cref{eq:gk_2d_odd_appendix_genericFe}, we find that the odd drift-kinetic equation for a mode with complex frequency \(\omega\) is
\begin{align}
	&-\omega \left(\hek^\text{(odd)} - \frac{2\vpar}{\vthe}\frac{\dApark}{\rhoe B} \Fe \right) + \omegade\left(\frac{2\vpar^2}{\vthe^2} + \frac{\vperp^2}{\vthe^2}\right)\hek^\text{(odd)} - \omegaste \frac{2\vpar}{\vthe}\frac{\dApark}{\rhoe B}\Fe\nonumber \\
	&\quad - \omegaTe \frac{2\vpar}{\vthe}\frac{\dApark}{\rhoe B} \frac{\ne v_0}{4} \delta(v_x)\delta(v_y) \partd{}{\vpar} \left[\delta(\vpar - v_0) - \delta(\vpar + v_0)\right] = 0.
	\label{eq:gk_2d_odd_beams_appendix}
\end{align}
Substituting \(\hek^\text{(odd)}\) from \cref{eq:gk_2d_odd_beams_appendix} into Amp\`ere's law \cref{eq:parallel_amperes_mdm_appendix}, we obtain the following dispersion relation:
\begin{equation}
	-\kperpdesq - \frac{\omega -\omegaste -\omegaTe}{\omega - \omegade} + \frac{\omegaTe\omegade}{(\omega - \omegade)^2} = 0.
	\label{eq:disp_beams}
\end{equation}
It is straightforward to show that \cref{eq:disp_beams} has unstable solutions when
\begin{equation}
	(\kappan + \kappaT - 1)^2 + 4\kappaT (1+\kperpdesq) < 0,
	\label{eq:beams_threshold}
\end{equation}
which is possible only if \(\kappaT < 0\). Note that in the case of zero magnetic drifts, viz., \(\omegade = 0\), \cref{eq:disp_beams} reproduces the magnetic drift wave \cref{eq:magnetic_drift_wave_freq}.

\Cref{fig:mdm_beams} shows the unstable solutions of \cref{eq:disp_beams} for the same parameters as \cref{fig:mdm_omega,fig:mdm_omega_nogradB}. The qualitative similarity between these is evident, in particular the fact that the unstable solutions of \cref{eq:disp_beams} can have either sign of \(\re\:\omega\). The main difference is that, in the limit \(\kperpdesq \gg 1\), the stability boundary associated with the \(\re\:\omega > 0\), i.e., co-propagating, modes asymptotes to \(\kappaT \to 0^-\), as can be seen directly from \cref{eq:beams_threshold}. This is consistent with the hypothesis that the stabilisation of the MDM at positive real frequency is due to the Landau-like damping arising from the magnetic drifts. The discontinuous nature of the distribution function \cref{eq:Fe_beams} implies that only a single \(\omega\), viz., \(\omega=\omegade\), is resonant, unlike in the continuous case, where every \(\omega > 0\) resonates with some particles.

\bibliographystyle{jpp}
\bibliography{mdm_paper.bib}

\begin{thebibliography}{75}
\expandafter\ifx\csname natexlab\endcsname\relax\def\natexlab#1{#1}\fi
\def\au#1{#1} \def\ed#1{#1} \def\yr#1{#1}\def\at#1{#1}\def\jt#1{\textit{#1}}
  \def\bt#1{#1}\def\bvol#1{\textbf{#1}} \def\vol#1{#1} \def\pg#1{#1}
  \def\publ#1{#1}\def\arxiv#1{#1}\def\org#1{#1}\def\st#1{\textit{#1}}

\bibitem[Abel {\em et~al.\/}(2013)Abel, Plunk, Wang, Barnes, Cowley, Dorland \&
  Schekochihin]{abel13}
{\sc \au{Abel, I.~G.}, \au{Plunk, G.~G.}, \au{Wang, E.}, \au{Barnes, M.},
  \au{Cowley, S.~C.}, \au{Dorland, W.} \& \au{Schekochihin, A.~A.}} \yr{2013}
  \at{Multiscale gyrokinetics for rotating tokamak plasmas: fluctuations,
  transport and energy flows}.  \jt{Rep. Prog. Phys.}  \bvol{76}~(11),
  \pg{116201}.

\bibitem[Adam {\em et~al.\/}(1973)Adam, Laval \& Pellat]{adam73}
{\sc \au{Adam, J.~C.}, \au{Laval, G.} \& \au{Pellat, R.}} \yr{1973}
  \at{Localized drift dissipative modes in tokamaks}.  \jt{Nucl. Fusion}
  \bvol{13}~(1),  \pg{47--53}.

\bibitem[Adam {\em et~al.\/}(1976)Adam, Tang \& Rutherford]{adam76}
{\sc \au{Adam, J.~C.}, \au{Tang, W.~M.} \& \au{Rutherford, P.~H.}} \yr{1976}
  \at{Destabilization of the trapped-electron mode by magnetic curvature drift
  resonances}.  \jt{Phys. Fluids}  \bvol{19}~(4),  \pg{561--566}.

\bibitem[Adkins {\em et~al.\/}(2023)Adkins, Ivanov \& Schekochihin]{adkins23}
{\sc \au{Adkins, T.}, \au{Ivanov, P.~G.} \& \au{Schekochihin, A.~A.}} \yr{2023}
   \at{Scale invariance and critical balance in electrostatic drift-kinetic
  turbulence}.  \jt{J. Plasma Phys.}  \bvol{89}~(4),  \pg{905890406}.

\bibitem[Adkins {\em et~al.\/}(2022)Adkins, Schekochihin, Ivanov \&
  Roach]{adkins22}
{\sc \au{Adkins, T.}, \au{Schekochihin, A.~A.}, \au{Ivanov, P.~G.} \&
  \au{Roach, C.~M.}} \yr{2022}  \at{Electromagnetic instabilities and plasma
  turbulence driven by electron-temperature gradient}.  \jt{J. Plasma Phys.}
  \bvol{88}~(4),  \pg{905880410}.

\bibitem[Bateman \& Peng(1977)]{bateman77}
{\sc \au{Bateman, G.~S.} \& \au{Peng, Y. K.~M.}} \yr{1977}
  \at{Magnetohydrodynamic stability of flux-conserving tokamak equilibria}.
  \jt{Phys. Rev. Lett.}  \bvol{38}~(15),  \pg{829--832}.

\bibitem[Beer(1995)]{beer95_thesis}
{\sc \au{Beer, M.~A.}} \yr{1995}  \at{Gyrofluid models of turbulent transport
  in tokamaks}. PhD thesis, Princeton University.

\bibitem[Beer {\em et~al.\/}(1995)Beer, Cowley \& Hammett]{beer95}
{\sc \au{Beer, M.~A.}, \au{Cowley, S.~C.} \& \au{Hammett, G.~W.}} \yr{1995}
  \at{Field-aligned coordinates for nonlinear simulations of tokamak
  turbulence}.  \jt{Phys. Plasmas}  \bvol{2}~(7),  \pg{2687--2700}.

\bibitem[Biglari {\em et~al.\/}(1989)Biglari, Diamond \& Rosenbluth]{biglari89}
{\sc \au{Biglari, H.}, \au{Diamond, P.~H.} \& \au{Rosenbluth, M.~N.}} \yr{1989}
   \at{Toroidal ion-pressure-gradient-driven drift instabilities and transport
  revisited}.  \jt{Phys. Fluids B}  \bvol{1}~(1),  \pg{109--118}.

\bibitem[Callen(1977)]{callen77}
{\sc \au{Callen, J.~D.}} \yr{1977}  \at{Drift-wave turbulence effects on
  magnetic structure and plasma transport in tokamaks}.  \jt{Phys. Rev. Lett.}
  \bvol{39}~(24),  \pg{1540--1543}.

\bibitem[Candy {\em et~al.\/}(2007)Candy, Waltz, Fahey \& Holland]{candy07}
{\sc \au{Candy, J.}, \au{Waltz, R.~E.}, \au{Fahey, M.~R.} \& \au{Holland, C.}}
  \yr{2007}  \at{The effect of ion-scale dynamics on
  electron-temperature-gradient turbulence}.  \jt{Plasma Phys. Control. Fusion}
   \bvol{49}~(8),  \pg{1209--1220}.

\bibitem[Catto(2019)]{catto19}
{\sc \au{Catto, P.~J.}} \yr{2019}  \at{Practical gyrokinetics}.  \jt{J. Plasma
  Phys.}  \bvol{85}~(03),  \pg{925850301}.

\bibitem[Catto \& Tsang(1978)]{catto78a}
{\sc \au{Catto, P.~J.} \& \au{Tsang, K.~T.}} \yr{1978}  \at{Trapped electron
  instability in tokamaks: Analytic solution of the two-dimensional eigenvalue
  problem}.  \jt{Phys. Fluids}  \bvol{21}~(8),  \pg{1381--1388}.

\bibitem[Chandran \& Schekochihin(2024)]{chandran24}
{\sc \au{Chandran, B. D.~G.} \& \au{Schekochihin, A.~A.}} \yr{2024}  \at{The
  gyrokinetic dispersion relation of microtearing modes in collisionless
  toroidal plasmas}.  \jt{J. Plasma Phys.}  \bvol{90}~(2),  \pg{905900204}.

\bibitem[Cheng \& Chen(1981)]{cheng81}
{\sc \au{Cheng, C.~Z.} \& \au{Chen, L.}} \yr{1981}  \at{Ballooning-mode theory
  of trapped-electron instabilities in tokamaks}.  \jt{Nucl. Fusion}
  \bvol{21}~(3),  \pg{403--408}.

\bibitem[Connor {\em et~al.\/}(1978)Connor, Hastie \& Taylor]{connor78}
{\sc \au{Connor, J.~W.}, \au{Hastie, R.~J.} \& \au{Taylor, J.~B.}} \yr{1978}
  \at{Shear, periodicity, and plasma ballooning modes}.  \jt{Phys. Rev. Lett.}
  \bvol{40}~(6),  \pg{396--399}.

\bibitem[Coppi {\em et~al.\/}(1966)Coppi, Furth, Rosenbluth \&
  Sagdeev]{coppi66}
{\sc \au{Coppi, B.}, \au{Furth, H.~P.}, \au{Rosenbluth, M.~N.} \& \au{Sagdeev,
  R.~Z.}} \yr{1966}  \at{Drift instability due to impurity ions}.  \jt{Phys.
  Rev. Lett.}  \bvol{17}~(7),  \pg{377--379}.

\bibitem[Coppi {\em et~al.\/}(1967)Coppi, Rosenbluth \& Sagdeev]{coppi67}
{\sc \au{Coppi, B.}, \au{Rosenbluth, M.~N.} \& \au{Sagdeev, R.~Z.}} \yr{1967}
  \at{Instabilities due to temperature gradients in complex magnetic field
  configurations}.  \jt{Phys. Fluids}  \bvol{10}~(3),  \pg{582--587}.

\bibitem[Cowley {\em et~al.\/}(1991)Cowley, Kulsrud \& Sudan]{cowley91}
{\sc \au{Cowley, S.~C.}, \au{Kulsrud, R.~M.} \& \au{Sudan, R.}} \yr{1991}
  \at{Considerations of ion-temperature-gradient-driven turbulence}.  \jt{Phys.
  Fluids B}  \bvol{3}~(10),  \pg{2767--2782}.

\bibitem[D’haeseleer {\em et~al.\/}(1991)D’haeseleer, Hitchon, Callen \&
  Shohet]{d’haeseleer91_book}
{\sc \au{D’haeseleer, W.~D.}, \au{Hitchon, W. N.~G.}, \au{Callen, J.~D.} \&
  \au{Shohet, J.~L.}} \yr{1991} {\em Flux Coordinates and Magnetic Field
  Structure\/}.  \publ{Berlin, Heidelberg: Springer-Verlag}.

\bibitem[Faddeeva \& Terent'ev(1954)]{faddeeva54}
{\sc \au{Faddeeva, V.~N.} \& \au{Terent'ev, N.~M.}} \yr{1954} {\em Tablicy
  zna\v ceni\u\i\ funkcii {$w(z)=e^{-z^{2}}(1+\frac{2i}{\sqrt\pi}\int^z_0
  e^{t^{2}}dt)$} ot kompleksnogo argumenta\/}.  \publ{Moscow: Gosudarstv.
  Izdat. Tehn.-Teor. Lit.}

\bibitem[Frei {\em et~al.\/}(2022)Frei, Ernst \& Ricci]{frei22}
{\sc \au{Frei, B.~J.}, \au{Ernst, S.} \& \au{Ricci, P.}} \yr{2022}
  \at{Numerical implementation of the improved sugama collision operator using
  a moment approach}.  \jt{Phys. Plasmas}  \bvol{29}~(9),  \pg{093902}.

\bibitem[Frei {\em et~al.\/}(2023)Frei, Hoffmann, Ricci, Brunner \&
  Tecchioll]{frei23}
{\sc \au{Frei, B.~J.}, \au{Hoffmann, A. C.~D.}, \au{Ricci, P.}, \au{Brunner,
  S.} \& \au{Tecchioll, Z.}} \yr{2023}  \at{Moment-based approach to the
  flux-tube linear gyrokinetic model}.  \jt{J. Plasma Phys.}  \bvol{89}~(4),
  \pg{905890414}.

\bibitem[{Fried} \& {Conte}(1961)]{fried61}
{\sc \au{{Fried}, B.~D.} \& \au{{Conte}, S.~D.}} \yr{1961} {\em The Plasma
  Dispersion Function\/}.  \publ{New York, London: Academic Press Inc.}

\bibitem[Fyfe \& Montgomery(1976)]{fyfe76}
{\sc \au{Fyfe, D.} \& \au{Montgomery, D.}} \yr{1976}  \at{High-beta turbulence
  in two-dimensional magnetohydrodynamics}.  \jt{J. Plasma Phys.}
  \bvol{16}~(2),  \pg{181--191}.

\bibitem[Giacomin {\em et~al.\/}(2024)Giacomin, Kennedy, Casson, C.~J.,
  Dickinson, Patel \& Roach]{giacomin24}
{\sc \au{Giacomin, M.}, \au{Kennedy, D.}, \au{Casson, F.~J.}, \au{C.~J., Ajay},
  \au{Dickinson, D.}, \au{Patel, B.~S.} \& \au{Roach, C.~M.}} \yr{2024}  \at{On
  electromagnetic turbulence and transport in step}.  \jt{Plasma Phys. Control.
  Fusion}  \bvol{66}~(5),  \pg{055010}.

\bibitem[Guzdar {\em et~al.\/}(1983)Guzdar, Chen, Tang \& Rutherford]{guzdar83}
{\sc \au{Guzdar, P.~N.}, \au{Chen, L.}, \au{Tang, W.~M.} \& \au{Rutherford,
  P.~H.}} \yr{1983}  \at{Ion-temperature-gradient instability in toroidal
  plasmas}.  \jt{Phys. Fluids}  \bvol{26}~(3),  \pg{673--677}.

\bibitem[Hallenbert \& Plunk(2022)]{hallenbert22}
{\sc \au{Hallenbert, A.} \& \au{Plunk, G.~G.}} \yr{2022}  \at{Predicting the
  z-pinch dimits shift through gyrokinetic tertiary instability analysis of the
  entropy mode}.  \jt{J. Plasma Phys.}  \bvol{88}~(4),  \pg{905880402}.

\bibitem[Hammett {\em et~al.\/}(1993)Hammett, Beer, Dorland, Cowley \&
  Smith]{hammett93}
{\sc \au{Hammett, G.~W.}, \au{Beer, M.~A.}, \au{Dorland, W.}, \au{Cowley,
  S.~C.} \& \au{Smith, S.~A.}} \yr{1993}  \at{Developments in the gyrofluid
  approach to tokamak turbulence simulations}.  \jt{Plasma Phys. Control.
  Fusion}  \bvol{35}~(8),  \pg{973--985}.

\bibitem[Hardman {\em et~al.\/}(2020)Hardman, Barnes \& Roach]{hardman20}
{\sc \au{Hardman, M.~R.}, \au{Barnes, M.} \& \au{Roach, C.~M.}} \yr{2020}
  \at{Stabilisation of short-wavelength instabilities by parallel-to-the-field
  shear in long-wavelength {E × B} flows}.  \jt{J. Plasma Phys.}
  \bvol{86}~(6),  \pg{905860601}.

\bibitem[Hardman {\em et~al.\/}(2019)Hardman, Barnes, Roach \&
  Parra]{hardman19}
{\sc \au{Hardman, M.~R.}, \au{Barnes, M.}, \au{Roach, C.~M.} \& \au{Parra,
  F.~I.}} \yr{2019}  \at{A scale-separated approach for studying coupled ion
  and electron scale turbulence}.  \jt{Plasma Phys. Control. Fusion}
  \bvol{61}~(6),  \pg{065025}.

\bibitem[Helander {\em et~al.\/}(2011)Helander, Mishchenko, Kleiber \&
  Xanthopoulos]{helander11}
{\sc \au{Helander, P.}, \au{Mishchenko, A.}, \au{Kleiber, R.} \&
  \au{Xanthopoulos, P.}} \yr{2011}  \at{Oscillations of zonal flows in
  stellarators}.  \jt{Plasma Phys. Control. Fusion}  \bvol{53}~(5),
  \pg{054006}.

\bibitem[Helander {\em et~al.\/}(2013)Helander, Proll \& Plunk]{helander13}
{\sc \au{Helander, P.}, \au{Proll, J. H.~E.} \& \au{Plunk, G.~G.}} \yr{2013}
  \at{Collisionless microinstabilities in stellarators. i. analytical theory of
  trapped-particle modes}.  \jt{Phys. Plasmas}  \bvol{20}~(12),  \pg{122505}.

\bibitem[Hoffmann {\em et~al.\/}(2023)Hoffmann, Frei \& Ricci]{hoffmann23}
{\sc \au{Hoffmann, A. C.~D.}, \au{Frei, B.~J.} \& \au{Ricci, P.}} \yr{2023}
  \at{Gyrokinetic simulations of plasma turbulence in a z-pinch using a
  moment-based approach and advanced collision operators}.  \jt{J. Plasma
  Phys.}  \bvol{89}~(2),  \pg{905890214}.

\bibitem[Ivanov \& Adkins(2023)]{ivanov23}
{\sc \au{Ivanov, P.~G.} \& \au{Adkins, T.}} \yr{2023}  \at{An analytical form
  of the dispersion function for local linear gyrokinetics in a curved magnetic
  field}.  \jt{J. Plasma Phys.}  \bvol{89}~(2),  \pg{905890213}.

\bibitem[Ivanov {\em et~al.\/}(2022)Ivanov, Schekochihin \& Dorland]{ivanov22}
{\sc \au{Ivanov, P.~G.}, \au{Schekochihin, A.~A.} \& \au{Dorland, W.}}
  \yr{2022}  \at{Dimits transition in three-dimensional
  ion-temperature-gradient turbulence}.  \jt{J. Plasma Phys.}  \bvol{88}~(5),
  \pg{905880506}.

\bibitem[Ivanov {\em et~al.\/}(2020)Ivanov, Schekochihin, Dorland, Field \&
  Parra]{ivanov20}
{\sc \au{Ivanov, P.~G.}, \au{Schekochihin, A.~A.}, \au{Dorland, W.}, \au{Field,
  A.~R.} \& \au{Parra, F.~I.}} \yr{2020}  \at{Zonally dominated dynamics and
  dimits threshold in curvature-driven itg turbulence}.  \jt{J. Plasma Phys.}
  \bvol{86}~(5),  \pg{855860502}.

\bibitem[Jian {\em et~al.\/}(2021)Jian, Holland, Candy, Ding, Belli, Chan,
  Staebler, Garofalo, Mcclenaghan \& Snyder]{jian21}
{\sc \au{Jian, X.}, \au{Holland, C.}, \au{Candy, J.}, \au{Ding, S.}, \au{Belli,
  E.}, \au{Chan, V.}, \au{Staebler, G.~M.}, \au{Garofalo, A.~M.},
  \au{Mcclenaghan, J.} \& \au{Snyder, P.}} \yr{2021}  \at{Role of microtearing
  mode in diii-d and future high-$\beta_p$ core plasmas}.  \jt{Phys. Plasmas}
  \bvol{28}~(4),  \pg{042501}.

\bibitem[Kennedy {\em et~al.\/}(2023)Kennedy, Giacomin, Casson, Dickinson,
  Hornsby, Patel \& Roach]{kennedy23}
{\sc \au{Kennedy, D.}, \au{Giacomin, M.}, \au{Casson, F.~J.}, \au{Dickinson,
  D.}, \au{Hornsby, W.~A.}, \au{Patel, B.~S.} \& \au{Roach, C.~M.}} \yr{2023}
  \at{Electromagnetic gyrokinetic instabilities in step}.  \jt{Nucl. Fusion}
  \bvol{63}~(12),  \pg{126061}.

\bibitem[Kim {\em et~al.\/}(1994)Kim, Kishimoto, Horton \& Tajima]{kim94}
{\sc \au{Kim, J.~Y.}, \au{Kishimoto, Y.}, \au{Horton, W.} \& \au{Tajima, T.}}
  \yr{1994}  \at{Kinetic resonance damping rate of the toroidal ion temperature
  gradient mode}.  \jt{Phys. Plasmas}  \bvol{1}~(4),  \pg{927--936}.

\bibitem[Kobayashi {\em et~al.\/}(2015)Kobayashi, Gürcan \&
  Diamond]{kobayashi15}
{\sc \au{Kobayashi, S.}, \au{Gürcan, Ö.~D.} \& \au{Diamond, P.~H.}} \yr{2015}
   \at{Direct identification of predator-prey dynamics in gyrokinetic
  simulations}.  \jt{Phys. Plasmas}  \bvol{22}~(9),  \pg{090702}.

\bibitem[Kobayashi \& Rogers(2012)]{kobayashi12}
{\sc \au{Kobayashi, S.} \& \au{Rogers, B.~N.}} \yr{2012}  \at{The quench rule,
  dimits shift, and eigenmode localization by small-scale zonal flows}.
  \jt{Phys. Plasmas}  \bvol{19}~(1),  \pg{012315}.

\bibitem[Kruskal \& Kulsrud(1958)]{kruskal58}
{\sc \au{Kruskal, M.~D.} \& \au{Kulsrud, R.~M.}} \yr{1958}  \at{Equilibrium of
  a magnetically confined plasma in a toroid}.  \jt{Phys. Fluids}
  \bvol{1}~(4),  \pg{265--274}.

\bibitem[Kuroda {\em et~al.\/}(1998)Kuroda, Sugama, Kanno, Okamoto \&
  Horton]{kuroda98}
{\sc \au{Kuroda, T.}, \au{Sugama, H.}, \au{Kanno, R.}, \au{Okamoto, M.} \&
  \au{Horton, W.}} \yr{1998}  \at{Initial value problem of the toroidal ion
  temperature gradient mode}.  \jt{J. Phys. Soc. Jpn.}  \bvol{67}~(11),
  \pg{3787--3787}.

\bibitem[Loureiro {\em et~al.\/}(2016)Loureiro, Dorland, Fazendeiro, Kanekar,
  Mallet, Vilelas \& Zocco]{loureiro16}
{\sc \au{Loureiro, N.~F.}, \au{Dorland, W.}, \au{Fazendeiro, L.}, \au{Kanekar,
  A.}, \au{Mallet, A.}, \au{Vilelas, M.~S.} \& \au{Zocco, A.}} \yr{2016}
  \at{Viriato: A fourier-hermite spectral code for strongly magnetized
  fluid-kinetic plasma dynamics}.  \jt{Comput. Phys. Comm.}  \bvol{206},
  \pg{45--63}.

\bibitem[Maeyama {\em et~al.\/}(2015)Maeyama, Idomura, Watanabe, Nakata, Yagi,
  Miyato, Ishizawa \& Nunami]{maeyama15}
{\sc \au{Maeyama, S.}, \au{Idomura, Y.}, \au{Watanabe, T.-H.}, \au{Nakata, M.},
  \au{Yagi, M.}, \au{Miyato, N.}, \au{Ishizawa, A.} \& \au{Nunami, M.}}
  \yr{2015}  \at{Cross-scale interactions between electron and ion scale
  turbulence in a tokamak plasma}.  \jt{Phys. Rev. Lett.}  \bvol{114}~(25),
  \pg{255002}.

\bibitem[Maeyama {\em et~al.\/}(2017)Maeyama, Watanabe, Idomura, Nakata,
  Ishizawa \& Nunami]{maeyama17}
{\sc \au{Maeyama, S.}, \au{Watanabe, T.-H.}, \au{Idomura, Y.}, \au{Nakata, M.},
  \au{Ishizawa, A.} \& \au{Nunami, M.}} \yr{2017}  \at{Cross-scale interactions
  between turbulence driven by electron and ion temperature gradients via
  sub-ion-scale structures}.  \jt{Nucl. Fusion}  \bvol{57}~(6),  \pg{066036}.

\bibitem[Mandell {\em et~al.\/}(2018)Mandell, Dorland \& Landreman]{mandell18}
{\sc \au{Mandell, N. R.}, \au{Dorland, W.} \& \au{Landreman, M.}} \yr{2018}
  \at{Laguerre–hermite pseudo-spectral velocity formulation of gyrokinetics}.
   \jt{J. Plasma Phys.}  \bvol{84}~(1),  \pg{905840108}.

\bibitem[Manheimer \& Cook(1978)]{manheimer78}
{\sc \au{Manheimer, W.~M.} \& \au{Cook, I.}} \yr{1978}  \bt{A theory of
  particle and energy flux from the magnetic flutter of drift waves}. {\em
  Tech. Rep.\/} NRL Memorandum Report 3752.  \org{Naval Research Laboratory,
  Washington, DC}.

\bibitem[Mischenko(2024)]{mischenko24_thesis}
{\sc \au{Mischenko, N.}} \yr{2024} Hyperbolicity of a {Hermite-Laguerre} moment
  model of the plasma periphery in slab geometry. Master's thesis, Eindhoven
  University of Technology.

\bibitem[Mishchenko {\em et~al.\/}(2018)Mishchenko, Plunk \&
  Helander]{mishchenko18}
{\sc \au{Mishchenko, A.}, \au{Plunk, G.~G.} \& \au{Helander, P.}} \yr{2018}
  \at{Electrostatic stability of electron–positron plasmas in dipole
  geometry}.  \jt{J. Plasma Phys.}  \bvol{84}~(2),  \pg{905840201}.

\bibitem[Newton {\em et~al.\/}(2010)Newton, Cowley \& Loureiro]{newton10}
{\sc \au{Newton, S.~L.}, \au{Cowley, S.~C.} \& \au{Loureiro, N.~F.}} \yr{2010}
  \at{Understanding the effect of sheared flow on microinstabilities}.
  \jt{Plasma Phys. Control. Fusion}  \bvol{52}~(12),  \pg{125001}.

\bibitem[Parisi {\em et~al.\/}(2020)Parisi, Parra, Roach, Giroud, Dorland,
  Hatch, Barnes, Hillesheim, Aiba, Ball, Ivanov \& {JET
  contributors}]{parisi20}
{\sc \au{Parisi, J.~F.}, \au{Parra, F.~I.}, \au{Roach, C.~M.}, \au{Giroud, C.},
  \au{Dorland, W.}, \au{Hatch, D.~R.}, \au{Barnes, M.}, \au{Hillesheim, J.~C.},
  \au{Aiba, N.}, \au{Ball, J.}, \au{Ivanov, P.~G.} \& \au{{JET contributors}}}
  \yr{2020}  \at{Toroidal and slab {ETG} instability dominance in the linear
  spectrum of {JET-ILW} pedestals}.  \jt{Nucl. Fusion}  \bvol{60}~(12),
  \pg{126045}.

\bibitem[Patel(2021)]{patel21_thesis}
{\sc \au{Patel, B.}} \yr{2021}  \at{Confinement physics for a steady state
  netelectric burning spherical tokamak}. PhD thesis, University of York.

\bibitem[Plunk {\em et~al.\/}(2010)Plunk, Cowley, Schekochihin \&
  Tatsuno]{plunk10}
{\sc \au{Plunk, G.~G.}, \au{Cowley, S.~C.}, \au{Schekochihin, A.~A.} \&
  \au{Tatsuno, T.}} \yr{2010}  \at{Two-dimensional gyrokinetic turbulence}.
  \jt{J. Fluid Mech.}  \bvol{664},  \pg{407--435}.

\bibitem[Plunk \& Helander(2023)]{plunk23}
{\sc \au{Plunk, G.~G.} \& \au{Helander, P.}} \yr{2023}  \at{Energetic bounds on
  gyrokinetic instabilities. {Part} 3. {Generalized} free energy}.  \jt{J.
  Plasma Phys.}  \bvol{89}~(4),  \pg{905890419}.

\bibitem[Pogutse(1968)]{pogutse68}
{\sc \au{Pogutse, O.~P.}} \yr{1968}  \at{Magnetic drift instability in a
  collisionless plasma}.  \jt{Plasma Phys.}  \bvol{10}~(7),  \pg{649--664}.

\bibitem[Pouquet(1978)]{pouquet78}
{\sc \au{Pouquet, A.}} \yr{1978}  \at{On two-dimensional magnetohydrodynamic
  turbulence}.  \jt{J. Fluid Mech.}  \bvol{88}~(1),  \pg{1}.

\bibitem[Ricci {\em et~al.\/}(2006)Ricci, Rogers \& Dorland]{ricci06}
{\sc \au{Ricci, P.}, \au{Rogers, B.~N.} \& \au{Dorland, W.}} \yr{2006}
  \at{Small-scale turbulence in a closed-field-line geometry}.  \jt{Phys. Rev.
  Lett.}  \bvol{97}~(24),  \pg{245001}.

\bibitem[Rodríguez {\em et~al.\/}(2024)Rodríguez, Helander \&
  Goodman]{rodriguez24}
{\sc \au{Rodríguez, E.}, \au{Helander, P.} \& \au{Goodman, A.~G.}} \yr{2024}
  \at{The maximum-{J} property in quasi-isodynamic stellarators}.  \jt{J.
  Plasma Phys.}  \bvol{90}~(2),  \pg{905900212}.

\bibitem[Romanelli(1989)]{romanelli89}
{\sc \au{Romanelli, F.}} \yr{1989}  \at{Ion temperature-gradient-driven modes
  and anomalous ion transport in tokamaks}.  \jt{Phys. Fluids B}  \bvol{1}~(5),
   \pg{1018--1025}.

\bibitem[Rudakov \& Sagdeev(1961)]{rudakov61}
{\sc \au{Rudakov, L.~I.} \& \au{Sagdeev, R.~Z.}} \yr{1961}  \at{On the
  instability of a nonuniform rarefied plasma in a strong magnetic field}.
  \jt{Dokl. Akad. Nauk SSSR}  \bvol{138}~(3),  \pg{581--583}.

\bibitem[Schekochihin(2022)]{schekochihin22}
{\sc \au{Schekochihin, A.~A.}} \yr{2022}  \at{{MHD} turbulence: a biased
  review}.  \jt{J. Plasma Phys.}  \bvol{88}~(5),  \pg{155880501}.

\bibitem[Schekochihin {\em et~al.\/}(2009)Schekochihin, Cowley, Dorland,
  Hammett, Howes, Quataert \& Tatsuno]{schekochihin09}
{\sc \au{Schekochihin, A.~A.}, \au{Cowley, S.~C.}, \au{Dorland, W.},
  \au{Hammett, G.~W.}, \au{Howes, G.~G.}, \au{Quataert, E.} \& \au{Tatsuno,
  T.}} \yr{2009}  \at{Astrophysical gyrokinetics: Kinetic and fluid turbulent
  cascades in magnetized weakly collisional plasmas}.  \jt{Astrophys. J. Suppl.
  Ser.}  \bvol{182}~(1),  \pg{310--377}.

\bibitem[Smith(1997)]{smith97_thesis}
{\sc \au{Smith, S.~A.}} \yr{1997}  \at{Dissipative closures for statistical
  moments, fluid moments, and subgrid scales in plasma turbulence}. PhD thesis,
  Princeton University.

\bibitem[Sugama(1999)]{sugama99}
{\sc \au{Sugama, H.}} \yr{1999}  \at{Damping of toroidal ion temperature
  gradient modes}.  \jt{Phys. Plasmas}  \bvol{6}~(9),  \pg{3527--3535}.

\bibitem[Terry {\em et~al.\/}(1982)Terry, Anderson \& Horton]{terry82}
{\sc \au{Terry, P.}, \au{Anderson, W.} \& \au{Horton, W.}} \yr{1982}
  \at{Kinetic effects on the toroidal ion pressure gradient drift mode}.
  \jt{Nucl. Fusion}  \bvol{22}~(4),  \pg{487--497}.

\bibitem[Tholerus {\em et~al.\/}(2024)Tholerus, Casson, Marsden, Wilson,
  Brunetti, Fox, Freethy, Hender, Henderson, Hudoba, Kirov, Koechl, Meyer,
  Muldrew, Olde, Patel, Roach, Saarelma \& Xia]{tholerus24}
{\sc \au{Tholerus, E.}, \au{Casson, F.~J.}, \au{Marsden, S.~P.}, \au{Wilson,
  T.}, \au{Brunetti, D.}, \au{Fox, P.}, \au{Freethy, S.~J.}, \au{Hender,
  T.~C.}, \au{Henderson, S.~S.}, \au{Hudoba, A.}, \au{Kirov, K.~K.},
  \au{Koechl, F.}, \au{Meyer, H.}, \au{Muldrew, S.~I.}, \au{Olde, C.},
  \au{Patel, B.~S.}, \au{Roach, C.~M.}, \au{Saarelma, S.} \& \au{Xia, G.}}
  \yr{2024}  \at{Flat-top plasma operational space of the step power plant}.
  \jt{Nucl. Fusion}  \bvol{64}~(10),  \pg{106030}.

\bibitem[Waltz {\em et~al.\/}(2007)Waltz, Candy \& Fahey]{waltz07}
{\sc \au{Waltz, R.~E.}, \au{Candy, J.} \& \au{Fahey, M.}} \yr{2007}
  \at{Coupled ion temperature gradient and trapped electron mode to electron
  temperature gradient mode gyrokinetic simulations}.  \jt{Phys. Plasmas}
  \bvol{14}~(5),  \pg{056116}.

\bibitem[Watanabe \& Sugama(2004)]{watanabe04}
{\sc \au{Watanabe, T.-H.} \& \au{Sugama, H.}} \yr{2004}  \at{Kinetic simulation
  of steady states of ion temperature gradient driven turbulence with weak
  collisionality}.  \jt{Phys. Plasmas}  \bvol{11}~(4),  \pg{1476--1483}.

\bibitem[White(2014)]{white14_book}
{\sc \au{White, R.~B.}} \yr{2014} {\em The theory of toroidally confined
  plasmas\/}, 3rd edn.  \publ{London: Imperial College Press}.

\bibitem[Xu \& Rosenbluth(1991)]{xu91}
{\sc \au{Xu, X.~Q.} \& \au{Rosenbluth, M.~N.}} \yr{1991}  \at{Numerical
  simulation of ion-temperature-gradient-driven modes}.  \jt{Phys. Fluids B}
  \bvol{3}~(3),  \pg{627--643}.

\bibitem[Zocco {\em et~al.\/}(2015)Zocco, Loureiro, Dickinson, Numata \&
  Roach]{zocco15}
{\sc \au{Zocco, A.}, \au{Loureiro, N.~F.}, \au{Dickinson, D.}, \au{Numata, R.}
  \& \au{Roach, C.~M.}} \yr{2015}  \at{Kinetic microtearing modes and
  reconnecting modes in strongly magnetised slab plasmas}.  \jt{Plasma Phys.
  Control. Fusion}  \bvol{57}~(6),  \pg{065008}.

\bibitem[Zocco \& Schekochihin(2011)]{zocco11}
{\sc \au{Zocco, A.} \& \au{Schekochihin, A.~A.}} \yr{2011}  \at{Reduced
  fluid-kinetic equations for low-frequency dynamics, magnetic reconnection,
  and electron heating in low-beta plasmas}.  \jt{Phys. Plasmas}
  \bvol{18}~(10),  \pg{102309}.

\bibitem[Zocco {\em et~al.\/}(2018)Zocco, Xanthopoulos, Doerk, Connor \&
  Helander]{zocco18}
{\sc \au{Zocco, A.}, \au{Xanthopoulos, P.}, \au{Doerk, H.}, \au{Connor, J.~W.}
  \& \au{Helander, P.}} \yr{2018}  \at{Threshold for the destabilisation of the
  ion-temperature-gradient mode in magnetically confined toroidal plasmas}.
  \jt{J. Plasma Phys.}  \bvol{84}~(1),  \pg{715840101}.

\end{thebibliography}

\end{document}